\documentclass[british,nofootinbib, 11pt,a4paper]{revtex4-2}
\usepackage[T1]{fontenc}
\usepackage[latin9]{inputenc}
\usepackage{mathrsfs}
\usepackage{mathtools}
\usepackage{amsmath}
\usepackage{amsthm}
\usepackage{amssymb}
\usepackage{esint}
\providecommand\notationname{Remark}
\theoremstyle{remark}
\newtheorem*{notation*}{\protect\notationname}
\providecommand\remarkname{Remark}
\newtheorem*{rem*}{\protect\remarkname}
\usepackage{babel}
\begin{document}
\title{Comments on Celestial CFT and $AdS_{3}$ String Theory}
\author{Igor Mol}
\affiliation{Federal University of Juiz de Fora, Minas Gerais, Brazil}
\email{igormol@ime.unicamp.br}

\date{15 October 2026}
\begin{abstract}
In a recent work, \citet{ogawa2024celestial} proposed a model for
celestial conformal field theory (CFT) based on the $\mathrm{H}^{+}_{3}$-Wess-Zumino-Novikov-Witten
(WZNW) model. In this paper, we extend the model advanced by \citet{ogawa2024celestial},
demonstrating how it can holographically generate tree-level MHV scattering
amplitudes for both gluons and gravitons when analytically continued
to the ultra-hyperbolic four-dimensional Klein space $\mathbf{K}\cong\mathbf{R}^{2,2}$,
thereby offering an alternative to celestial Liouville theory. We
construct a holographic dictionary in which vertex operators and conformal
primaries in celestial CFT are derived from their worldsheet counterparts
in Euclidean $\mathrm{AdS}_{3}$ (bosonic) string theory. Within this
dictionary, we derive the celestial stress-energy tensor, compute
the two- and three-point functions, and determine the celestial operator
product expansion (OPE). Additionally, we derive a system of partial
differential equations that characterises the celestial amplitudes
of our model, utilising the Knizhnik--Zamolodchikov (KZ) equations
and worldsheet Ward identities. In the Appendix, we provide a concise
introduction to the $\mathrm{H}^{+}_{3}$-WZNW model, with emphasis
on its connection to Euclidean $\mathrm{AdS}_{3}$ string theory.
\end{abstract}
\maketitle
\newpage{}

\tableofcontents{}

\section{Introduction}

It has been recently observed in a series of works by \citet{stieberger2023yang,stieberger2023celestial,giribet2024remarks,melton2024celestial,mol2024holographic},
within the framework of what may be termed ``celestial Liouville theory,''
that one can holographically derive the celestial amplitudes for gluons
in pure Yang-Mills theory and for gravitons in General Relativity
(GR), formulated perturbatively on a flat spacetime background, from
the correlation functions of Liouville vertex operators in the semiclassical
limit $b\rightarrow0$ of Liouville field theory. These Liouville
vertex operators are appropriately dressed by an $SO\left(2N\right)$
level-one Kac-Moody current algebra on the celestial sphere, with
the holographic correspondence understood in the large-$N$ limit.

However, drawing upon the work of \citet{teschner1997conformal,teschner1999mini,teschner1999structure,teschner2000operator}
and \citet{ribault2005h3+}, it is well known that the semiclassical
limit of Liouville field theory correlation functions coincides with
the mini-superspace limit of the $H^{+}_{3}$-WZNW model. In the context
of celestial conformal field theory (CFT), this raises the question
of whether this coincidence hints at a deeper connection or whether
there exists a more compelling model for celestial CFT based on the
$H^{+}_{3}$-WZNW model instead of Liouville theory.

The purpose of this work is to explore this question, and we shall
affirmatively demonstrate that the $H^{+}_{3}$-WZNW model provides
a more robust foundation for celestial CFT. Building on the insights
of \citet{de1999string,giveon1998comments}, and interpreting the
$H^{+}_{3}$-WZNW model as the Euclidean continuation of $AdS_{3}$
(bosonic) string theory, we propose a holographic dictionary wherein
the vertex operators, correlation functions, and stress-energy tensor
of celestial CFT are unambiguously derived from their counterparts
in Euclidean $AdS_{3}$ string theory. We will demonstrate in detail
how, via this holographic dictionary, one can derive the tree-level
celestial amplitudes for the scattering of MHV gluons and gravitons,
analytically continued to the ultra-hyperbolic Klein space $\mathbf{R}^{2}_{2}$,
within the framework of celestial leaf amplitudes as introduced by
\citet{casali2022celestial,melton2023celestial}\footnote{A point requiring special attention is the incorporation of helicity.
We will show that the leading-trace, large-level limit of the correlators
of vertex operators, constructed by dressing the scalar conformal
primaries of the $H^{+}_{3}$-WZNW model with a level-one $\mathrm{SO}(N)$
Kac-Moody current algebra, reproduces a scalar $\mathrm{AdS}_{3}$
Feynman-Witten contact diagram multiplied by the cyclic Parke-Taylor
denominator. However, to account for the helicity numerator, we follow
a suggestion of \citet{nair2005note} and further dress these vertex
operators with \emph{auxiliary anti-commuting helicity variables}.
Acting on the resulting leading-trace, large-level limit of the correlation
functions with a \emph{helicity projector} then yields the desired
tree-level MHV celestial leaf amplitudes for gluons and gravitons.}. Furthermore, as an application of our dictionary, we shall investigate
the exact celestial two- and three-point functions, as well as the
structure constants of the celestial operator product expansion (OPE).
Additionally, we will derive a system of partial differential equations
(PDEs) characterising the celestial amplitudes in the frequency-momentum
representation, following from our dictionary via the Knizhnik--Zamolodchikov
(KZ) equations and worldsheet Ward identities. 

It is worth noting that a similar model based on the $H^{+}_{3}$-WZNW
model was recently proposed by \citet{ogawa2024celestial}, and we
have benefited greatly from their insights. However, there are significant
differences between their model and the one we present. Specifically,
the hyperboloid $H^{+}_{3}$ in \citet{ogawa2024celestial}'s work
corresponds to the hyperbolic foliation of momentum space, whereas
in our construction, the hyperboloid associated with the string target
space is identified with the hyperbolic foliation of Klein space $\mathbf{R}^{2}_{2}$,
as discussed in \citet{casali2022celestial,melton2023celestial}.
This distinction introduces a crucial difference: in \citet{ogawa2024celestial}'s
model, the authors were compelled to define the celestial conformal
primaries such that the celestial three-point amplitude vanishes,
a constraint imposed by Minkowski-space kinematics. In contrast, since
we work in the split signature of Klein space, we are not subject
to this vanishing constraint, and indeed, we shall analyse the three-point
function in our celestial CFT and its relationship to the structure
constants of the celestial OPE.

Another notable distinction is that \citet{ogawa2024celestial} omitted
consideration of the worldsheet variables in their construction of
the final form of the correlation functions. In contrast, our holographic
dictionary prescribes that one must integrate over the worldsheet
coordinates to obtain a celestial amplitude from the correlation functions
of $AdS_{3}$ string theory. As we shall demonstrate in detail, the
factors involving products of gamma functions, arising from worldsheet
integrals, play a key role in determining the structure constants
of the celestial OPE.

This paper is structured as follows. In Section \ref{sec:Holographic-Reconstruction-of},
we begin by stating the postulates of our holographic dictionary,
and move to the derivation of its consequences, starting with the
holographic derivation of tree-level MHV amplitudes for both pure
Yang-Mills theory and Einstein's gravity. In Section \ref{sec:Correlation-Functions.-Operator},
we will demonstrate how our construction uniquely determines the two-
and three-point functions as well as the celestial OPE. Additionally,
we will examine a system of partial differential equations characterising
the celestial amplitudes, derived from our holographic dictionary
in frequency space, and, in Section \ref{sec:Discussion}, we summarise
our findings and discuss potential avenues for further research stemming
from this work.

\textcompwordmark{}

\paragraph*{Postscript: Note added in proof}

After the first version of the present manuscript had been completed,
\citet{donnay2025parafermions} proposed an alternative formulation
to generate the tree-level MHV leaf amplitudes for gluons utilising
non-compact parafermions appertaining to the $\mathrm{SL}(2,\mathbf{R})/\mathrm{U}(1)$
coset model, dressed by a level-one $\mathrm{SO}(N_{c})$ Kac-Moody
current algebra. In the semiclassical, large-$N_{c}$ limits, the
construction proposed by \citet{donnay2025parafermions} reproduces
the dressed-Liouville theory description by means of the $\mathrm{H}^{+}_{3}\text{-WZNW}$/Liouville
theory correspondence, where the maximal spectral-flow violation plays
a key role in their construction. The gluon helicities are represented
with the same current algebra, without the need to introduce additional
negative-weight fermions, as used in the dressed-Liouville theory
model for celestial CFT. Therefore, the proposal of \citet{donnay2025parafermions}
gives a complementary construction of the MHV leaf amplitudes in the
framework provided by the WZNW model.

\subsubsection{Outline of the Holographic Dictionary}

According to the holographic dictionary proposed in the present manuscript,
the translation between the observables belonging to gauge theory
and general relativity, defined perturbatively on four-dimensional
Minkowski spacetime, and the observables constructed from the data
provided by the $\mathrm{H}^{+}_{3}\text{-WZNW}$ model, inspired
by Euclidean $\mathrm{AdS}_{3}$ string theory, is accomplished in
two parts. The first part concerns the application of the formalism
of celestial leaf amplitudes to recast the entries of the four-dimensional
$\mathcal{S}$-matrix into a form that can be reproduced as a certain
limit (which we shall precisely specify in the subsequent discussion)
from the correlators of the $\mathrm{H}^{+}_{3}\text{-WZNW}$ model.
The second part consists in identifying the celestial vertex operators
for gluons and gravitons in the perturbative celestial CFT conjecturally
dual to gauge theory and general relativity at tree level from the
objects naturally defined from the $\mathrm{H}^{+}_{3}\text{-WZNW}$
model, such as its conformal primary fields and boundary Kac-Moody
current algebras. The principal claim of the present manuscript is
a proposed correspondence between a subset of the observables of the
celestial CFT and the minisuperspace limit of the helicity-projected
correlators of appropriately dressed, worldsheet-integrated vertex
operators constructed from the conformal primaries of the $\mathrm{H}^{+}_{3}\text{-WZNW}$
model.

Let us begin by specifying the first part of the dictionary. The initial
data consists of four-dimensional scattering amplitudes in the MHV
sub-sectors of gauge theory and general relativity at tree level in
Minkowski spacetime $\mathbf{M}\cong\mathbf{R}^{1,3}$. We then translate
these amplitudes from their momentum-space representation to the representation
spanned by the conformal primary basis, which, for the scattering
of massless particles, takes the form of a Mellin transform with respect
to the frequencies of the scattered particles. We then perform the
analytic continuation from Lorentzian signature to neutral, or Kleinian,
signature, thereby arriving at Kleinian celestial amplitudes. We subsequently
foliate the timelike $\mathbf{W}^{\mathrm{T}}$ and spacelike $\mathbf{W}^{\mathrm{S}}$
wedges of Klein space $\mathbf{K}\cong\mathbf{R}^{2,2}$, away from
the null cone, by hyperbolic leaves, each of which is diffeomorphic
to three-dimensional Lorentzian anti-de Sitter space with periodically
identified time, given by the manifold $\mathrm{AdS}_{3}/\mathbf{Z}$.
In each wedge, this allows us to decompose the original Kleinian celestial
amplitudes in four-dimensional space into a radial integral and three-dimensional
integrals over the leaves of the hyperbolic foliation of Klein space.
These three-dimensional integrals, together with the appropriate kinematical
factors, define the celestial leaf amplitudes, which admit a representation
as conformal correlators on the two-dimensional conformal boundary
of $\mathrm{AdS}_{3}/\mathbf{Z}$, namely the Lorentzian celestial
torus $\mathcal{CT}$. 

However, since our final objective is to reproduce those observables
in terms of quantities naturally defined by the $\mathrm{H}^{+}_{3}\text{-WZNW}$
model, we must further apply a second analytic continuation to the
celestial amplitudes from the Lorentzian signature of $\mathrm{AdS}_{3}/\mathbf{Z}$
to arrive at the Euclidean signature of the three-dimensional Lobachevsky
space $\mathrm{H}^{+}_{3}$, which is realised as the upper sheet
of the unit hyperboloid embedded in $\mathbf{R}^{1,3}$. This second
analytic continuation requires a careful consideration of the choices
of branches, integration contours, phases and $\mathtt{i}\varepsilon$-prescriptions.
For clarity, the leaf amplitudes analytically continued to the Lobachevsky
space $\mathrm{H}^{+}_{3}$ will be called \emph{Euclidean leaf correlators}. 

The Euclidean leaf correlators are thereby the observables that we
desire to recover from the data appertaining to the $\mathrm{H}^{+}_{3}\text{-WZNW}$
model. Before proceeding to the second part of the dictionary, let
us recall from \citet{melton2023celestial} that the original distributional
Kleinian celestial amplitudes are reconstructed from their leaf counterparts
by performing a radial integration of the leaf amplitudes and by summing
the contributions originating from the timelike $\mathbf{W}^{\mathrm{T}}$
and spacelike $\mathbf{W}^{\mathrm{S}}$ wedges of Klein space.

Now we can discuss the second part of the dictionary, which consists
in identifying the celestial vertex operators from the data belonging
to the $\mathrm{H}^{+}_{3}\text{-WZNW}$ model, and in introducing
a procedure by means of which the Euclidean celestial amplitudes can
be generated from their correlators. A spinless celestial vertex operator
$\mathcal{O}_{\Delta}$ with conformal dimension $\Delta=1+\mathtt{i}\lambda$
is constructed from a conformal primary field $\Phi^{j}$ of the $\mathrm{H}^{+}_{3}\text{-WZNW}$
model with isospin $j$, and the celestial conformal dimension is
determined by the $\mathrm{H}^{+}_{3}\text{-WZNW}$ isospin by $\Delta_{j}=-2j=1-2\mathtt{i}s$.
The primary field of the $\mathrm{H}^{+}_{3}\text{-WZNW}$ model is
subsequently dressed by a boundary Kac-Moody current algebra $J^{\mathsf{a}}(x)$,
where the boundary means the boundary of the Lobachevsky space $\mathrm{H}^{+}_{3}$,
which is diffeomorphic to $\mathrm{S}^{2}$ and charted by the holomorphic
coordinate $x$. Furthermore, we also dress the conformal primaries
by Grassmann-odd auxiliary variables modelling the helicity state
of the boson under consideration. We implement this part of the proposed
holographic dictionary in subsection II.B.2, where we identify the
celestial gluon operators, and in subsection II.C.2, where we construct
the celestial graviton operators. Finally, the desired Euclidean celestial
leaf amplitudes for gluons and gravitons in the MHV sub-sectors of
gauge theory and general relativity are generated by taking the minisuperspace
limit of the correlators of the celestial vertex operators. It is
also necessary to apply a projector operator to project the helicity
states of the scattered bosons, which we encoded by the Grassmann-valued
auxiliary variables.

We can therefore summarise the limits in which the proposed holographic
dictionary works as follows. From the viewpoint of four-dimensional
bulk physics, we are working in the MHV sub-sectors of Yang-Mills
theory and general relativity at tree level. In the case of gauge
theory, we should also add that we are working in the large-$N_{c}$
regime. From the viewpoint of the $\mathrm{H}^{+}_{3}\text{-WZNW}$
model, we are taking the large-level, or minisuperspace, limit, and
applying a helicity projector operator. Let us close our outline of
the character of the proposed dictionary by adding that, at present,
we are currently finalising a forthcoming manuscript in which we have
made a significant development according to which quantum-mechanical
corrections to gauge-theory amplitudes, namely, loops in the planar
regime of the theory, can be generated from the finite-$k$ contribution
of the correlators of the $\mathrm{H}^{+}_{3}\text{-WZNW}$ model,
thus confirming our intuition in \citet{mol2024partial}.

\subsubsection{The Question of Unitarity}

\paragraph{Note.}

The following section was originally written in response to my referee
at \emph{The European Physical Journal C}, and I must thank the referee
for encouraging me to include it in my manuscript.

\textcompwordmark{}

Now, let me explain how the non-unitarity expected from the celestial
CFT is compatible with the proposed holographic dictionary based on
the $\mathrm{H}^{+}_{3}\text{-WZNW}$ model. To make my argument unambiguous,
I believe it should be helpful to begin by distinguishing some closely
related but nonetheless distinct notions: the unitarity of the worldsheet-matter
CFT appertaining to the $\mathrm{H}^{+}_{3}\text{-WZNW}$ model; the
absence of reflection positivity underlying the celestial CFT, in
the sense expected from conventional radial quantisation; and the
unitarity of the action of $\mathrm{SL}(2,\mathbf{C})$ on the zero-mode
Hilbert space $L^{2}(\mathrm{H}^{+}_{3})$ of the $\mathrm{H}^{+}_{3}\text{-WZNW}$
model in the group-theoretic sense.

First, consider the following statement concerning the target space
of the $\mathrm{H}^{+}_{3}\text{-WZNW}$ model:
\begin{equation}
\mathrm{H}^{+}_{3}\cong\frac{\mathrm{SO}^{+}(1,3)}{\mathrm{SO}(3)}\cong\frac{\mathrm{SL}(2,\mathbf{C})}{\mathrm{SU}(2)}\,.\label{eq:-8}
\end{equation}
Geometrically, the meaning of the above assertion is that the target
space underlying the $\mathrm{H}^{+}_{3}\text{-WZNW}$ model is a
homogeneous space. Nonetheless, we should emphasise that the $\mathrm{H}^{+}_{3}\text{-WZNW}$
model belongs to a \emph{different} class of conformal field theories
than that defined by the conventional $G/H$ coset construction. 

In fact, in the ordinary vector-gauged realisation of a $G/H$ coset
CFT, we begin with a WZNW model with target space $G$ and a global
symmetry $G_{\mathrm{L}}\times G_{\mathrm{R}}$. Subsequently, we
gauge a diagonally embedded subgroup $H\subset G_{\mathrm{L}}\times G_{\mathrm{R}}$.
Now, by applying this construction with $G=\mathrm{SO}^{+}(1,3)$
and $H=\mathrm{SO}(3)$, we arrive at a model that does not have an
$\mathrm{SO}^{+}(1,3)$ symmetry, because the symmetry described by
$\mathrm{SO}^{+}(1,3)_{L}\times\mathrm{SO}^{+}(1,3)_{\mathrm{R}}$
is broken by gauging the diagonally embedded $\mathrm{SO}(3)$ subgroup.
In other words, \emph{the full original Lorentz symmetry does not
survive as a global symmetry of the gauged model}. 

On the other hand, the model defined by performing the analytic continuation
of the $\mathrm{SU}(2)$ WZNW model is a sigma-model possessing $\mathrm{H}^{+}_{3}$
as target space, which \emph{does} have an $\mathrm{SO}^{+}(1,3)$
symmetry. Therefore, even if the $\mathrm{H}^{+}_{3}\text{-WZNW}$
model is a sigma-model whose target space is a homogeneous space $\mathrm{SO}^{+}(1,3)/\mathrm{SO}(3)$
or $\mathrm{SL}(2,\mathbf{C})/\mathrm{SU}(2)$, it is not an $\mathrm{SO}^{+}(1,3)/\mathrm{SO}(3)$
or $\mathrm{SL}(2,\mathbf{C})/\mathrm{SU}(2)$ coset model in the
sense of the conventional vector-gauged realisation of the coset construction,
explained, for example, in \citet{francesco1997conformal}. See, for
example, subsection 5.2 of \citet{murthy2026localization} or the
introduction of \citet{gawedzki1992non}.

The physical significance of the preceding observations is that the
familiar unitarity properties underlying the coset CFTs, originating
from compact WZNW models at positive integer levels, should not be
inferred merely from the fact that the target manifold is a homogeneous
space. In especial, the worldsheet-matter CFT appertaining to the
$\mathrm{H}^{+}_{3}\text{-WZNW}$ model, on which our proposed holographic
dictionary is based, is \emph{non-unitary}. The reason is that the
$\mathrm{H}^{+}_{3}\text{-WZNW}$ model can be understood as originating
from the analytic continuation of the $\mathrm{SU}(2)$ WZNW model,
and the procedure of analytic continuation multiplies the Wess-Zumino
term by the imaginary unity, thereby making that contribution imaginary
when the worldsheet has Lorentzian signature, a point emphasised in
section 5 of \citet{murthy2026localization}. The resulting worldsheet-matter
CFT thereby lacks the reality properties required for conventional
unitarity.

Note that there exists no contradiction between the above remarks
and the positivity of the zero-mode sector of the $\mathrm{H}^{+}_{3}\text{-WZNW}$
model. Recall that the Hilbert state-space of the zero-mode sector
is
\begin{equation}
\mathscr{H}_{\text{zero-mode}}=L^{2}(\mathrm{H}^{+}_{3},\mathsf{d}h)\,,
\end{equation}
and carries a unitary $\mathrm{SL}(2,\mathbf{C})$ representation.
However, the contribution originating from the WZ term vanishes in
the zero-mode truncation. Thus, positivity of the zero-mode inner
product does not imply reflection positivity of the complete worldsheet-matter
CFT.

\subsubsection{The Puzzle of the Central Charge}

Now we should address the following puzzle. First, recall that the
existence of the extended BMS algebra (see \citet{BarnichTroessaert2010}),
possessing two centre-less Virasoro sub-algebras, served as a motivation
(see \citet{PasterskiShao2017}) for the reformulation of the scattering
amplitudes in perturbative quantum gravity as the correlation functions
of a putative CFT residing on the celestial sphere at null infinity,
referred to as the ``celestial CFT.'' A significant consequence
of this reformulation was the re-interpretation (see \citet{KapecEtAl2014Virasoro}
and \citet{KapecEtAl2017StressTensor}) of the sub-leading soft gravity
theorem as a conformal Ward identity for the physical celestial stress-tensor
$\boldsymbol{T}^{\mathrm{CCFT}}_{zz}$. 

On the other hand, in the present manuscript, we shall propose that
the large-level limit of the correlation functions of a set of appropriately
dressed vertex operators, constructed from the conformal primary fields
of the $\mathrm{H}^{+}_{3}\text{-WZNW}$ model, reproduces the tree-level
MHV leaf amplitudes for gravitons, thereby suggesting a candidate
model for the perturbative celestial CFT approximating the tree-level
MHV sub-sector of general relativity defined perturbatively around
flat space. However, the boundary-spacetime Virasoro algebra realised
by the string theory construction on Euclidean $\mathrm{AdS}_{3}$,
which resides in the boundary of the conformal completion of the three-dimensional
Lobachevsky space $\mathrm{H}^{+}_{3}$, has a non-zero central charge
related to the three-dimensional Newton constant. 

Therefore, we must clarify how our proposal is consistent with the
fact that, as argued by \citet{donnay2021bms}, under a set of boundary
conditions and flux prescriptions, the classical BMS flux algebra
can be shown to close under the standard Peierls bracket without a
central term, and that the corresponding celestial stress-energy tensor
OPE has a vanishing Virasoro central charge.

Our answer to this question is based on an observation due to \citet{Nguyen2022},
and in order to explain it precisely, we should briefly recall some
geometric notions about asymptotically flat spacetimes.

Let $u$ denote a coordinate function associated to a generator of
null infinity $\mathscr{I}$, and consider an oriented system of real
coordinates $x^{A}$ charting the corresponding celestial cuts. We
may accordingly represent the metric tensor on the celestial cuts
by $q_{AB}=\mathrm{e}^{2\psi}q_{0,AB}$, where $q_{0,AB}$ is a reference
metric possessing constant curvature, whereas $\partial_{u}\psi=0$.
In addition, let $\mathrm{D}_{A}$ be the Levi-Civita covariant derivative
operator, metrically compatible with $q_{AB}$, and for any given
tensor $X_{AB}$, we define its trace-free part by
\begin{equation}
(X_{AB})^{\mathrm{TF}}\coloneqq X_{AB}-\frac{1}{2}q_{AB}q^{CD}X_{CD}\,.
\end{equation}
We now introduce the lift to null infinity of a symmetric, trace-free
tensor residing on the celestial cuts, given by
\[
\mathcal{T}_{AB}\coloneqq2\,(\mathrm{D}_{A}\mathrm{D}_{B}\psi+\mathrm{D}_{A}\psi\,\mathrm{D}_{B}\psi)^{\mathrm{TF}}\,,\quad\mathcal{T}_{u\mu}\coloneqq0\,.
\]
Thus, given a local complex conformal coordinate $z$ charting a coordinate
neighbourhood of the celestial sphere, in terms of which $q_{AB}\mathrm{d}x^{A}\mathrm{d}x^{B}=\mathrm{e}^{2\sigma}\mathrm{d}z\mathrm{d}\overline{z}$,
we have
\begin{equation}
\mathcal{T}_{zz}=2\,(\partial^{2}_{z}\psi-2(\partial_{z}\sigma)(\partial_{z}\psi)+(\partial_{z}\psi)^{2})\,.
\end{equation}

Next, let $\boldsymbol{N}^{zz}$ denote the covariant news tensor
with indices raised with respect to $q_{AB}$, so that $\boldsymbol{N}^{zz}=q^{zA}q^{zB}\boldsymbol{N}_{AB}$,
which describes gravitational radiation through $\mathscr{I}$. In
what follows, we shall specialise our discussion to the round-sphere
conformal frame and set the super-translation mode to zero. We denote
by $\gamma$ the conformal factor appertaining to the round-sphere
metric, so that $q_{AB}\mathrm{d}x^{A}\mathrm{d}x^{B}=\gamma^{2}\mathrm{d}z\mathrm{d}\overline{z}$.
Then, the physical celestial stress-tensor can be stated as
\begin{equation}
\boldsymbol{T}^{\text{CCFT}}_{zz}=-\frac{1}{32\pi G_{\mathrm{N}}}\int_{\mathscr{C}}\frac{\mathrm{d}\overline{z}}{2\pi\mathrm{i}}\int_{\mathbf{R}}\mathrm{d}u\;u\gamma^{2}\,\bigl(\mathrm{D}^{3}_{z}-2\mathcal{T}_{zz}\mathrm{D}_{z}-(\mathrm{D}_{z}\mathcal{T}_{zz})\bigr)\,\boldsymbol{N}^{zz}\,.
\end{equation}
Note that, in the preceding equality, the contour integral $\int_{\mathscr{C}}\frac{\mathrm{d}\overline{z}}{2\pi\mathrm{i}}$
is utilised in order to extract the holomorphic celestial current
density. 

Now, recall that the holographic dictionary which we carefully explained
in the preceding subsection exists within the context of the uplifting
of the $\mathrm{AdS}_{3}/\mathrm{CFT}_{2}$ correspondence to flat-space
holography, in the sense of \citet{BallEtAl2019}. Accordingly, following
\citet{Nguyen2022}, we shall argue that there is nothing in the geometry
underlying this uplift which compels us to identify the physical celestial
stress-tensor $\boldsymbol{T}^{\mathrm{CCFT}}_{zz}$ with the boundary-spacetime
Virasoro stress-tensor $\boldsymbol{t}_{\mathrm{H}^{+}_{3}}$ originating
from the string theory construction on Euclidean $\mathrm{AdS}_{3}$
space. To see why, let $c_{\mathrm{H}^{+}_{3}}$ be the central charge
of the boundary-spacetime Virasoro algebra of the underlying string-theory
construction, and let $|\Psi\rangle$ be a boundary state whose associated
semi-classical geometry belongs to the family of locally super-rotated
vacua. Then, following \citet{BalasubramanianKraus1999} and \citet{Nguyen2022},
it can be shown that
\begin{equation}
\frac{\langle\Psi|\boldsymbol{t}_{\mathrm{H}^{+}_{3}}|\Psi\rangle}{\langle\Psi|\Psi\rangle}\sim\frac{c_{\mathrm{H}^{+}_{3}}}{12}\mathcal{T}_{zz}[q_{AB},\psi]\,,\label{eq:-9}
\end{equation}
which is non-zero for the non-trivial Schwarzian profiles, in the
family of states to which $|\Psi\rangle$ belongs. In Eq. (\ref{eq:-9}),
we should evaluate $\mathcal{T}_{zz}$ on the geometric data $\{q_{AB},\psi\}$,
associated with the locally super-rotated vacuum corresponding to
$|\Psi\rangle$, which we introduced above.

Therefore, as emphasised by \citet{Nguyen2022}, the preceding result
may be physically interpreted as saying that, under the uplifting
of the $\mathrm{AdS}_{3}/\mathrm{CFT}_{2}$ correspondence to the
flat-space holography, the classical expectation value of the boundary
stress tensor $\boldsymbol{t}_{\mathrm{H}^{+}_{3}}$, with respect
to boundary states whose semiclassical geometry can be realised as
a locally super-rotated vacuum, is mapped into $\mathcal{T}_{zz}$
rather than the physical celestial stress-tensor $\boldsymbol{T}^{\mathrm{CCFT}}_{zz}$
. Furthermore, the shear of the super-rotated vacuum is $C^{\text{vac.}}_{zz}=u\mathcal{T}_{zz}$,
so that the covariant news becomes $\boldsymbol{N}_{zz}=\partial_{u}C^{\text{vac.}}_{zz}-\mathcal{T}_{zz}=0$,
since $\partial_{u}\mathcal{T}_{zz}=0$. Hence $\mathcal{T}_{zz}$
may be non-zero whereas the vacuum remains non-radiative. (Remember
that we specialised our discussion to the round-sphere conformal frame,
and that we assumed that the super-translation mode is zero.)

Therefore, we have no reason compelling us to identify the holographic
uplifting of the stress-tensor underlying the boundary-spacetime Virasoro
algebra, originating from the string-theory construction, with the
physical celestial stress-energy tensor. Consequently, no contradiction
follows merely from the difference between their central charges. 

\section{Holographic Construction of Tree-Level MHV Amplitudes\label{sec:Holographic-Reconstruction-of}}

We shall start our discussion in Subsection \ref{subsec:The-Holographic-Dictionary}
by postulating the entries of our holographic dictionary, which maps
correlation functions, vertex operators, and conformal primaries in
Euclidean $AdS_{3}$ (bosonic) string theory to their counterparts
in celestial CFT. This construction builds upon the model advanced
by \citet{ogawa2024celestial}. Subsequently, in Subsections \ref{subsec:Yang-Mills-Theory}
and \ref{subsec:Einstein's-Gravity}, we will demonstrate that our
dictionary enables the holographic derivation of tree-level MHV scattering
amplitudes in both Yang-Mills theory and Einstein's gravity. The results
presented herein build upon a series of works developed in the context
of $AdS_{3}/CFT_{2}$, especially those by \citet{de1999string,giveon1998comments,maldacena2001strings,maldacena2001strings2,giribet2001correlators,bars1999string},
and celestial Liouville theory, including those by \citet{stieberger2023yang,stieberger2023celestial,giribet2024remarks,melton2024celestial,mol2024holographic}.

\subsection{The Holographic Dictionary\label{subsec:The-Holographic-Dictionary}}

The first entry of our holographic dictionary draws inspiration from
the analysis presented by \citet[Section 4]{de1999string}, and posits
that to each sequence of worldsheet vertex operators $V_{1}\left(x_{1};z_{1}\right)$,
..., $V_{n}\left(x_{n};z_{n}\right)$, there corresponds a family
of celestial vertex operators $\mathcal{V}_{1}\left(x_{1}\right)$,
..., $\mathcal{V}_{n}\left(x_{n}\right)$, such that the correlation
function of the latter is expressed in terms of the correlation function
of the former as follows:
\begin{equation}
\left\langle \mathcal{V}_{1}\left(x_{1}\right)...\mathcal{V}_{n}\left(x_{n}\right)\right\rangle =\frac{1}{W}\prod^{n}_{i=1}\int d^{2}z_{i}\left\langle \left\langle V_{1}\left(x_{1};z_{1}\right)...V_{n}\left(x_{n};z_{n}\right)\right\rangle \right\rangle ,\label{eq:Zero-Entry}
\end{equation}
where $W$ represents a normalisation constant associated with the
worldsheet ``area,'' ensuring that the worldsheet integrals are well-defined
and finite. Here, $\left\langle ...\right\rangle $ denotes the correlation
function in celestial CFT, while $\left\langle \left\langle ...\right\rangle \right\rangle $
denotes the correlation function in the worldsheet CFT. This formulation
is consistent with the conventional framework of string perturbation
theory (see \citet{d1988geometry} and \citet{polchinski1998string}).
The worldsheet integrals in Eq. (\ref{eq:Zero-Entry}) represent a
key distinction between our holographic dictionary and that proposed
by \citet{ogawa2024celestial}. In contrast to their approach, where
the worldsheet variables $z_{i},\bar{z}_{i}$ are omitted in the final
form of the correlation function, we incorporate these variables by
performing explicit integration over the worldsheet. As will be demonstrated
in Subsection \ref{subsec:Operator-Product-Expansion}, the product
of gamma functions that arises from the worldsheet integrals in Eq.
(\ref{eq:Zero-Entry}) plays an important role in determining the
structure constants of the celestial operator product expansion.

The second entry of our holographic dictionary derives its inspiration
from the model proposed by \citet{ogawa2024celestial}, yet it introduces
a significant distinction arising from the split signature of Klein
space $\mathbf{R}^{2}_{2}$, to which we shall analytically continue
the scattering amplitudes of Yang-Mills theory and Einstein's gravity
in Subsections \ref{subsec:Yang-Mills-Theory} and \ref{subsec:Einstein's-Gravity}.
We propose that each conformal primary $\Phi^{j}\left(x;z\right)$
with spin $j$ in the $H^{+}_{3}$-WZNW model corresponds to a celestial
(scalar) wavefunction $\phi_{\Delta}\left(x\right)$ with conformal
weight $\Delta$ in the celestial CFT, such that the $H^{+}_{3}$-WZNW
spin and the celestial scaling dimension are related by the equation
$j+\Delta/2=0$, as suggested by \citet{ogawa2024celestial} based
on an analysis of the transformation properties of $\Phi^{j}\left(x;z\right)$
and $\phi_{\Delta}\left(x\right)$ under conformal transformations.
Thus, the principal distinction between our proposal and that of \citet{ogawa2024celestial}
lies in the signature of the spacetime: while \citet{ogawa2024celestial}
works in Minkowski signature, where Lorentzian kinematics necessitated
selecting a celestial conformal primary as a linear combination of
$\Phi^{j}\left(x;z\right)$ and its shadow transform--resulting in
a vanishing three-point function--our model works in the split signature,
freeing us from the constraint of the vanishing three-point function.
In fact, as we shall demonstrate in Subsections \ref{subsec:Three-Point-Function}
and \ref{subsec:Operator-Product-Expansion}, the three-point function
in our model is closely related to the structure constants of the
celestial operator product expansion.

Having now articulated the entries of our holographic dictionary,
we shall proceed to derive their implications. In the subsequent Subsections
\ref{subsec:Yang-Mills-Theory} and \ref{subsec:Einstein's-Gravity},
we will derive the tree-level MHV celestial amplitudes for gluons
and gravitons from the proposed dictionary. Thereafter, in Section
\ref{sec:Correlation-Functions.-Operator}, we will address the completeness
and consistency of these postulates with the mathematical formalism
of conformal field theory by computing the two- and three-point functions
and determining the structure constants of the OPE.

\subsection{Yang-Mills Theory\label{subsec:Yang-Mills-Theory}}

In this subsection, we begin by succinctly reviewing the formalism
of celestial leaf amplitudes introduced by \citet{melton2023celestial},
which prescribes an analytic continuation of celestial amplitudes
from Minkowski spacetime to the ultra-hyperbolic Klein space $\mathbf{R}^{2}_{2}$.
Following this, we discuss the holographic construction of these Klein
space amplitudes for pure Yang-Mills theory within the framework of
the holographic dictionary proposed in these notes, which establishes
a correspondence between Euclidean $AdS_{3}$ (bosonic) string theory
and celestial CFT.

\subsubsection{Celestial Leaf Amplitudes for Yang-Mills Theory}
\begin{notation*}
In what follows, we denote points on the celestial sphere by $x_{i},\bar{x}_{i}\in\mathbf{CP}^{1}$,
referred to as \emph{celestial coordinates}. Let $\nu^{A}\coloneqq\left(x,1\right)^{T}$
and $\bar{\nu}^{\dot{A}}\coloneqq\left(\bar{x},1\right)^{T}$ be a
pair of two-component spinors parametrised by the celestial coordinates
$x,\bar{x}$. The \emph{standard null vector }is then defined by $q^{\mu}\left(x,\bar{x}\right)\coloneqq(\sigma^{\mu})_{A\dot{B}}\nu^{A}\bar{\nu}^{\dot{B}}$.
For each family of two-component spinors $\mu^{A}_{i}\coloneqq\sqrt{\omega_{i}}\left(x_{i},1\right)^{T}$
and $\bar{\mu}^{\dot{A}}_{i}\coloneqq\sqrt{\omega_{i}}\left(\bar{x}_{i},1\right)^{T}$,
indexed by $i=1,...,n$, and describing the momenta of a sequence
of gluons with frequencies $\omega_{1},...,\omega_{n}$, the associated
four-momenta are given by $p^{\mu}_{i}=(\sigma^{\mu})_{A\dot{B}}\mu^{A}_{i}\bar{\mu}^{\dot{B}}_{i}=\omega_{i}q^{\mu}\left(x_{i},\bar{x}_{i}\right)$.
The spinor-helicity brackets are then defined as $\left\langle ij\right\rangle \coloneqq\varepsilon_{AB}\mu^{A}_{i}\mu^{B}_{j}$
and $\left[ij\right]\coloneqq\varepsilon_{\dot{A}\dot{B}}\bar{\mu}^{\dot{A}}_{i}\bar{\mu}^{\dot{B}}_{i}$,
such that, in our conventions, $2p_{i}\cdot p_{j}=\left\langle ij\right\rangle \left[ij\right]$,
which follows from $\text{tr}\left(\sigma^{\mu}\bar{\sigma}^{\nu}\right)=2\eta^{\mu\nu}$.
\end{notation*}
The starting point of our analysis is the Parke-Taylor formula (cf.
\citet{parke1986amplitude}; for a contemporary pedagogical review,
see \citet{elvang2013scattering,badger2024scattering}) for the tree-level
scattering amplitude $\mathcal{A}_{n}$ of MHV gluons $1^{-}$, $2^{-}$,
$3^{+}$, ..., $n^{+}$, which is expressed in spinor-helicity variables
as follows:
\begin{equation}
\mathcal{A}_{n}=\frac{\left\langle 12\right\rangle ^{4}}{\left\langle 12\right\rangle \left\langle 23\right\rangle ...\left\langle n1\right\rangle }ig^{n-2}_{\text{YM}}\delta^{\left(4\right)}\left(\sum^{n}_{i=1}p^{\mu}_{i}\right),\label{eq:Parke-Taylor}
\end{equation}
where $p^{\mu}_{i}$ denotes the four-momentum of the $i^{\text{th}}$
gluon and $g_{\text{YM}}$ is the Yang-Mills coupling constant. Utilising
the celestial holography dictionary, as elaborated upon in detail
by \citet{pasterski2021lectures,raclariu2021lectures,strominger2018lectures,puhmcelestial},
the spinor-helicity variables are connected to the frequencies $\omega_{i}$
and the complex parametrisation $x_{i},\bar{x}_{i}\in\mathbf{CP}^{1}$
of the celestial sphere through the relation: 
\begin{equation}
\left\langle ij\right\rangle =\sqrt{\omega_{i}\omega_{j}}x_{ij}.
\end{equation}

Therefore, recalling the integral representation of the $4d$ Dirac
delta function, $\delta^{\left(4\right)}\left(p\right)=1/\left(2\pi\right)^{4}\int d^{4}Ye^{ip\cdot Y}$,
and substituting the frequencies $\omega_{i}$ and celestial coordinates
$x_{i},\bar{x}_{i}$ into Eq. (\ref{eq:Parke-Taylor}), we arrive
at the expression: 
\begin{equation}
\mathcal{A}_{n}=ig^{n-2}_{\text{YM}}\frac{x^{4}_{12}}{x_{12}x_{23}...x_{n1}}\int\frac{d^{4}Y}{\left(2\pi\right)^{2}}\prod^{n}_{i=1}\omega^{\alpha_{i}}_{i}e^{i\omega_{i}q\left(x_{i},\bar{x}_{i}\right)\cdot Y},\label{eq:Step-19}
\end{equation}
where we introduce constants to streamline the notation: $\alpha_{1}=\alpha_{2}=1$
and $\alpha_{3}=...=\alpha_{n}=-1$.

Using the prescriptions outlined in the celestial holography dictionary,
the celestial amplitude corresponding to $\mathcal{A}_{n}$ is given
by the $\varepsilon$-regulated Mellin transform of Eq. (\ref{eq:Step-19}):
\begin{equation}
\widehat{\mathcal{A}}_{n}=\prod^{n}_{i=1}\int^{\infty}_{0}d\omega_{i}\omega^{\Delta_{i}-1}_{i}e^{-\varepsilon\omega_{i}}\mathcal{A}_{n},
\end{equation}
which results in:
\begin{equation}
\widehat{\mathcal{A}}_{n}=ig^{n-2}_{\text{YM}}\frac{x^{4}_{12}}{x_{12}x_{23}...x_{n1}}\int\frac{d^{4}Y}{\left(2\pi\right)^{4}}\prod^{n}_{i=1}\frac{\Gamma\left(2\rho_{i}\right)}{\left(\varepsilon-iq\left(x_{i},\bar{x}_{i}\right)\cdot Y\right)^{2\rho_{i}}},\label{eq:Step-21}
\end{equation}
where we define $\rho_{i}\coloneqq\left(\Delta_{i}+\alpha_{i}\right)/2$.

By analytically continuing the spacetime integral in Eq. (\ref{eq:Step-21})
to Klein space $\mathbf{R}^{2}_{2}$, and employing the approach described
by \citet{melton2023celestial}, we derive the expression:
\begin{align}
 & \int\frac{d^{4}Y}{\left(2\pi\right)^{4}}\prod^{n}_{i=1}\frac{\Gamma\left(2\rho_{i}\right)}{\left(\varepsilon-iq\left(x_{i},\bar{x}_{i}\right)\cdot Y\right)^{2\rho_{i}}}\\
 & =\frac{\delta\left(4-2\sum_{i}\rho_{i}\right)}{\left(2\pi\right)^{3}}\int_{\hat{y}^{2}=1}d^{3}\hat{y}\prod^{n}_{i=1}\frac{\Gamma\left(2\rho_{i}\right)}{\left(\varepsilon-iq\left(x_{i},\bar{x}_{i}\right)\cdot\hat{y}\right)}+\left(\bar{x}_{i}\leftrightarrow-\bar{x}_{i}\right).
\end{align}
Here, the notation $\left(\bar{x}_{i}\leftrightarrow-\bar{x}_{i}\right)$
denotes the repetition of the preceding term with the variables $\bar{x}_{i}$
substituted by $-\bar{x}_{i}$. Consequently, Eq. (\ref{eq:Step-21})
can be reformulated as:
\begin{equation}
\widehat{\mathcal{A}}_{n}=ig^{n-2}_{\text{YM}}\frac{\delta\left(4-2\sum_{i}\rho_{i}\right)}{\left(2\pi\right)^{3}}\frac{x^{4}_{12}}{x_{12}x_{23}...x_{n1}}\int_{\hat{y}^{2}=1}d^{3}\hat{y}\prod^{n}_{i=1}\frac{\Gamma\left(2\rho_{i}\right)}{\left(\varepsilon-iq\left(x_{i},\bar{x}_{i}\right)\cdot\hat{y}\right)}+\left(\bar{x}_{i}\leftrightarrow-\bar{x}_{i}\right).\label{eq:YM-Celestial-Amplitude}
\end{equation}
This expression for the celestial amplitude $\widehat{\mathcal{A}}_{n}$
is key in establishing that our holographic dictionary can reproduce
the tree-level MHV scattering amplitude for gluons in an appropriate
limit.

\subsubsection{Celestial Vertex Operators for Gluons\label{subsec:Celestial-Vertex-Operators}}

We now turn our attention to the vertex operators within celestial
CFT whose correlation functions generate the celestial amplitude $\widehat{\mathcal{A}}_{n}$
as derived in Eq. (\ref{eq:YM-Celestial-Amplitude}). To achieve this,
it is necessary to introduce a ``dressing'' of the worldsheet conformal
primaries $\Phi^{j}\left(x;z\right)$ in the $H^{+}_{3}$-WZNW model
by the level-one Kac-Moody currents $K^{a}\left(x\right)$ of $SO\left(2N\right)$,
residing on the boundary of $H^{+}_{3}$, which we identify with the
celestial sphere. Hereafter, we shall refer to $K^{a}\left(x\right)$
as the \emph{celestial Kac-Moody currents}.

The generators $K^{a}\left(x\right)$ of the celestial Kac-Moody current
algebra are characterised by the OPE:
\begin{equation}
K^{a}\left(x_{1}\right)K^{b}\left(x_{2}\right)\sim\frac{\delta^{ab}}{x^{2}_{12}}+\frac{if^{abc}}{x_{12}}K^{c}\left(x_{2}\right),\label{eq:OPE-3}
\end{equation}
where $f^{abc}$ denote the structure constants of $\mathfrak{so}\left(2N\right)\simeq D_{N}$.

Recalling that the dual Coxeter number of the Lie algebra $D_{N}$
is $2N-2$, we find that the KZ equations for the currents $K^{a}\left(x\right)$
take the form:
\begin{equation}
\left(\frac{\partial}{\partial x_{i}}+\frac{1}{2N-1}\sum_{j\neq i}\frac{T^{c}_{i}T^{c}_{j}}{x_{i}-x_{j}}\right)\left\langle K^{a_{1}}\left(x_{1}\right)...K^{a_{n}}\left(x_{n}\right)\right\rangle =0,\label{eq:KZ-1}
\end{equation}
where $T^{a}$ are the generators of the fundamental representation
of $SO\left(2N\right)$. Decomposing the correlation function $\left\langle K^{a_{1}}\left(x_{1}\right)...K^{a_{n}}\left(x_{n}\right)\right\rangle $
into the canonical $SO\left(2N\right)$ invariants, $I_{1}=\delta_{m_{1},m_{2}}\delta_{m_{3},m_{4}}$
and $I_{2}=\delta_{m_{1},m_{3}}\delta_{m_{2},m_{4}}$, and retaining
only the terms $T^{a}_{i}T^{a}_{j}I_{1}$ and $T^{a}_{i}T^{a}_{j}I_{2}$
proportional to $N$, we take the limit $N\rightarrow\infty$ of the
KZ equations. The asymptotic solution is:
\begin{equation}
\lim_{N\rightarrow\infty}\left\langle \mathcal{O}^{a_{1}}\left(x_{1}\right)...\mathcal{O}^{a_{n}}\left(x_{n}\right)\right\rangle =\text{tr}\left(T^{a_{1}}...T^{a_{n}}\right)\frac{1}{x_{12}x_{23}...x_{n1}}.\label{eq:Parke-Taylor-Correlation-Function}
\end{equation}

Thus, in the large-$N$ limit, the correlation functions of the celestial
Kac-Moody currents $K^{a}\left(x\right)$ generate the colour-ordered
Parke-Taylor factor (cf. \citet{parke1986amplitude}), which is associated
with the tree-level MHV scattering amplitudes of gluons. 

We then define the level-one $SO\left(2N\right)$ vertex operators
as:
\begin{equation}
V^{a}_{\rho}\left(x\right)\coloneqq\Gamma\left(\rho\right)K^{a}\left(x\right)\int d^{2}z\;\Phi^{\rho}\left(x;z\right),\label{eq:-1}
\end{equation}
where, from the perspective of celestial CFT, $\rho$ is understood
as a label associated with the vertex operators $V^{a}_{\rho}\left(x\right)$
rather than a scaling dimension. Consequently, the correlation function
of the vertex operators in the large-$N$ limit yields: 
\begin{equation}
\lim_{N\rightarrow\infty}\left\langle V^{a_{1}}_{\rho_{1}}\left(x_{1}\right)...V^{a_{n}}_{\rho_{n}}\left(x_{n}\right)\right\rangle =\frac{\text{tr}\left(T^{a_{1}}...T^{a_{n}}\right)}{x_{12}x_{23}...x_{n1}}\frac{1}{W}\int d^{2}z_{1}...d^{2}z_{n}\left\langle \left\langle \Phi^{\rho_{1}}\left(x_{1};z_{1}\right)...\Phi^{\rho_{n}}\left(x_{n};z_{n}\right)\right\rangle \right\rangle ,
\end{equation}
where $W$ is a renormalisation constant associated with the worldsheet
``area,'' required to render the worldsheet integrals well-defined
and finite. Finally, in the mini-superspace limit (see Eq. (\ref{eq:Mini-Superspace-Limit})),
wherein:
\begin{equation}
\lim_{k\rightarrow\infty}\left\langle \left\langle \Phi^{\rho_{1}}\left(x_{1};z_{1}\right)...\Phi^{\rho_{n}}\left(x_{n};z_{n}\right)\right\rangle \right\rangle =\mathcal{N}\int\frac{d\rho d\gamma d\bar{\gamma}}{\rho^{3}}\prod^{n}_{i=1}\left(\frac{\rho}{\rho^{2}+\left|x_{i}-\gamma\right|^{2}}\right)^{2\rho_{i}},\label{eq:Step-B}
\end{equation}
we arrive at:
\begin{equation}
\lim_{N\rightarrow\infty}\lim_{k\rightarrow\infty}\left\langle V^{a_{1}}_{\rho_{1}}\left(x_{1}\right)...V^{a_{n}}_{\rho_{n}}\left(x_{n}\right)\right\rangle =\mathcal{N}\frac{\text{tr}\left(T^{a_{1}}...T^{a_{n}}\right)}{x_{12}x_{23}...x_{n1}}\int_{AdS_{3}}d^{3}\hat{y}\prod^{n}_{i=1}\frac{\Gamma\left(2\rho_{i}\right)}{\left(\varepsilon-iq\left(x_{i},\bar{x}_{i}\right)\cdot\hat{y}\right)^{2\rho_{i}}}.\label{eq:Step-C}
\end{equation}

The right-hand side of the preceding equation may be identified with
a scalar $\mathrm{AdS}_{3}$ Feynman-Witten contact diagram, multiplied
by the Chan-Paton factor $\text{tr}(T^{a_{1}}...T^{a_{n}})$, and
the cyclic Parke-Taylor denominator. Nonetheless, in order to reproduce
the full celestial leaf amplitude for gluons arranged in an MHV configuration,
whose first two gluons have negative helicity and whose remaining
gluons have positive helicity, it remains necessary to account for
the helicity numerator $x^{4}_{12}$. The strategy that we now describe
for reproducing this numerator is based on an observation due to \citet{nair2005note},
namely a Berezin integral identity that we shall now recall.

Let $\{\theta^{\alpha}_{A}\}^{4}_{\alpha=1}$ be a family of Grassmann-valued
undotted van der Waerden spinors, normalised so that their Berezin
integrals satisfy
\begin{equation}
\int\mathrm{d}^{2}\theta^{\alpha}\;\theta^{\alpha}_{A}\theta^{\alpha}_{B}=\varepsilon_{AB}\,,\label{eq:}
\end{equation}
for all $\alpha=1,2,3,4$. In what follows, we denote the contraction
of $\theta^{\alpha}_{A}$ with an undotted spinor $\nu^{A}$ by $\nu\cdot\theta^{\alpha}=\nu^{A}\theta^{\alpha}_{A}$.
Let us further define a spinor $\nu^{A}(x)=(x,1)$, depending on the
holomorphic coordinate $x$ on the celestial sphere, and a family
of Grassmann-valued variables
\begin{equation}
\eta^{\alpha}(x)\coloneqq\nu(x)\cdot\theta^{\alpha}\,.
\end{equation}
Therefore, Eq. (\ref{eq:}) implies the Berezin integral identity
\begin{equation}
\int\mathrm{d}^{8}\theta\;\prod^{4}_{\alpha=1}\eta^{\alpha}(x_{r})\,\prod^{4}_{\beta=1}\eta^{\beta}(x_{s})=x^{4}_{rs}\,.\label{eq:-2}
\end{equation}
This formula naturally suggests that, to obtain the helicity numerator,
we may introduce the negative-helicity celestial gluon operators $\boldsymbol{O}^{a-}_{\rho}(x)$
by dressing the stringy vertex operators $V^{a}_{\rho}(x)$, defined
in Eq. (\ref{eq:-1}), with the Grassmann-valued factor $\prod^{4}_{\alpha=1}\eta^{\alpha}(x)$.
The positive-helicity celestial gluon operators $\boldsymbol{O}^{a+}_{\rho}(x)$
are given simply by the original vertex operators. Thus, we set
\begin{equation}
\boldsymbol{O}^{a+}_{\rho}(x)\coloneqq V^{a}_{\rho}(x),\qquad\boldsymbol{O}^{a-}_{\rho}(x)\coloneqq V^{a}_{\rho}(x)\,\prod^{4}_{\alpha=1}\eta^{\alpha}(x)\,.
\end{equation}

Let us observe that the anti-commuting quantities $\eta^{\alpha}(x)$,
and the Grassmann-valued spinors $\theta^{\alpha}_{A}$, are nothing
more than \emph{auxiliary helicity bookkeeping variables}. We emphasise
that $\eta^{\alpha}(x)$ should not be interpreted as dynamical physical
degrees of freedom, or as components of a superfield propagating on
the worldsheet. Analogously, the anti-commuting spinors $\theta^{\alpha}_{A}$
should not be identified with the fermionic directions of a target
supermanifold. In the present manuscript, we are working entirely
within the framework of \emph{bosonic} Euclidean $\mathrm{AdS}_{3}$
string theory. Therefore, the bookkeeping devices that we have just
introduced may be physically interpreted as polarisation-like variables
whose only purpose is to project the desired helicity information
of the external states.

Keeping this in mind, and following Nair's suggestion, we identify
the Berezin integration as the\emph{ helicity projector}, defined
by
\begin{equation}
\Pi^{\text{YM}}_{\text{Nair}}[\Psi]\coloneqq\int\mathrm{d}^{8}\theta\;\Psi(\theta)\,,
\end{equation}
where $\Psi$ is any function of the Grassmann spinors $\theta^{\alpha}_{A}$.

Therefore, from Eqs. (\ref{eq:YM-Celestial-Amplitude}, \ref{eq:Step-C},
\ref{eq:-2}), the tree-level MHV celestial amplitude for gluon scattering
in pure Yang-Mills theory, associated with the helicity configuration
$1^{-},2^{-},3^{+},\dots,n^{+}$, may be expressed as:
\begin{align}
\widehat{\mathcal{A}}_{n} & =ig^{n-2}_{\text{YM}}\frac{\delta\left(4-2\sum_{i}\rho_{i}\right)}{\left(2\pi\right)^{3}\mathcal{N}}\lim_{N\rightarrow\infty}\lim_{k\rightarrow\infty}\Pi^{\mathrm{YM}}_{\mathrm{Nair}}\left\langle \boldsymbol{O}^{-a_{1}}_{\Delta_{1}+\alpha_{1}}\left(x_{1}\right)\boldsymbol{O}^{-a_{2}}_{\Delta_{2}+\alpha_{2}}\left(x_{2}\right)\prod^{n}_{k=3}\boldsymbol{O}^{+a_{k}}_{\Delta_{k}+\alpha_{k}}\left(x_{k}\right)\right\rangle \label{eq:-4}\\
 & \;+\left(\bar{x}_{i}\leftrightarrow-\bar{x}_{i}\right)\,.
\end{align}

\paragraph{Remarks.}

Let us now observe that, in the preceding expression (a key entry
in our dictionary, relating the tree-level MHV celestial amplitudes
for gluons in gauge theory to the helicity projection of the leading-trace,
large-level limit of appropriately dressed celestial vertex operators),
the Yang-Mills coupling constant $g_{\text{YM}}$ enters only through
the proportionality factor. Therefore, while Eq. (\ref{eq:-4}) shows
that the kinematical structure and conformal dependence of the gluonic
celestial amplitudes can be reproduced from data pertaining to the
$H^{+}_{3}$-WZNW model, at the present stage of our proposed holographic
dictionary, the coupling parameter $g_{\text{YM}}$ has a \emph{phenomenological}
character. We may, for example, absorb $g_{\text{YM}}$ into the normalisation
of the positive-helicity celestial gluon operators $\boldsymbol{O}^{a+}_{\rho}(x)$.

Deriving $g_{\text{YM}}$ from the physical constants characterising
the $H^{+}_{3}$-WZNW model, such as the fundamental string length
$\ell_{0}$ or the $\mathrm{AdS}_{3}$ radius $\ell$, however, lies
beyond the scope of the present bottom-up approach to perturbative
celestial CFT. It would certainly be very satisfactory and mathematically
elegant to obtain the gauge-theoretic coupling constant $g_{\text{YM}}$
in the bulk Klein space $\mathbf{R}^{2}_{2}$ from critical string
theory on $\mathrm{AdS}_{3}\times X$. We expect that such a derivation,
if possible, would require detailed knowledge of the spacetime CFT
associated with $X$, and it remains a long-standing goal of our research
programme.\footnote{I am grateful to the referee for drawing my attention to the need
to clarify this important point.}

\subsection{Einstein's Gravity\label{subsec:Einstein's-Gravity}}

In this subsection, we shall explain how the preceding argument extends
to gravity. The formulation by which tree-level celestial amplitudes
for MHV gravitons can be represented as a correlation function of
\emph{Liouville} vertex operators, ``dressed'' by an $SO\left(2N\right)$
level-one Kac-Moody current algebra on the celestial sphere, has been
thoroughly analysed by \citet{mol2024holographic}. This construction
is framed within the large-$N$ semiclassical limit of Liouville field
theory, where $b\rightarrow0$. Our objective here is to demonstrate
how this formalism may be adapted to the holographic dictionary introduced
in Subsection \ref{subsec:The-Holographic-Dictionary}.

\subsubsection{Mathematical Preliminaries}

We must first undertake a brief preparatory discussion to introduce
the mathematical objects needed for our construction. For further
elaboration, the interested reader is referred to \citet{mol2024holographic}.
It is important to note that we shall employ a slightly different
notation in this discussion: the variables $z,\bar{z}$ denote coordinates
on the string worldsheet, while $x,\bar{x}$ correspond to coordinates
on the celestial sphere. To begin, let us recall from \citet{pasterski2017conformal}
that the scalar celestial wavefunction with conformal weight $\Delta=2\sigma$
is given by:
\begin{equation}
\phi_{2\sigma}\left(x;z\right)\coloneqq\frac{\Gamma\left(2\sigma\right)}{\left(\varepsilon-iq\left(x_{i},\bar{x}_{i}\right)\cdot x\right)^{2\sigma}}.
\end{equation}
Let $\mathsf{P}_{i}$ denote the operator acting on the space of celestial
wavefunctions $\{\phi_{2\sigma}\}$ by shifting the conformal weights
(with respect to the celestial CFT, and not to be confused with the
$H^{+}_{3}$-WZNW model) by one-half, such that:
\begin{equation}
\mathsf{P}_{i}\phi_{2\sigma_{j}}\left(x_{i};z_{i}\right)\coloneqq\delta_{ij}\phi_{2\left(\sigma_{j}+1/2\right)}\left(x_{i};z_{i}\right).
\end{equation}

We must also introduce a pair of chiral quasi-primary free fermions,
$\chi\left(x\right)$ and $\chi^{\dagger}\left(x\right)$, defined
on the celestial sphere by their respective mode expansions:
\begin{equation}
\chi\left(x\right)\coloneqq\sum^{\infty}_{n=0}b_{n}x^{n},\,\,\,\chi^{\dagger}\left(x\right)\coloneqq\sum^{\infty}_{n=0}b^{\dagger}_{n}x^{-n-1},
\end{equation}
where $b_{n}$ and $b^{\dagger}_{n}$ are fermionic annihilation and
creation operators, satisfying the anti-commutation relations $\{b_{m},b^{\dagger}_{n}\}=\delta_{mn}$
for all $m,n\geq0$. These operators act on the vacuum state $\big|0\rangle$
such that $b_{n}\big|0\rangle=0$ for each non-negative integer $n$.
The two-point function for this doublet is given by:
\begin{equation}
\langle\chi\left(x_{1}\right)\chi^{\dagger}\left(x_{2}\right)\rangle\coloneqq\frac{1}{x_{1}-x_{2}}.
\end{equation}
Finally, let $\nu^{A}_{i}\coloneqq\left(x_{i},1\right)^{T}$ and $\bar{\nu}^{\dot{A}}_{i}\coloneqq\left(\bar{x}_{i},1\right)^{T}$
be a pair of two-component spinors parametrising the four-momentum
of the $i$-th graviton, and let $\lambda^{A}\coloneqq\left(\lambda,1\right)^{T}\in\mathbf{C}^{2}$
be an auxiliary two-component spinor associated with the state $\big|\lambda\rangle\coloneqq\chi^{\dagger}\left(\lambda\right)\big|0\rangle$.
We have shown in \citet{mol2024holographic} that, by defining the
operators:
\begin{equation}
\mathsf{A}_{i}\coloneqq\exp\left(q\left(x,\bar{x}\right)\cdot y\,\mathsf{P}_{i}\right)\chi^{\dagger}\left(x\right)\chi\left(x\right),\mathsf{B}_{i}\coloneqq\frac{\bar{\nu}^{\dot{A}}_{i}\lambda^{A}}{\left\langle \nu_{i},\lambda\right\rangle }\frac{\partial}{\partial y^{A\dot{A}}}\exp\left(q\left(x,\bar{x}\right)\cdot y\,\mathsf{P}_{i}\right),
\end{equation}
where the four-vector $\left(y^{\mu}\right)\in\mathbf{R}^{4}$ should
be interpreted as an index rather than spacetime coordinates, the
tree-level MHV celestial amplitude $\widehat{\mathcal{M}}_{n}$ for
the scattering of $n$-gravitons can be expressed as:
\begin{align}
\widehat{\mathcal{M}}_{n} & =\left(\frac{\kappa}{2}\right)^{n-2}\frac{x^{8}_{12}}{x_{12}x_{23}...x_{n1}}\int\frac{d^{2}\lambda}{\pi}\int dy\delta\left(y\right)\langle\lambda\big|\mathsf{A}_{1}\left(\prod^{n-2}_{i=2}\mathsf{B}_{i}\right)\mathsf{A}_{n-1}\mathsf{A}_{n}\bar{\partial}\big|\lambda\rangle\\
 & \int\frac{d^{4}Y}{\left(2\pi\right)^{4}}\prod^{n}_{j=1}\phi_{2\sigma_{j}}\left(x_{j};Y\right)+\mathcal{P}\left(2,...,n-2\right),\label{eq:Celestial-1}
\end{align}
where $\mathcal{P}\left(2,...,n-2\right)$ denotes the permutation
of the symbols within the set $\left\{ 2,...,n-2\right\} $, and we
define the celestial conformal weights as follows:
\begin{equation}
\sigma_{i}\coloneqq\frac{1}{2}\left(\Delta_{i}+\alpha_{i}-1\right)\,\,\,\text{for \ensuremath{i\in\left\{ 1,n-1,n\right\} },}\label{eq:Sigma}
\end{equation}
\begin{equation}
\sigma_{i}\coloneqq\frac{1}{2}\left(\Delta_{i}+\alpha_{i}\right)\,\,\,\text{for \ensuremath{2\leq i\leq n-2}}.\label{eq:Sigma-1}
\end{equation}
Furthermore, 
\begin{equation}
\kappa^{2}=32\pi G_{N}\,,
\end{equation}
where $G_{N}$ is the four-dimensional Newton's constant on the bulk
Klein space.

\subsubsection{Celestial Vertex Operators for Gravitons}

We are now prepared to demonstrate that Eq. (\ref{eq:Celestial-1})
can be reformulated using the celestial vertex operators constructed
from $AdS_{3}$ string theory, as detailed in Subsection \ref{subsec:First-Entry:-Kac-Moody}.
We first define the ``dressed'' vertex operators in celestial CFT
for gravitons:
\begin{equation}
G_{i}\left(x_{i}\right)\coloneqq\Gamma\left(2\sigma_{i}\right)T^{a_{i}}K^{a_{i}}\left(x_{i}\right)\mathsf{A}_{i}\int d^{2}z_{i}\Phi^{2\sigma_{i}}\left(x_{i};z_{i}\right),\label{eq:Definition-2}
\end{equation}
for $i=1,n-1,n$, and:
\begin{equation}
H_{j}\left(x_{j}\right)\coloneqq\Gamma\left(2\sigma_{i}\right)T^{a_{j}}K^{a_{j}}\left(x_{j}\right)\mathsf{B}_{j}\int d^{2}z_{j}\Phi^{2\sigma_{j}}\left(x_{j};z_{j}\right),\label{eq:Definition-3}
\end{equation}
for $j=2,...,n-2$, where $K^{a}\left(x\right)$ denotes the celestial
Kac-Moody currents introduced in Subsection \ref{subsec:Celestial-Vertex-Operators},
and obeying the OPE (\ref{eq:OPE-3}). We emphasise that the order
of the operators $\mathsf{A}_{i}$ and $\mathsf{B}_{j}$ in Eqs. (\ref{eq:Definition-2},
\ref{eq:Definition-3}) is important, as the worldsheet conformal
primaries $\Phi^{2\sigma_{i}}\left(x_{i};z_{i}\right)$ inherit the
action of the operator $\mathsf{P}_{i}$ due to their dependence on
the celestial conformal weights, as given by Eqs. (\ref{eq:Sigma},
\ref{eq:Sigma-1}).

As in the preceding subsection, it is also necessary to introduce
auxiliary anti-commuting variables to generate the helicity numerator
$x^{8}_{12}$ required by the BGK formula. In the gravitational case,
we shall employ a collection $\{\xi^{\hat{\alpha}}_{A}\}^{8}_{\hat{\alpha}=1}$
of Grassmann-valued van der Waerden spinors normalised by
\begin{equation}
\int\mathrm{d}^{2}\xi^{\hat{\alpha}}\;\xi^{\hat{\alpha}}_{A}\xi^{\hat{\alpha}}_{B}=\varepsilon_{AB}\,,\label{eq:-3}
\end{equation}
where the indices $\hat{\alpha}=1,\dots,8$ in the gravitational case
are adorned with a hat in order to distinguish them unambiguously
from their gluonic counterparts. From these variables we define the
auxiliary anti-commuting helicity variables
\begin{equation}
\psi^{\hat{\alpha}}(x)\coloneqq\nu(x)\cdot\xi^{\hat{\alpha}}\,,
\end{equation}
where $\nu^{A}(x)=(x,1)$ and $x$ is the holomorphic coordinate on
the celestial sphere, precisely as in the preceding subsection. Let
us emphasise that $\psi^{\hat{\alpha}}(x)$ should not be regarded
as a fermionic physical degree of freedom propagating on the worldsheet,
nor should the underlying $\xi^{\hat{\alpha}}_{A}$ be identified
with the anti-commuting coordinates parameterising the ``soul''
of a target superspace. These variables play an entirely \emph{auxiliary
role}. Their only purpose is to project the helicity component of
the external graviton states.

Accordingly, Nair's observation for gravitons follows from Eq. (\ref{eq:-3})
and reads
\begin{equation}
\int\mathrm{d}^{16}\xi\;\prod^{8}_{\alpha=1}\psi^{\alpha}(x_{r})\,\prod^{8}_{\beta=1}\psi^{\beta}(x_{s})=x^{8}_{rs}\,.
\end{equation}
The helicity projector for gravitons is then defined by
\begin{equation}
\Pi^{\mathrm{GR}}_{\text{Nair}}[\Psi]\coloneqq\int\mathrm{d}^{16}\xi\;\Psi(\xi)\,,
\end{equation}
for any function $\Psi=\Psi(\xi)$. Now, as in the gluonic construction,
the celestial graviton operators must be defined by dressing the vertex
operators with the Grassmann-valued factors $\prod^{8}_{\hat{\alpha}=1}\psi^{\hat{\alpha}}(x)$.
On the other hand, in the gravitational case, there are two kinds
of vertex operators, $G_{i}$ and $H_{j}$, defined respectively in
Eqs. (\ref{eq:Definition-2}) and (\ref{eq:Definition-3}). Therefore,
gravitons with positive helicity are associated with the operators
\begin{equation}
\boldsymbol{G}^{+}_{i}(x_{i})\coloneqq G_{i}(x_{i}),\qquad\boldsymbol{H}^{+}_{j}(x_{j})\coloneqq H_{j}(x_{j})\,,
\end{equation}
whereas gravitons with negative helicity are associated with
\begin{equation}
\boldsymbol{G}^{-}_{i}(x_{i})\coloneqq G_{i}(x_{i})\,\prod^{8}_{\alpha=1}\psi^{\alpha}(x_{i}),\qquad\boldsymbol{H}^{-}_{j}(x_{j})\coloneqq H_{j}(x_{j})\,\prod^{8}_{\alpha=1}\psi^{\alpha}(x_{j})\,.
\end{equation}

Thus, by employing reasoning analogous to that leading to Eqs. (\ref{eq:Step-B},
\ref{eq:Step-C}) in the previous section, we obtain
\begin{align}
\widehat{\mathcal{M}}_{n} & =\frac{1}{\left(2\pi\right)^{3}\mathcal{N}\gamma_{n}}\left(\frac{\kappa}{2}\right)^{n-2}\lim_{N\rightarrow\infty}\lim_{k\rightarrow\infty}\Pi^{\text{GR}}_{\text{Nair}}\int\frac{d^{2}\lambda}{\pi}\int dy\delta\left(y\right)\label{eq:-5}\\
 & \quad\times\langle\lambda\big|\boldsymbol{G}^{-}_{1}\boldsymbol{H}^{-}_{2}\left(\prod^{n-2}_{i=3}\boldsymbol{H}^{+}_{i}\right)\boldsymbol{G}^{+}_{n-1}\boldsymbol{G}^{+}_{n}\bar{\partial}\big|\lambda\rangle+\left(\bar{x}_{i}\leftrightarrow-\bar{x}_{i}\right)+\mathcal{P}\left(2,...,n-2\right),
\end{align}
where $\gamma_{n}\coloneqq\text{tr}\left(T^{a_{1}}...T^{a_{n}}\right)^{2}$.

In conclusion, we have demonstrated that the tree-level celestial
amplitude for MHV graviton scattering in Einstein gravity, in the
helicity configuration $1^{-},2^{-},3^{+},\dots,n^{+}$, can be obtained
as the large-$N$, large-level ($k\to\infty$) limit of the correlation
function of the vertex operators constructed from the Kac-Moody currents
and conformal primaries of the $H^{+}_{3}$-WZNW model, appropriately
dressed by auxiliary Grassmann-valued helicity variables.

\paragraph{Remark.}

Let us note, as at the end of Subsection \ref{subsec:Celestial-Vertex-Operators},
that the four-dimensional gravitational coupling constant $\kappa$
in the bulk Klein space enters Eq. (\ref{eq:-5}) only through the
proportionality factor. We may absorb the factor $\kappa/2$ into
the normalisation of the positive-helicity celestial graviton operators.
However, a top-down derivation of $\kappa$ in terms of the physical
parameters describing the $H^{+}_{3}$-WZNW model, such as the radius
of three-dimensional anti-de Sitter space or the fundamental string
length, remains a long-standing goal of our programme. Accordingly,
at present, our candidate model for perturbative celestial CFT, which
approximates the tree-level MHV sector of Einstein gravity, reproduces
the kinematical structure and conformal dependence of the BGK formula
when expressed in the celestial conformal-primary basis, but treats
the gravitational coupling parameter $\kappa$ as a phenomenological
ingredient.

\section{Correlation Functions, Celestial OPE, Knizhnik--Zamolodchikov Equations
and Stress-Energy Tensor\label{sec:Correlation-Functions.-Operator}}

In the preceding sections, we have introduced a holographic dictionary
establishing a correspondence between Euclidean $AdS_{3}$ (bosonic)
string theory and celestial CFT. Now, we shall compute the two- and
three-point functions in Subsections \ref{subsec:Two-point-function}
and \ref{subsec:Three-Point-Function}, respectively, and determine
the structure constants of the celestial operator product expansion
(OPE) in Subsection \ref{subsec:Operator-Product-Expansion}. Finally,
in Subsection \ref{subsec:Differential-Equations-for}, we will derive
a system of partial differential equations (PDEs) characterising the
celestial amplitudes in the energy-momentum representation, drawing
from the Knizhnik--Zamolodchikov (KZ) equations and the worldsheet
Ward identities.

\subsection{Two-point function\label{subsec:Two-point-function}}

In this subsection, we derive the celestial $2$-point function by
employing the exact form of the corresponding correlation function
in the $H^{+}_{3}$-WZNW model, studied in detail by \citet{teschner1997conformal}.
For readers seeking a comprehensive exposition of the mathematical
subtleties inherent in the model, we direct them to Teschner's subsequent
works (\citet{teschner1999mini,teschner1999structure,teschner2000operator}).

Our aim is to compute the correlation function of two scalar conformal
primaries, $\phi_{\Delta_{1}}\left(x_{1}\right)$ and $\phi_{\Delta_{2}}\left(x_{2}\right)$,
within the celestial CFT, characterised by conformal weights $\Delta_{1}$
and $\Delta_{2}$, respectively. According to our proposed holographic
dictionary between Euclidean $AdS_{3}$ bosonic string theory and
celestial CFT, these scalar wavefunctions correspond to the worldsheet
conformal primaries $\Phi^{j_{1}}\left(x_{1};z_{1}\right)$ and $\Phi^{j_{2}}\left(x_{2};z_{2}\right)$.
In this correspondence, the spins in the $H^{+}_{3}$-WZNW model are
related to the conformal weights in the celestial CFT by $j_{1}=-\Delta_{1}/2$
and $j_{2}=-\Delta_{2}/2$.

According to \citet{teschner1999mini,teschner1997conformal}, the
$2$-point function of these conformal primaries is given by:
\begin{align}
 & \left\langle \Phi^{j_{1}}\left(x_{1};z_{1}\right)\Phi^{j_{2}}\left(x_{2};z_{2}\right)\right\rangle =\\
 & \frac{1}{\left|z_{12}\right|^{2\beta}}\left(\delta^{\left(2\right)}\left(x_{1}-x_{2}\right)\delta\left(j_{1}+j_{2}+1\right)+\frac{B\left(j_{1}\right)}{\left|x_{12}\right|^{-2\left(j_{1}+j_{2}\right)}}\delta\left(j_{2}-j_{1}\right)\right),\label{eq:2-Point-Function}
\end{align}
where the exponents and proportionality constants are defined as:
\begin{equation}
\beta\coloneqq\frac{b^{2}}{2}\left(2-\Delta_{1}\right)\left(2-\Delta_{2}\right),\,\,\,b^{2}=\frac{1}{k-2},
\end{equation}
and $B\left(j_{1}\right)$ denotes a normalisation constant that depends
on the choice of normalisation of the ``plane-wave'' basis $\Phi^{j}\left(x;z\right)$.
For the conventions adopted by \citet{giveon1998comments}, which
we adhere to in these notes, $B\equiv1$. Nonetheless, to preserve
generality and maintain consistency with \citet{teschner1997conformal},
we retain the explicit dependence on the constant $B\left(j_{1}\right)$
throughout this subsection.

Applying our holographic dictionary, which maps correlation functions
in celestial CFT to those in $AdS_{3}$ string theory, we obtain:
\begin{align}
 & \left\langle \phi_{\Delta_{1}}\left(x_{1}\right)\phi_{\Delta_{2}}\left(x_{2}\right)\right\rangle _{\text{CCFT}}\\
 & =\frac{1}{N}\int d^{2}z_{1}\int d^{2}z_{2}\left\langle \Phi^{j_{1}}\left(x_{1};z_{1}\right)\Phi^{j_{2}}\left(x_{2};z_{2}\right)\right\rangle \\
 & =\left(\delta^{\left(2\right)}\left(x_{1}-x_{2}\right)\delta\left(j_{1}+j_{2}+1\right)+\frac{B\left(j_{1}\right)}{\left|x_{12}\right|^{-2\left(j_{1}+j_{2}\right)}}\delta\left(j_{2}-j_{1}\right)\right)\frac{1}{N}\int d^{2}z_{1}\int d^{2}z_{2}\frac{1}{\left|z_{12}\right|^{2\beta}}\label{eq:Step-9}
\end{align}
Here, $N$ denotes the worldsheet ``area.'' The factor $1/N$ serves
as a renormalisation of the worldsheet integration measure, ensuring
that the correlation function remains well-defined and finite. This
renormalisation is a standard procedure in string perturbation theory
(cf. \citet{d1988geometry}). 

As demonstrated in the Appendix \ref{subsec:Generalised-Dirac-Delta},
Eq. (\ref{eq:Integral-4}), the worldsheet integral evaluates to:
\begin{equation}
\frac{1}{N}\int d^{2}z_{1}\int d^{2}z_{2}\frac{1}{\left|z_{12}\right|^{2\beta}}=\frac{\pi^{2}}{2}\boldsymbol{\delta}\left(\beta-1\right),\label{eq:Identity-4}
\end{equation}
where, following \citet{donnay2020asymptotic}, $\boldsymbol{\delta}\left(i\left(\Delta-z\right)\right)$
represents the generalised Dirac delta ``function,'' analytically
continued to the complex plane via:
\begin{equation}
\boldsymbol{\delta}\left(i\left(\Delta-z\right)\right)\coloneqq\frac{1}{2\pi}\int^{\infty}_{0}d\tau\,\tau^{\Delta-z-1},\label{eq:Dirac-1}
\end{equation}
such that the following identity holds:
\begin{equation}
\varphi\left(\Delta\right)=-i\int_{\mathcal{C}}dz\boldsymbol{\delta}\left(i\left(\Delta-z\right)\right)\varphi\left(z\right),
\end{equation}
for the contour $\mathcal{C}\coloneqq-1/2+i\mathbf{R}$.

Substituting this result into Eq. (\ref{eq:Step-9}) and expressing
the final result in terms of the conformal weights, we arrive at:
\begin{align}
 & \left\langle \phi_{\Delta_{1}}\left(x_{1}\right)\phi_{\Delta_{2}}\left(x_{2}\right)\right\rangle _{\text{CCFT}}=\\
 & \pi^{2}\boldsymbol{\delta}\left(\beta-1\right)\left(\delta^{\left(2\right)}\left(x_{1}-x_{2}\right)\delta\left(2-\Delta_{1}-\Delta_{2}\right)+\frac{B\left(-\Delta_{1}/2\right)}{\left|x_{12}\right|^{\Delta_{1}+\Delta_{2}}}\delta\left(\Delta_{1}-\Delta_{2}\right)\right)\label{eq:Celestial-Two-Point-Function}
\end{align}

The $2$-point function in our celestial CFT for the scalar conformal
primaries $\phi_{\Delta_{1}}\left(x_{1}\right)$ and $\phi_{\Delta_{2}}\left(x_{2}\right)$
takes the form of a linear combination involving a delta function
localised at $x_{1}=x_{2}$ and a conventional $1/\left|x_{12}\right|^{\Delta_{1}+\Delta_{2}}$
power-law dependence on the conformal weights. This result stands
in contrast to those reported by \citet{furugori2024celestial}. One
might, as suggested by \citet{ogawa2024celestial}, explore the possibility
of constructing linear combinations of these conformal primaries to
generate a $2$-point function that solely includes the delta function
$\delta^{\left(2\right)}\left(x_{1}-x_{2}\right)$, the power-law
term $1/\left|x_{12}\right|^{\Delta_{1}+\Delta_{2}}$, or their shadow
transforms.

However, we prefer to follow an approach that avoids the imposition
of intuition derived from conventional CFTs when examining celestial
CFTs. Instead, we maintain that the mathematical structures we have
uncovered merit serious consideration on their own terms, and their
consequences should be rigorously examined. The ability of the mini-superspace
limit of the $H^{+}_{3}$-WZNW model to holographically reconstruct
the tree-level MHV scattering amplitudes for gluons and gravitons
is a compelling reason for adhering to the mathematical framework
of this model in deriving the properties of celestial CFT. Thus, in
our view, one must be prepared to encounter correlation functions--and,
as will be demonstrated in the subsequent sections, structure constants
of celestial OPEs--manifesting in a non-conventional distributional
form.

It is also instructive to compute the shadow transform of the $2$-point
function given by Eq. (\ref{eq:2-Point-Function}):
\begin{align}
 & \int d^{2}y_{1}\frac{1}{\left|x_{1}-y_{1}\right|^{4\left(1+j_{1}\right)}}\int d^{2}y_{2}\frac{1}{\left|x_{2}-y_{2}\right|^{4\left(1+j_{2}\right)}}\left\langle \Phi^{j_{1}}\left(y_{1};z_{1}\right)\Phi^{j_{2}}\left(y_{2};z_{2}\right)\right\rangle \\
 & =\frac{1}{\left|z_{12}\right|^{2\beta}}\delta\left(j_{1}+j_{2}+1\right)\int d^{2}y_{1}d^{2}y_{2}\frac{\delta^{\left(2\right)}\left(y_{1}-y_{2}\right)}{\left|x_{1}-y_{1}\right|^{4\left(1+j_{1}\right)}\left|x_{2}-y_{2}\right|^{4\left(1+j_{2}\right)}}\\
 & +\frac{1}{\left|z_{12}\right|^{2\beta}}B\left(j_{1}\right)\delta\left(j_{1}-j_{2}\right)\int d^{2}y_{1}d^{2}y_{2}\frac{\delta^{\left(2\right)}\left(y_{1}-y_{2}\right)}{\left|x_{1}-y_{1}\right|^{4\left(1+j_{1}\right)}\left|x_{2}-y_{2}\right|^{4\left(1+j_{2}\right)}\left|x_{12}\right|^{-2\left(j_{1}+j_{2}\right)}}.\label{eq:Step-10}
\end{align}
In the Appendix \ref{subsec:Worldsheet-Integrals-Related}, we gave
a detailed analysis of the following integral over the complex plane:
\begin{equation}
\mathcal{I}_{0}\left(\tau_{1},\tau_{2}\big|x_{1},x_{2}\right)\coloneqq\int d^{2}y\frac{1}{\left|x_{1}-y\right|^{2\tau_{1}}\left|y-x_{2}\right|^{2\tau_{2}}}.\label{eq:Integral-3}
\end{equation}
We found in Eq. (\ref{eq:Integral-2}) that this integral evaluates,
in the distributional sense, to:
\begin{equation}
\mathcal{I}_{0}=\frac{\pi}{\left|x_{1}-x_{2}\right|^{2\left(\tau_{1}+\tau_{2}-1\right)}}\frac{\Gamma\left(1-\tau_{1}\right)}{\Gamma\left(\tau_{1}\right)}\frac{\Gamma\left(1-\tau_{2}\right)}{\Gamma\left(\tau_{2}\right)}\frac{\Gamma\left(\tau_{1}+\tau_{2}-1\right)}{\Gamma\left(2-\tau_{1}-\tau_{2}\right)}.\label{eq:Identity-2}
\end{equation}
From Eq. (\ref{eq:Identity-2}), we demonstrated in Appendix \ref{subsec:Worldsheet-Integrals-Related}
(cf. Eq. (\ref{eq:Integral-5})) that, by taking the limit $\tau_{1}+\tau_{2}\rightarrow2$,
one recovers the well-known identity from $2d$ CFT (cf. \citet{simmons2014projectors}):
\begin{equation}
\int d^{2}y\frac{1}{\left|x_{1}-y\right|^{2\tau}\left|y-x_{2}\right|^{2\left(2-\tau\right)}}=\frac{4\pi^{2}}{\nu^{2}}\delta^{\left(2\right)}\left(x_{1}-x_{2}\right),\label{eq:Identity-3}
\end{equation}
where $\tau\eqqcolon1+i\nu$. 

Consequently, using Eqs. (\ref{eq:Identity-2}, \ref{eq:Identity-3}),
we arrive at the following results:
\begin{equation}
\lim_{j_{1}+j_{2}+1\rightarrow0}\int d^{2}y_{1}d^{2}y_{2}\frac{\delta^{\left(2\right)}\left(y_{1}-y_{2}\right)}{\left|x_{1}-y_{1}\right|^{4\left(1+j_{1}\right)}\left|x_{2}-y_{2}\right|^{4\left(1+j_{2}\right)}}=-\frac{4\pi^{2}}{\left(1+2j_{1}\right)^{2}}\delta^{\left(2\right)}\left(x_{1}-x_{2}\right),
\end{equation}
and, similarly,
\begin{align}
 & \lim_{j_{1}-j_{2}\rightarrow0}\int d^{2}y_{1}d^{2}y_{2}\frac{\delta^{\left(2\right)}\left(y_{1}-y_{2}\right)}{\left|x_{1}-y_{1}\right|^{4\left(1+j_{1}\right)}\left|x_{2}-y_{2}\right|^{4\left(1+j_{2}\right)}\left|x_{1}-x_{2}\right|^{-2\left(j_{1}+j_{2}\right)}}\\
 & =-\frac{\pi^{2}}{\left(1+2j_{1}\right)^{2}}\frac{1}{\left|x_{1}-x_{2}\right|^{2\left(2+j_{1}+j_{2}\right)}}.
\end{align}
Thus, Eq. (\ref{eq:Step-10}) can be rewritten as:
\begin{align}
 & \frac{1}{2\pi}\int d^{2}y_{1}\frac{1}{\left|x_{1}-y_{1}\right|^{4\left(1+j_{1}\right)}}\frac{1}{2\pi}\int d^{2}y_{2}\frac{1}{\left|x_{2}-y_{2}\right|^{4\left(1+j_{2}\right)}}\left\langle \Phi^{j_{1}}\left(y_{1};z_{1}\right)\Phi^{j_{2}}\left(y_{2};z_{2}\right)\right\rangle \\
 & =-\frac{1}{\left(1+2j_{1}\right)^{2}}\frac{1}{\left|z_{12}\right|^{2\beta}}\left(\delta^{\left(2\right)}\left(x_{1}-x_{2}\right)\delta\left(j_{1}+j_{2}+1\right)+\frac{B\left(j_{1}\right)}{4}\frac{1}{\left|x_{12}\right|^{2\left(2+j_{1}+j_{2}\right)}}\delta\text{\ensuremath{\left(j_{1}-j_{2}\right)}}\right).\label{eq:Step-11}
\end{align}

It is then natural to define the shadow transform of the worldsheet
conformal primary $\Phi^{j}\left(x;z\right)$ as:
\begin{equation}
\widetilde{\Phi}^{j}\left(x;z\right)\coloneqq\frac{1}{2\pi}\int d^{2}y\frac{1}{\left|x-y\right|^{4\left(1+j\right)}}\Phi^{j}\left(y;z\right),
\end{equation}
so that Eq. (\ref{eq:Step-11}) can be reorganised as:
\begin{align}
 & \left\langle \widetilde{\Phi}^{j_{1}}\left(x_{1};z_{1}\right)\widetilde{\Phi}^{j_{2}}\left(x_{2};z_{2}\right)\right\rangle \\
 & =-\frac{1}{\left(1+2j_{1}\right)^{2}}\frac{1}{\left|z_{12}\right|^{2\beta}}\left(\delta^{\left(2\right)}\left(x_{1}-x_{2}\right)\delta\left(j_{1}+j_{2}+1\right)+\frac{B\left(j_{1}\right)}{4}\frac{1}{\left|x_{12}\right|^{2\left(2+j_{1}+j_{2}\right)}}\delta\text{\ensuremath{\left(j_{1}-j_{2}\right)}}\right).\label{eq:Step-12}
\end{align}

We conclude by extending the holographic dictionary, postulating that
the shadow-transformed worldsheet conformal primaries $\widetilde{\Phi}^{j}\left(x;z\right)$,
characterised by spin $j$, correspond to the celestial scalar conformal
primaries $\widetilde{\phi}_{\Delta}\left(x\right)$, with conformal
weight $\Delta=-2j$. Consequently, integrating Eq. (\ref{eq:Step-12})
over the worldsheet coordinates $z_{1},z_{2}$ yields:
\begin{align}
 & \left\langle \widetilde{\phi}_{\Delta_{1}}\left(x_{1}\right)\widetilde{\phi}_{\Delta_{2}}\left(x_{2}\right)\right\rangle _{\text{CCFT}}\\
 & =\frac{1}{N}\int d^{2}z_{1}\int d^{2}z_{2}\left\langle \widetilde{\Phi}^{j_{1}}\left(x_{1};z_{1}\right)\widetilde{\Phi}^{j_{2}}\left(x_{2};z_{2}\right)\right\rangle \\
 & =\frac{\pi^{2}}{\lambda^{2}}\boldsymbol{\delta}\left(\beta-1\right)\left(\delta^{\left(2\right)}\left(x_{1}-x_{2}\right)\delta\left(2-\Delta_{1}-\Delta_{2}\right)+\frac{B\left(-\Delta_{1}/2\right)}{4}\frac{1}{\left|x_{12}\right|^{4-\Delta_{1}-\Delta_{2}}}\delta\left(\Delta_{1}-\Delta_{2}\right)\right),\label{eq:Celestial-Two-Point-Function-1}
\end{align}
where $\Delta_{1}\eqqcolon1+i\lambda$.

The celestial $2$-point function of the shadow-transformed conformal
primaries exhibits a similar structure to that of Eq. (\ref{eq:Celestial-Two-Point-Function}),
namely, a linear combination of the delta function, $\delta^{\left(2\right)}\left(x_{1}-x_{2}\right)$,
and a standard power-law dependence, $1/\left|x_{12}\right|^{2-\Delta_{1}-\Delta_{2}}$.
However, the functional dependence on the conformal weights $\Delta_{1},\Delta_{2}$
in Eqs. (\ref{eq:Celestial-Two-Point-Function} , \ref{eq:Celestial-Two-Point-Function-1})--aside
from the prefactor $1/\lambda^{2}_{1}=1/\left(1-\Delta_{1}\right)^{2}$--is
distinct. Furthermore, it is essential to recall that, within celestial
CFT, the conformal weights $\Delta_{1},\Delta_{2}$ are dynamical
physical variables, corresponding to the Mellin conjugates of the
frequencies of the particles involved in the scattering process. Therefore,
the functional dependencies on $\Delta_{1},\Delta_{2}$ encode non-trivial
physical information about the scattering events. However, as demonstrated
by \citet{teschner1999mini}, the bases $\Phi^{j}\left(x;z\right)$
and $\widetilde{\Phi}^{-1-j}\left(x;z\right)$ yield the same representation
within the framework of the $H^{+}_{3}$-WZNW model, a result that
becomes apparent from the structure of the shadow-transformed $2$-point
function derived earlier. Consequently, this equivalence of bases
in the $H^{+}_{3}$-WZNW model leads to non-trivial physical implications
for celestial CFT, an important matter that we will leave unresolved
for the time being, to be addressed in future investigations.

\subsection{Three-Point Function\label{subsec:Three-Point-Function}}

In this subsection, we employ the exact form of the $3$-point function
in the $H^{+}_{3}$-WZNW model, as determined by \citet{teschner1997conformal},
to evaluate the celestial amplitude corresponding to the scattering
of three massless scalar particles within the framework of our holographic
dictionary. To structure our analysis, we adhere to the formalism
introduced by \citet{teschner1999structure,teschner2000operator}
in his subsequent works concerning the $H^{+}_{3}$-WZNW model. For
a more comprehensive account, we direct the reader to these works.
It is important to note, however, that we adopt the normalisation
conventions for the worldsheet primaries as defined by \citet{giveon1998comments}.

Let us denote the conformal weights associated with the celestial
scalar conformal primaries $\phi_{\Delta_{1}}\left(x_{1}\right)$,
$\phi_{\Delta_{2}}\left(x_{2}\right)$ and $\phi_{\Delta_{3}}\left(x_{3}\right)$
as $\Delta_{1}$, $\Delta_{2}$ and $\Delta_{3}$, respectively. Correspondingly,
we introduce the quantities $j_{1}\coloneqq-\Delta_{1}/2$, $j_{2}\coloneqq-\Delta_{2}/2$
and $j_{3}\coloneqq-\Delta_{3}/2$, which represent the spins of the
worldsheet primaries within the $H^{+}_{3}$-WZNW model. For the sake
of notational simplicity, we temporarily omit the anti-holomorphic
variables from our arguments, thereby denoting the worldsheet primaries
as $\Phi^{j_{1}}\left(x_{1};z_{1}\right)$, $\Phi^{j_{2}}\left(x_{2};z_{2}\right)$
and $\Phi^{j_{3}}\left(x_{3};z_{3}\right)$.

It is important to emphasise that the correspondence $\phi_{\Delta_{k}}\left(x_{k}\right)\leftrightarrow\Phi^{j_{k}}\left(x_{k};z_{k}\right)$
of our holographic dictionary maps a \emph{scalar} celestial wavefunction
to a worldsheet conformal primary with \emph{spin} $j_{k}$. The relation
$j_{k}\leftrightarrow-\Delta_{k}/2$ summarises the mapping between
the notion of spin in the $H^{+}_{3}$-model and the conformal weight
in the celestial CFT.

The scaling dimensions of the worldsheet primaries are related to
their respective spins through the relations:
\begin{equation}
h_{1}=-b^{2}j_{1}\left(j_{1}+1\right),\,\,\,h_{2}=-b^{2}j_{2}\left(j_{2}+1\right),\,\,\,h_{3}=-b^{2}j_{3}\left(j_{3}+1\right),
\end{equation}
where $b^{2}\coloneqq1/\left(k-2\right)$. 

According to \citet[Eq. (22)]{teschner2000operator}, the $3$-point
correlation function in the $H^{+}_{3}$-WZNW model is given by:
\begin{equation}
\left\langle \Phi^{j_{1}}\left(x_{1};z_{1}\right)\Phi^{j_{2}}\left(x_{2};z_{2}\right)\Phi^{j_{3}}\left(x_{3};z_{3}\right)\right\rangle =\frac{D\left(j_{1},j_{2},j_{3}\right)C\left(j_{1},j_{2},j_{3}\big|x_{1},x_{2},x_{3}\right)}{\left|z_{12}\right|^{2\sigma_{3}}\left|z_{23}\right|^{2\sigma_{1}}\left|z_{31}\right|^{2\sigma_{2}}},
\end{equation}
where the Clebsch-Gordan distributional kernel is defined by:
\begin{equation}
C\left(j_{1},j_{2},j_{3}\big|x_{1},x_{2},x_{3}\right)=\frac{1}{\left|x_{12}\right|^{2\lambda_{3}}\left|x_{23}\right|^{2\lambda_{1}}\left|x_{31}\right|^{2\lambda_{2}}}.
\end{equation}
The exponents associated with the worldsheet coordinates $z_{1}$,
$z_{2}$ and $z_{3}$ are defined as:
\begin{equation}
\sigma_{1}\coloneqq-h_{1}+h_{2}+h_{3},\,\,\,\sigma_{2}\coloneqq h_{1}-h_{2}+h_{3},\,\,\,\sigma_{3}\coloneqq h_{1}+h_{2}-h_{3},\label{eq:Exponents}
\end{equation}
while the exponents corresponding to the celestial coordinates $x_{1}$,
$x_{2}$ and $x_{3}$ are similarly given by:
\begin{equation}
\lambda_{1}\coloneqq j_{1}-j_{2}-j_{3},\,\,\,\lambda_{2}\coloneqq-j_{1}+j_{2}-j_{3},\,\,\,\lambda_{3}\coloneqq-j_{1}-j_{2}+j_{3}.\label{eq:Exponents-1}
\end{equation}

In accordance with the holographic dictionary proposed in Section
2, the celestial $3$-point amplitude corresponding to the scalar
wavefunctions $\phi_{\Delta_{1}}\left(x_{1}\right)$, $\phi_{\Delta_{2}}\left(x_{2}\right)$
and $\phi_{\Delta_{3}}\left(x_{3}\right)$ is expressed as:
\begin{align}
 & \left\langle \phi_{\Delta_{1}}\left(x_{1}\right)\phi_{\Delta_{2}}\left(x_{2}\right)\phi_{\Delta_{3}}\left(x_{3}\right)\right\rangle _{\text{CCFT}}\\
 & =\frac{1}{N}\int d^{2}z_{1}\int d^{2}z_{2}\int d^{2}z_{3}\left\langle \Phi^{j_{1}}\left(x_{1};z_{1}\right)\Phi^{j_{2}}\left(x_{2};z_{2}\right)\Phi^{j_{3}}\left(x_{3};z_{3}\right)\right\rangle \\
 & =D\left(j_{1},j_{2},j_{3}\right)C\left(j_{1},j_{2},j_{3}\big|x_{1},x_{2},x_{3}\right)\frac{1}{N}\int d^{2}z_{1}\int d^{2}z_{2}\int d^{2}z_{3}\frac{1}{\left|z_{12}\right|^{2\sigma_{3}}\left|z_{23}\right|^{2\sigma_{1}}\left|z_{31}\right|^{2\sigma_{2}}}.\label{eq:Step-5}
\end{align}
Here, $N$ denotes the worldsheet area, and the factor $1/N$ serves
as a renormalisation of the integration measure to ensure that the
correlation function is well-defined and finite--a standard procedure
in the computation of scattering amplitudes within string perturbation
theory (cf. \citet{d1988geometry,polchinski1998string,polchinski2001string}).

As demonstrated in Appendix \ref{subsec:Triple-Worldsheet-Integral},
the integral over the worldsheet coordinates evaluates to:
\begin{align}
 & \frac{1}{V}\int d^{2}z_{1}\int d^{2}z_{2}\int d^{2}z_{3}\frac{1}{\left|z_{12}\right|^{2\sigma_{3}}\left|z_{23}\right|^{2\sigma_{1}}\left|z_{31}\right|^{2\sigma_{2}}}\\
 & =2\pi^{3}\boldsymbol{\delta}\left(2-\sigma_{1}-\sigma_{2}-\sigma_{3}\right)\frac{\Gamma\left(1-\sigma_{1}\right)\Gamma\left(1-\sigma_{2}\right)\Gamma\left(1-\sigma_{3}\right)}{\Gamma\left(\sigma_{1}\right)\Gamma\left(\sigma_{2}\right)\Gamma\left(\sigma_{3}\right)}.\label{eq:Integral}
\end{align}
It is noteworthy that the triple integral over the worldsheet yields
the \emph{generalised }delta ``function'' $\boldsymbol{\delta}\left(2-\sum_{i}\sigma_{i}\right)$
associated with the conformal weights. This form is characteristic
of celestial amplitudes, as discussed in previous works (e.g., \citet{arkani2021celestial,gu2022note,stieberger2018strings}).
Importantly, this delta function would not be present had we chosen
to disregard the worldsheet coordinates $z_{1}$, $z_{2}$ and $z_{3}$,
as was done by \citet{ogawa2024celestial}.

Consequently, employing the previously defined equations (Eqs. (\ref{eq:Exponents},
\ref{eq:Step-5}, \ref{eq:Integral})), we arrive at:
\begin{align}
 & \left\langle \phi_{\Delta_{1}}\left(x_{1}\right)\phi_{\Delta_{2}}\left(x_{2}\right)\phi_{\Delta_{3}}\left(x_{3}\right)\right\rangle _{\text{CCFT}}\\
 & =2\pi^{3}\frac{\boldsymbol{\delta}\left(2-h_{1}-h_{2}-h_{3}\right)D\left(j_{1},j_{2},j_{3}\right)}{\left|x_{12}\right|^{2\left(-j_{1}-j_{2}+j_{3}\right)}\left|x_{23}\right|^{2\left(j_{1}-j_{2}-j_{3}\right)}\left|x_{31}\right|^{2\left(-j_{1}+j_{2}-j_{3}\right)}}\frac{\Gamma\left(2h_{1}-1\right)\Gamma\left(2h_{2}-1\right)\Gamma\left(2h_{3}-1\right)}{\Gamma\left(2-2h_{1}\right)\Gamma\left(2-2h_{2}\right)\Gamma\left(2-2h_{3}\right)}.
\end{align}

This exact form of the $3$-point celestial amplitude, derived as
the correlation function within Euclidean $AdS_{3}$ string theory
via our proposed holographic dictionary, explicitly exhibits the distributional
structure anticipated in celestial amplitudes. Moreover, the presence
of gamma functions of the scaling dimensions is indicative of an amplitude
derived from worldsheet integrations, aligning with the general expectations
from calculations in string perturbation theory.

\subsection{Celestial Operator Product Expansion\label{subsec:Operator-Product-Expansion}}

In this subsection, we determine the celestial operator product expansion
utilising the structure constants of the $H^{+}_{3}$-WZNW model as
derived by \citet{teschner1999structure,teschner2000operator}. However,
for consistency, we adopt the normalisation conventions used by \citet{giveon1998comments}.

Let $\Delta_{1}$ and $\Delta_{2}$ denote the conformal weights associated
with the scalar celestial wavefunctions $\phi_{\Delta_{1}}\left(x_{1}\right)$
and $\phi_{\Delta_{2}}\left(x_{2}\right)$, respectively. According
to the proposed holographic dictionary, these wavefunctions correspond
to the worldsheet primaries $\Phi^{j_{1}}\left(x_{1};z_{1}\right)$
and $\Phi^{j_{2}}\left(x_{2};z_{2}\right)$, where the spins in the
$H^{+}_{3}$-WZNW model are related to the celestial conformal weights
by $j_{1}=-\Delta_{1}/2$ and $j_{2}=-\Delta_{2}/2$.

According to \citet{teschner2000operator}, the worldsheet OPE for
the primaries $\Phi^{j_{1}}\left(x_{1};z_{1}\right)$ and $\Phi^{j_{2}}\left(x_{2};z_{2}\right)$
is then expressed as:
\begin{align}
 & \Phi^{j_{1}}\left(x_{1};z_{1}\right)\Phi^{j_{2}}\left(x_{2};z_{2}\right)\sim\\
 & -\frac{1}{2}\int_{-\frac{1}{2}+i\mathbf{R}^{+}}dj_{3}\frac{\left(1+2j_{3}\right)^{2}}{\left|z_{12}\right|^{\sigma_{3}}}D\left(j_{1},j_{2},j_{3}\right)\int d^{2}x_{3}\frac{1}{\left|x_{12}\right|^{2\lambda_{3}}\left|x_{23}\right|^{2\lambda_{1}}\left|x_{31}\right|^{2\lambda_{2}}}\Phi^{-1-j_{3}}\left(x_{3};z_{2}\right).\label{eq:OPE-1}
\end{align}
Here, the exponents $\lambda_{1}$, $\lambda_{2}$ and $\lambda_{3}$
associated with the celestial coordinates $x_{1}$, $x_{2}$ and $x_{3}$
are defined by Eq. (\ref{eq:Exponents}), and $\sigma_{3}=h_{1}+h_{2}-h_{3}$
as per Eq. (\ref{eq:Exponents-1}).

Our objective is to compute the celestial OPE involving two scalar
conformal primaries. To achieve this, we adopt the following strategy:
we investigate the behaviour of $\phi_{\Delta_{1}}\left(x_{1}\right)\phi_{\Delta_{2}}\left(x_{2}\right)$
within a correlation function of the celestial CFT. Utilising Eq.
(\ref{eq:OPE-1}), we express this as:
\begin{align}
 & \left\langle \phi_{\Delta_{1}}\left(x_{1}\right)\phi_{\Delta_{2}}\left(x_{2}\right)...\right\rangle _{\text{CCFT}}\sim-\frac{1}{2}\int_{-\frac{1}{2}+i\mathbf{R}^{+}}\left(1+2j_{3}\right)^{2}D\left(j_{1},j_{2},j_{3}\right)\\
 & \int d^{2}x_{3}\frac{1}{\left|x_{12}\right|^{2\lambda_{3}}\left|x_{23}\right|^{2\lambda_{1}}\left|x_{31}\right|^{2\lambda_{2}}}\int d^{2}z_{1}d^{2}z_{2}\frac{1}{\left|z_{12}\right|^{\sigma_{3}}}\left\langle \Phi^{-1-j_{3}}\left(x_{3};z_{2}\right)...\right\rangle _{\text{CCFT}}.\label{eq:OPE-2}
\end{align}

In the Appendix \ref{subsec:Shadow-Transform-of}, Eq. (\ref{eq:Integral-6}),
we have demonstrated that the worldsheet integral evaluates to:
\begin{equation}
\int d^{2}z_{1}\int d^{2}z_{2}\frac{1}{\left|z_{12}\right|^{\sigma_{3}}}\Phi^{-1-j_{3}}\left(x_{3};z_{2}\right)=\pi^{2}\boldsymbol{\delta}\left(h_{1}+h_{2}+h_{3}-2\right)\int d^{2}z_{2}\Phi^{-1-j_{3}}\left(x_{3};z_{2}\right).\label{eq:Integral-1}
\end{equation}
Utilising the relations between the scaling dimensions $h_{1}$, $h_{2}$,
$h_{3}$ and the spins $j_{1}$, $j_{2}$, $j_{3}$ in the $H^{+}_{3}$-WZNW
model:
\begin{equation}
h_{1}=-b^{2}j_{1}\left(j_{1}+1\right),\,\,\,h_{2}=-b^{2}j_{2}\left(j_{2}+1\right),\,\,\,h_{3}=-b^{2}j_{3}\left(j_{3}+1\right),
\end{equation}
where $b^{2}\coloneqq1/\left(k-2\right)$, we can decompose the generalised
Dirac delta function of the scaling dimensions appearing in Eq. (\ref{eq:Integral-1})
as follows:
\begin{equation}
\boldsymbol{\delta}\left(h_{1}+h_{2}+h_{3}-2\right)=\frac{\delta\left(j_{3}-j^{+}_{3}\right)}{b^{2}\left(1+2j^{+}_{3}\right)}+\frac{\delta\left(j_{3}-j^{-}_{3}\right)}{b^{2}\left(1+2j^{-}_{3}\right)},
\end{equation}
where:
\begin{equation}
j^{\pm}_{3}=-\frac{1}{2}\pm i\rho\left(j_{1},j_{2}\right),\,\,\,-\rho^{2}\left(j_{1},j_{2}\right)\coloneqq\frac{1}{4}+\frac{2}{b^{2}}+j_{1}\left(j_{1}+1\right)+j_{2}\left(j_{2}+1\right).\label{eq:Isospin}
\end{equation}
Substituting these relations into Eq. (\ref{eq:OPE-2}) yields:
\begin{align}
 & \left\langle \phi_{\Delta_{1}}\left(x_{1}\right)\phi_{\Delta_{2}}\left(x_{2}\right)...\right\rangle _{\text{CCFT}}\sim\\
 & -\frac{\pi^{2}}{2b^{2}}\left(1+2j^{+}_{3}\right)D\left(j_{1},j_{2},j_{3}\right)\mathcal{I}_{2}\left(\lambda_{1},\lambda_{2},\lambda_{3}\right)\left\langle \phi_{-1-\Delta_{3}\left(j^{+}_{3}\right)}\left(x_{2}\right)...\right\rangle _{\text{CCFT}},\label{eq:Step-8}
\end{align}
where we have introduced the new integral:
\begin{equation}
\mathcal{I}_{2}\left(\lambda_{1},\lambda_{2},\lambda_{3}\right)\coloneqq\frac{1}{\left|x_{12}\right|^{2\lambda_{3}}}\int d^{2}x_{3}\frac{1}{\left|x_{23}\right|^{2\lambda_{1}}\left|x_{31}\right|^{2\lambda_{2}}}.
\end{equation}

In the Appendix \ref{subsec:Worldsheet-Integral-with}, Eq. (\ref{eq:Integral-7}),
we have shown that this integral evaluates to:
\begin{equation}
\mathcal{I}_{2}=\frac{\pi}{\left|x_{12}\right|^{2\left(\lambda_{1}+\lambda_{2}+\lambda_{3}-1\right)}}\frac{\Gamma\left(\lambda_{1}+\lambda_{2}-1\right)}{\Gamma\left(2-\lambda_{1}-\lambda_{2}\right)}\frac{\Gamma\left(1-\lambda_{1}\right)\Gamma\left(1-\lambda_{2}\right)}{\Gamma\left(\lambda_{1}\right)\Gamma\left(\lambda_{2}\right)}.
\end{equation}
Recalling that the exponents $\lambda_{1}$, $\lambda_{2}$ and $\lambda_{3}$
are related to the spins $j_{1}$, $j_{2}$ and $j_{3}$ through the
relations provided in Eq. (\ref{eq:Exponents-1}), we find:
\begin{equation}
\mathcal{I}_{2}=\pi\left|x_{12}\right|^{2\left(j_{1}+j_{2}+j_{3}+1\right)}\frac{\Gamma\left(-1-2j_{3}\right)}{\Gamma\left(2+2j_{3}\right)}\frac{\Gamma\left(1-j_{1}+j_{2}+j_{3}\right)\Gamma\left(1+j_{1}-j_{2}+j_{3}\right)}{\Gamma\left(j_{1}-j_{2}-j_{3}\right)\Gamma\left(-j_{1}+j_{2}-j_{3}\right)}.
\end{equation}
This expression can be rewritten using Euler's beta functions as:
\begin{equation}
\mathcal{I}_{2}=-\frac{\pi}{1+2j_{2}}\left|x_{12}\right|^{2\left(j_{1}+j_{2}+j_{3}+1\right)}\frac{B\left(1-j_{1}+j_{2}+j_{3},1+j_{1}-j_{2}+j_{3}\right)}{B\left(j_{1}-j_{2}-j_{3},-j_{1}+j_{2}-j_{3}\right)}.
\end{equation}

Finally, substituting this result into Eq. (\ref{eq:Step-8}), we
obtain the celestial OPE following from our proposed holographic dictionary:
\begin{align}
 & \phi_{\Delta_{1}}\left(x_{1}\right)\phi_{\Delta_{2}}\left(x_{2}\right)\sim\label{eq:Final-OPE}\\
 & \frac{\pi^{3}}{2b^{2}}\left|x_{12}\right|^{2\left(j_{1}+j_{2}+j^{+}_{3}+1\right)}D\left(j_{1},j_{2},j^{+}_{3}\right)\frac{B\left(1-j_{1}+j_{2}+j_{3},1+j_{1}-j_{2}+j_{3}\right)}{B\left(j_{1}-j_{2}-j_{3},-j_{1}+j_{2}-j_{3}\right)}\phi_{-1-\Delta\left(j^{+}_{3}\right)}.
\end{align}
where $\Delta\left(j^{+}_{3}\right)=-2j^{+}_{3}$ with $j^{+}_{3}$
given by Eq. (\ref{eq:Isospin}). 

To appreciate the physical significance of this OPE, we recall that,
in celestial CFT, the conformal dimensions $\Delta$ acquire the interpretation
of dynamical physical quantities. This arises from the fact that $\Delta$
are Mellin conjugates to the energies of scattered particles. Therefore,
the product of gamma functions, which depends on the conformal weights--arising
from the worldsheet integrals and manifesting both the worldsheet
conformal invariance and the factorisation properties of the string
vertex operators--is not merely a proportionality constant. Rather,
it is a physical quantity that characterises the scattering processes.

It is noteworthy that even the much simpler celestial OPE for gluons
in pure Yang-Mills theory, given by: 
\begin{equation}
\mathcal{G}^{a+}_{\Delta_{1}}\left(x_{1}\right)\mathcal{G}^{b+}_{\Delta_{2}}\left(x_{2}\right)\sim\frac{if^{abc}}{z_{12}}\frac{\Gamma\left(-1-2j_{1}\right)\Gamma\left(-1-2j_{2}\right)}{\Gamma\left(-2j_{1}\right)\Gamma\left(-2j_{2}\right)}\mathcal{G}^{c+}_{\Delta_{1}+\Delta_{2}-1}\left(x_{2}\right),
\end{equation}
with $j_{1}=-\Delta_{1}/2$ and $j_{2}=-\Delta_{2}/2$, while considerably
less intricate than Eq. (\ref{eq:Final-OPE}), exhibits a similar
structure. In this expression, the product of gamma functions that
depends on the conformal weights encodes information about the scattering
processes of gluons in flat space. This observation leads one to infer
that the structure of the OPEs in the Euclidean $AdS_{3}$ string
model for celestial CFT naturally encapsulates the physics of these
scattering processes. We also note that this structure appears to
deviate from that of conventional CFTs.

\subsection{Differential Equations for the Celestial CFT\label{subsec:Differential-Equations-for}}

In the previous subsections, we have illustrated our holographic dictionary
by computing the two- and three-point celestial amplitudes and extracting
the structure constants of the celestial OPE, derived from the corresponding
quantities in the $H^{+}_{3}$-WZNW model. The analytic structure
of this model has been rigorously investigated by \citet{ribault2005h3+,teschner1997conformal,teschner1999mini,teschner1999structure,teschner2000operator}.
In this subsection, we conclude the exposition of our holographic
dictionary between Euclidean $AdS_{3}$ bosonic string theory and
celestial CFT in Klein space by deriving a system of partial differential
equations that characterise the celestial amplitudes from the Knizhnik--Zamolodchikov
(KZ) equations and the worldsheet Ward identities of the $H^{+}_{3}$-WZNW
model. 

Let $\widehat{G}_{n}=\widehat{G}_{n}\left(x_{1},...,x_{n};\Delta_{1},...,\Delta_{n};z_{1},...,z_{n}\right)$
denote the $n$-point function of the $H^{+}_{3}$-WZNW model, defined
as:
\begin{equation}
\widehat{G}_{n}\coloneqq\left\langle \Phi^{j_{1}}\left(x_{1};z_{1}\right)\Phi^{j_{2}}\left(x_{2};z_{2}\right)...\Phi^{j_{n}}\left(x_{n};z_{n}\right)\right\rangle ,\label{eq:N-Point-Function}
\end{equation}
where the spins in the $H^{+}_{3}$-WZNW model are related to the
conformal weights in the celestial CFT by $j_{k}=-\Delta_{k}/2$.
According to the holographic dictionary, the celestial amplitude corresponding
to $\widehat{G}_{n}$ is obtained by integrating Eq. (\ref{eq:N-Point-Function})
over the worldsheet variables $z_{1},...,z_{n}$.

The KZ equations for the $H^{+}_{3}$-WZNW model are deduced as follows.
As discussed in \citet{giveon1998comments}, the Sugawara stress tensor
$T_{\text{S}}\left(z\right)$ of the worldsheet CFT is given by:
\begin{equation}
T_{\text{S}}\left(z\right)=\frac{1}{k-2}\left(J^{+}J^{-}-\left(J^{3}\right)^{2}\right)=\frac{b^{2}}{2}\left(J\partial^{2}_{x}J-\frac{1}{2}\left(\partial_{x}J\right)^{2}\right).\label{eq:Sugawara}
\end{equation}
Utilising the Sugawara construction (reviewed in \citet{ginsparg1988applied,francesco1997conformal,ketov1995conformal}),
we find that the stress-energy tensor $T_{\text{S}}\left(z\right)$
satisfies the Ward identity:
\begin{align}
 & \left\langle T_{\text{S}}\left(z\right)\Phi^{j_{1}}\left(x_{1};z_{1}\right)\Phi^{j_{2}}\left(x_{2};z_{2}\right)...\Phi^{j_{n}}\left(x_{n};z_{n}\right)\right\rangle \\
 & =\sum^{n}_{i=1}\left(\frac{1}{z-z_{i}}\frac{\partial}{\partial z_{i}}+\frac{h_{i}}{\left(z-z_{i}\right)^{2}}\right)\left\langle \Phi^{j_{1}}\left(x_{1};z_{1}\right)\Phi^{j_{2}}\left(x_{2};z_{2}\right)...\Phi^{j_{n}}\left(x_{n};z_{n}\right)\right\rangle \label{eq:Ward-3}
\end{align}
Substituting Eq. (\ref{eq:Sugawara}) into the left-hand side of Eq.
(\ref{eq:Ward-3}) and applying Wick's theorem yields the KZ equations
for the $H^{+}_{3}$-WZNW model.

However, as \citet{ribault2005h3+} observed, it is more convenient
to express the KZ equations in terms of the so-called $\mu$-basis
$\Phi^{j}\left(\mu;z\right)$, defined by:
\begin{equation}
\Phi^{j}\left(\mu;z\right)\coloneqq\frac{\left|\mu\right|^{2j+2}}{\pi}\int d^{2}xe^{\mu x-\bar{\mu}\bar{x}}\Phi^{j}\left(x;z\right).
\end{equation}
In this coordinate system $\mu$ on the celestial sphere, the KZ equations
take the form (\citet{ribault2005h3+}):
\begin{equation}
\frac{\partial\widehat{G}_{n}}{\partial z_{k}}+b^{2}\sum_{k\neq\ell}\frac{\mu_{k}\mu_{\ell}}{z_{k}-z_{\ell}}\left(\frac{\partial}{\partial\mu_{k}}-\frac{\partial}{\partial\mu_{\ell}}\right)^{2}\widehat{G}_{n}=b^{2}\sum_{k\neq\ell}\frac{\mu_{k}\mu_{\ell}}{z_{k}-z_{\ell}}\left[\frac{j_{k}\left(j_{k}+1\right)}{\mu^{2}_{k}}+\frac{j_{\ell}\left(j_{\ell}+1\right)}{\mu^{2}_{\ell}}\right]\widehat{G}_{n}.\label{eq:KZ}
\end{equation}

Following the standard prescription in celestial holography (see reviews
by \citet{pasterski2021lectures,raclariu2021lectures,strominger2018lectures,puhmcelestial}),
the scattering amplitude $\mathcal{A}_{n}$ associated with $\widehat{G}_{n}$
in the energy-momentum representation is obtained by performing an
inverse Mellin transform of the worldsheet integral of $\widehat{G}_{n}$:
\begin{equation}
\mathcal{A}_{n}=\prod^{n}_{i=1}\int^{c+i\infty}_{c-i\infty}\frac{d\Delta_{i}}{2\pi i}\omega^{-\Delta_{i}}_{i}\prod^{n}_{j=1}\int d^{2}z_{j}\widehat{G}_{n}\left(x_{1},...,x_{n};\Delta_{1},...,\Delta_{n};z_{1},...,z_{n}\right).
\end{equation}

Since our primary goal is to derive a system of partial differential
equations (PDEs) characterising the celestial $n$-point functions,
we may omit the worldsheet integrals and define instead a quantity
$G_{n}\coloneqq\left(x_{1},...,x_{n};\omega_{1},...,\omega_{n};z_{1},...,z_{n}\right)$
given by the inverse Mellin transform of $\widehat{G}_{n}$:
\begin{equation}
G_{n}=\prod^{n}_{i=1}\int^{c+i\infty}_{c-i\infty}\frac{d\Delta_{i}}{2\pi i}\omega^{-\Delta_{i}}_{i}\widehat{G}_{n}\left(x_{1},...,x_{n};\Delta_{1},...,\Delta_{n};z_{1},...,z_{n}\right).
\end{equation}
We aim to use the KZ equations to derive a system of PDEs for $G_{n}$.
Assuming the resulting equations can be solved by traditional methods,
the final amplitude is then given by integrating $G_{n}$ over the
worldsheet coordinates $z_{1},...,z_{n}$. Applying the inverse Mellin
transform to Eq. (\ref{eq:KZ}), and using the identities:
\begin{equation}
\omega\frac{\partial}{\partial\omega}\int^{c+i\infty}_{c-i\infty}\frac{d\Delta}{2\pi i}\omega^{-\Delta}\hat{f}\left(\Delta\right)=-\int^{c+i\infty}_{c-i\infty}\frac{d\Delta}{2\pi i}\Delta\omega^{-\Delta}\hat{f}\left(\Delta\right),
\end{equation}
and:
\begin{equation}
\left(\omega\frac{\partial}{\partial\omega}\right)^{2}\int^{c+i\infty}_{c-i\infty}\frac{d\Delta}{2\pi i}\omega^{-\Delta}\hat{f}\left(\Delta\right)=\int^{c+i\infty}_{c-i\infty}\frac{d\Delta}{2\pi i}\Delta^{2}\omega^{-\Delta}\hat{f}\left(\Delta\right),
\end{equation}
we obtain the following differential equation:
\begin{align}
 & \frac{\partial G_{n}}{\partial z_{k}}+b^{2}\sum_{k\neq\ell}\frac{\mu_{k}\mu_{\ell}}{z_{k}-z_{\ell}}\left(\frac{\partial}{\partial\mu_{k}}-\frac{\partial}{\partial\mu_{\ell}}\right)^{2}G_{n}=\frac{b^{2}}{2}\sum_{k\neq\ell}\frac{\mu_{k}\mu_{\ell}}{z_{k}-z_{\ell}}\left(\frac{\omega_{k}}{\mu^{2}_{k}}\frac{\partial}{\partial\omega_{k}}+\frac{\omega_{\ell}}{\mu^{2}_{\ell}}\frac{\partial}{\partial\omega_{\ell}}\right)G_{n}\\
 & +\frac{b^{2}}{4}\sum_{k\neq\ell}\frac{\mu_{k}\mu_{\ell}}{z_{k}-z_{\ell}}\left[\frac{1}{\mu^{2}_{k}}\left(\omega_{k}\frac{\partial}{\partial\omega_{k}}\right)^{2}+\frac{1}{\mu^{2}_{\ell}}\left(\omega_{\ell}\frac{\partial}{\partial\omega_{\ell}}\right)^{2}\right]G_{n}.
\end{align}

Let us now consider the worldsheet Ward identity:
\begin{equation}
\sum^{n}_{k=1}z_{k}\frac{\partial\widehat{G}_{n}}{\partial z_{k}}+\sum^{n}_{k=1}h_{k}\widehat{G}_{n}=0.\label{eq:Ward-4}
\end{equation}
Recalling that the scaling dimension $h_{k}$ in the $H^{+}_{3}$-WZNW
model is related to the spin $j_{k}$ by the relation $h_{k}=-b^{2}j_{k}\left(j_{k}+1\right)$,
and that the holographic dictionary prescribes that the spin is connected
to the celestial conformal weight $\Delta_{k}$ via $j_{k}=-\Delta_{k}/2$,
Eq. (\ref{eq:Ward-4}) can be reformulated as follows:
\begin{equation}
\sum^{n}_{k=1}z_{k}\frac{\partial\widehat{G}_{n}}{\partial z_{k}}+\frac{b^{2}}{2}\sum^{n}_{k=1}\Delta_{k}\widehat{G}_{n}-\frac{b^{2}}{4}\sum^{n}_{k=1}\Delta^{2}_{k}\widehat{G}_{n}=0.
\end{equation}
Finally, by applying the inverse Mellin transform, we obtain the following
differential equation:
\begin{equation}
\sum^{n}_{k=1}z_{k}\frac{\partial G_{n}}{\partial z_{k}}=\frac{b^{2}}{2}\sum^{n}_{k=1}\omega_{k}\frac{\partial G_{n}}{\partial\omega_{k}}+\frac{b^{2}}{4}\sum^{n}_{k=1}\left(\omega_{k}\frac{\partial}{\partial\omega_{k}}\right)^{2}G_{n}.\label{eq:PDE}
\end{equation}

\subsection{Celestial Stress-Energy Tensor\label{subsec:Virasoro-Current-Algebra}}

In this subsection, we shall discuss the celestial stress-energy tensor,
$\mathcal{T}\left(x\right)$. As a preliminary observation, note that,
based on the anticipated scaling dimension of $\mathcal{T}\left(x\right)$,
it necessarily follows that it may be represented as a linear combination
of the worldsheet currents, $J$ and $\bar{J}$, together with the
primary $\Phi$:
\begin{equation}
\mathcal{T}\left(x\right)\coloneqq\frac{1}{2k}\int_{\Sigma}d^{2}z\left(c_{1}J\partial^{2}_{x}\Phi+c_{2}\partial_{x}J\partial_{x}\Phi+c_{3}\Phi\partial^{2}_{x}J\right)\bar{J}.\label{eq:Stress-Tensor}
\end{equation}
Since all terms within brackets are evaluated at the worldsheet coordinates
$\left(z,\bar{z}\right)$ and the celestial coordinates $\left(x,\bar{x}\right)$,
we will temporarily omit the explicit arguments of the currents $J$
and $\bar{J}$, and the primary $\Phi$. However, when analysing the
behaviour of $\mathcal{T}\left(x\right)$ or its derivatives within
correlation functions, we will restore the explicit dependencies on
these variables to avoid ambiguity.

Given that the central extension $\mathcal{I}$, introduced in Eq.
(\ref{eq:Central-Extension}), is independent of $x$, we have $\partial^{2}_{x}\mathcal{I}=0$.
It follows that:
\begin{equation}
\int_{\Sigma}d^{2}z\,J\bar{J}\partial^{2}_{x}\Phi=-\int_{\Sigma}d^{2}z\,\left(\Phi\partial^{2}_{x}J+2\partial_{x}J\partial_{x}\Phi\right)\bar{J}.
\end{equation}
Substituting this into Eq. (\ref{eq:Stress-Tensor}), we derive the
expression:
\begin{equation}
\mathcal{T}\left(x\right)=\frac{1}{2k}\int_{\Sigma}d^{2}z\,\left(\alpha_{1}\partial_{x}J\partial_{x}\Phi+2\alpha_{2}\Phi\partial^{2}_{x}J\right)\bar{J},
\end{equation}
where the new constants $\alpha_{1}\coloneqq c_{2}-2c_{1}$ and $2\alpha_{2}\coloneqq c_{3}-c_{1}$
have been introduced. 

Using the identity $\bar{J}\Phi=\left(k/\pi\right)\partial_{\bar{z}}\lambda$,
and applying the complex-divergence theorem (Eq. (\ref{eq:Divergence-Theorem})),
we can rewrite $\mathcal{T}\left(x\right)$ as:
\begin{align}
\mathcal{T}\left(x\right) & =\frac{1}{2\pi}\int_{\Sigma}d^{2}z\,\left(\alpha_{1}\partial_{x}J\partial_{x}\partial_{\bar{z}}\lambda+2\alpha_{2}\partial^{2}_{x}J\partial_{\bar{z}}\lambda\right)\\
 & =\frac{1}{2}\oint\frac{dz}{2\pi i}\left(\alpha_{1}\partial_{x}J\partial_{x}\lambda+2\alpha_{2}\lambda\partial^{2}_{x}J\right).
\end{align}

To constrain the coefficients $\alpha_{1}$ and $\alpha_{2}$, we
impose BRST invariance, requiring that the term inside the brackets
in Eq. (\ref{eq:Stress-Tensor}) transforms as a worldsheet primary.
A more efficient approach, however, stems from recognising that, by
construction, $\mathcal{O}^{a}\left(y\right)$ generates a level-$\hat{k}_{G}$
celestial Kac-Moody current algebra and satisfies the Ward identities.
In particular, $\mathcal{T}\left(x\right)$ can only function as a
genuine stress tensor if the OPE $\mathcal{T}\left(x\right)\mathcal{O}^{a}\left(y\right)$
includes the term:
\begin{equation}
\mathcal{T}\left(x\right)\mathcal{O}^{a}\left(y\right)\sim...+\frac{1}{x-y}\partial_{y}\mathcal{O}^{a}\left(y\right)+...
\end{equation}

Thus, we compute this OPE and impose the above condition as a constraint.
It is advantageous to work with the partial derivative of $\mathcal{T}\left(x\right)$
with respect to the anti-holomorphic variable $\bar{x}$:
\begin{align}
\partial_{\bar{x}}\mathcal{T}\left(x\right) & =\frac{1}{2}\oint\frac{dz}{2\pi i}\left(\alpha_{1}\partial_{x}J\partial_{x}\partial_{\bar{x}}\lambda+2\alpha_{2}\partial^{2}_{x}J\partial_{\bar{x}}\lambda\right)\\
 & =\frac{1}{2}\oint\frac{dz}{2i}\left(\alpha_{1}\partial_{x}J\partial_{x}\Phi+2\alpha_{2}\Phi\partial^{2}_{x}J\right),
\end{align}
and evaluate its contribution within correlation functions that include
an insertion of the current $\mathcal{O}^{a}\left(y\right)$. 

To simplify our calculations, we divide $\mathcal{T}\left(x\right)$
into the components:
\begin{equation}
\mathcal{T}_{1}\left(x\right)\coloneqq\frac{\alpha_{1}}{2}\oint\frac{dz}{2i}\partial_{x}J\partial_{x}\Phi,\,\,\,\mathcal{T}_{2}\left(x\right)\coloneqq\alpha_{2}\oint\frac{dz}{2i}\Phi\partial^{2}_{x}J.
\end{equation}

The contribution of the OPE $\partial_{\bar{x}}\mathcal{T}_{1}\left(x\right)\mathcal{O}^{a}\left(y\right)$
to the correlation function is:
\begin{align}
 & \left\langle \partial_{x}\mathcal{T}_{1}\left(x\right)\mathcal{O}^{a}\left(y\right)...\right\rangle _{\text{CCFT}}\\
 & =-\frac{\alpha_{1}}{2k}\int_{\Sigma}d^{2}w\oint\frac{dz}{2i}\langle j^{a}\left(w\right)\partial_{x}J\left(x;z\right)\partial_{x}\Phi\left(x,\bar{x};z,\bar{z}\right)\Phi\left(y,\bar{y};w,\bar{w}\right)\bar{J}\left(\bar{y};\bar{w}\right)...\rangle\\
 & \sim-\frac{\alpha_{1}}{2k}\partial_{x}\delta^{\left(2\right)}\left(x-y\right)\int_{\Sigma}d^{2}w\oint_{C_{\varepsilon}\left(w\right)}\frac{dz}{2i}\langle j^{a}\left(w\right)\partial_{x}J\left(x;z\right)\Phi\left(y,\bar{y};w,\bar{w}\right)\bar{J}\left(\bar{y};\bar{w}\right)...\rangle.\label{eq:Step-X}
\end{align}

Recalling (cf. \citet[Sec. 2]{giveon1998comments}) that the worldsheet
current operator $J\left(x;z\right)$ acts on the primary $\Phi\left(x,\bar{x};z,\bar{z}\right)$
as:
\begin{equation}
J\left(x;z\right)\Phi\left(y,\bar{y};w,\bar{w}\right)\sim\frac{1}{z-w}\left[\left(y-x\right)^{2}\partial_{y}+2\left(y-x\right)\right]\Phi\left(y,\bar{y};w,\bar{w}\right),
\end{equation}
we take the partial derivative with respect to $x$ to obtain:
\begin{equation}
\partial_{x}J\left(x;z\right)\Phi\left(y,\bar{y};w,\bar{w}\right)\sim\frac{2}{z-w}\left[\left(x-y\right)\partial_{y}-1\right]\Phi\left(y,\bar{y};w,\bar{w}\right).\label{eq:Identity}
\end{equation}
Therefore,
\begin{equation}
\langle\partial_{\bar{x}}\mathcal{T}_{1}\left(x\right)\mathcal{O}^{a}\left(y\right)...\rangle_{\text{CCFT}}\sim\frac{\partial}{\partial\bar{x}}\left(\frac{\alpha_{1}}{\left(x-y\right)^{2}}\left\langle \mathcal{O}^{a}\left(y\right)...\right\rangle _{\text{CCFT}}-\frac{\alpha_{1}}{x-y}\left\langle \partial_{y}\mathcal{O}^{a}\left(y\right)...\right\rangle _{\text{CCFT}}\right).\label{eq:Step-4}
\end{equation}

Next, consider the contribution from $\mathcal{T}_{2}\left(x\right)$.
Taking its partial derivative and inserting it into the correlation
function with the Kac-Moody current $\mathcal{O}^{a}\left(y\right)$
yields:
\begin{align}
 & \langle\partial_{\bar{x}}\mathcal{T}_{2}\left(x\right)\mathcal{O}^{a}\left(y\right)...\rangle_{\text{CCFT}}\\
 & \sim-\frac{\alpha_{2}}{k}\int d^{2}w\oint\frac{dz}{2i}\langle\partial^{2}_{x}J\left(x;z\right)\Phi\left(x,\bar{x};z,\bar{z}\right)\Phi\left(y,\bar{y};w,\bar{w}\right)\bar{J}\left(\bar{y};\bar{w}\right)j^{a}\left(w\right)...\rangle.\label{eq:Step-2}
\end{align}

Taking the partial derivative of Eq. (\ref{eq:Identity}) with respect
to $x$ gives:
\begin{equation}
\partial^{2}_{x}J\left(x;z\right)\Phi\left(y,\bar{y};w,\bar{w}\right)\sim\frac{2}{z-w}\partial_{y}\Phi\left(y,\bar{y};w,\bar{w}\right).
\end{equation}
Substituting this result into Eq. (\ref{eq:Step-2}) gives:
\begin{equation}
\langle\partial_{\bar{x}}\mathcal{T}_{2}\left(x\right)\mathcal{O}^{a}\left(y\right)...\rangle_{\text{CCFT}}\sim\frac{\partial}{\partial\bar{x}}\left(\frac{2\alpha_{2}}{\left(x-y\right)^{2}}\left\langle \mathcal{O}^{a}\left(y\right)...\right\rangle _{\text{CCFT}}+\frac{2\alpha_{2}}{x-y}\left\langle \partial_{y}\mathcal{O}^{a}\left(y\right)...\right\rangle _{\text{CCFT}}\right).\label{eq:Step-3}
\end{equation}

Consequently, combining the results obtained in Eqs. (\ref{eq:Step-4},
\ref{eq:Step-3}), and integrating with respect to $\bar{x},$ we
finally obtain:
\begin{equation}
\mathcal{T}\left(x\right)\mathcal{O}^{a}\left(y\right)\sim\frac{2\alpha_{2}+\alpha_{1}}{\left(x-y\right)^{2}}\mathcal{O}^{a}\left(y\right)+\frac{2\alpha_{2}-\alpha_{1}}{x-y}\partial_{y}\mathcal{O}^{a}\left(y\right).\label{eq:Ward-2}
\end{equation}

Thus, a necessary condition for $\mathcal{T}\left(x\right)$ to be
a valid stress tensor in the celestial CFT is that the coefficient
of $\left(x-y\right)^{-1}\partial_{y}\mathcal{O}^{a}\left(y\right)$
in Eq. (\ref{eq:Ward-2}) must equal one. This implies that the constants
$\alpha_{1}$ and $\alpha_{2}$ must satisfy the constraint $2\alpha_{2}-\alpha_{1}=1$.
A convenient choice is $\alpha_{1}=\alpha_{2}=1$, which leads to
the final form of the celestial stress tensor:
\[
\mathcal{T}\left(x\right)=\frac{1}{2}\oint\frac{dz}{2\pi i}\left(\partial_{x}J\partial_{x}\lambda+2\lambda\partial^{2}_{x}J\right).
\]

In conclusion, following the derivation presented in Section 3 of
\citet{kutasov1999more}, the Ward identity for the OPE of the celestial
stress-energy tensor $\mathcal{T}\mathcal{T}$ is
\begin{equation}
\mathcal{T}\left(x_{1}\right)\mathcal{T}\left(x_{2}\right)\sim\frac{c^{(\text{st})}/2}{\left(x_{1}-x_{2}\right)^{4}}+\frac{2\mathcal{T}\left(x_{2}\right)}{\left(x_{1}-x_{2}\right)^{2}}+\frac{\partial_{x_{2}}\mathcal{T}\left(x_{2}\right)}{x_{1}-x_{2}}.\label{eq:-6}
\end{equation}

The constant $c^{(\text{st})}$ entering Eq. (\ref{eq:-6}) is the
central charge of the spacetime Virasoro algebra associated with the
stress-energy tensor $\mathcal{T}(x)$. In the $\mathrm{AdS}_{3}$
string-theory construction of \citet{kutasov1999more}, this central
charge is determined by the so-called central-extension operator $I$,
which is given by
\begin{equation}
I=\frac{1}{k^{2}}\int\mathrm{d}^{2}z\;J(x;z)\overline{J}(\bar{x};\bar{z})\Phi_{1}(x,\bar{x};z,\bar{z})\,.
\end{equation}
We recall that \citet{kutasov1999more} demonstrated that the $I$-operator
is independent of $x$ and $\bar{x}$, and behaves as a spacetime
identity operator up to a proportionality factor. Let us denote by
$\mathcal{P}(g_{s})$ the eigenvalue such that
\begin{equation}
\langle\langle IV_{1}\cdot\cdot\cdot V_{n}\rangle\rangle=\mathcal{P}(g_{s})\langle\langle IV_{1}\cdot\cdot\cdot V_{n}\rangle\rangle\,,\label{eq:-7}
\end{equation}
where $g_{s}$ denotes the $\mathrm{AdS}_{3}$ string coupling constant.
The double-bracket notation in the preceding equation signifies that
the expression holds when the correlators on each side of the equality
receive corrections from all worldsheet topologies, including disconnected
ones, and we refer the reader to \citet{de1999string} for a careful
discussion of this point. Furthermore, \citet{kutasov1999more} showed
that the central charge $c^{(\text{st})}$ is related to the level
$k$ and the eigenvalue $\mathcal{P}(g_{s})$ by
\begin{equation}
c^{(\text{st})}=6k\mathcal{P}(g_{s})\,.
\end{equation}
For clarity, we emphasise that $c^{(\text{st})}$ should not be confused
with the worldsheet $H^{+}_{3}$-WZNW central charge, which is given
by
\begin{equation}
c_{H^{+}_{3}}=\frac{3k}{k-2}\,.
\end{equation}

We now discuss a possible physical interpretation of Eq. (\ref{eq:-6}),
and in particular of the central charge $c^{(\text{st})}$, from the
point of view of the four-dimensional bulk theory. To contextualise,
we first recall that \citet{barnich2011bms} analysed the general
structure of the BMS algebra, and the associated Dirac bracket, in
three- and four-dimensional flat spacetimes by considering the classical
symmetry transformations and charges. They found that the global sub-algebra
in four dimensions has no central charges. This result was subsequently
confirmed through the computation of the celestial $\mathcal{T}\mathcal{T}$
OPE by \citet{fotopoulos2020extended} and \citet{donnay2021bms}.
Interestingly, however, \citet{barnich2017centrally} and \citet{distler2019double}
observed that a nontrivial extension of the classical algebra by a
generalised $2$-cocycle exists when the BMS algebra is promoted to
include local, singular superrotations. This \emph{suggests} that,
in our model for perturbative celestial CFT based on string theory
on $\mathrm{AdS}_{3}\times X$, the central charge $c^{(\text{st})}$
of the boundary CFT should be related to the bulk theory through the
action of superrotations. Formulating this statement precisely, and
using it to constrain the kinds of critical string compactifications
that are consistent with the behaviour of four-dimensional flat-space
bulk theories under superrotations, remains a long-standing goal of
our programme.

\section{Discussion\label{sec:Discussion}}

In this work, we have introduced a novel holographic dictionary connecting
$AdS_{3}$ string theory with celestial CFT. Building upon the foundational
ideas of \citet{giveon1998comments,de1999string,maldacena2001strings,maldacena2001strings2},
and drawing inspiration from the model advanced by \citet{ogawa2024celestial},
we have demonstrated that the correlation function, stress-energy
tensor, Kac-Moody currents, and vertex operators of the celestial
CFT can be systematically derived from their counterparts in the $H^{+}_{3}$-WZNW
model. Moreover, by applying the results rigorously established by
the mathematical physicists \citet{ribault2005h3+,teschner1997conformal,teschner1999mini,teschner1999structure,teschner2000operator},
we have determined the two- and three-point functions, as well as
the structure constants of the celestial OPE. Additionally, we have
derived a system of partial differential equations that characterise
celestial amplitudes in the frequency-momentum space, and have offered
a brief analysis of the infrared and ultraviolet behaviour of these
amplitudes.

We should point out that the holographic dictionary developed in the
present manuscript is consistent with the absence of standard reflection
positivity and the associated non-unitarity expected in celestial
conformal field theories. For, as explained by \citet{gawedzki1992non}
and recently reviewed in subsection 5.2 of \citet{murthy2026localization},
the full worldsheet-matter CFT appertaining to the $\mathrm{H}^{+}_{3}\text{-WZNW}$
model is itself non-unitary, notwithstanding the positiveness of the
$\mathrm{SL}(2,\mathbf{C})$-invariant inner product underlying the
Hilbert state space $L^{2}(\mathrm{H}^{+}_{3},\mathsf{d}h)$ describing
the quantum mechanics of the zero-mode truncation of the $\mathrm{H}^{+}_{3}\text{-WZNW}$
model. Independently, in the dictionary between celestial CFT and
the $\mathrm{H}^{+}_{3}\text{-WZNW}$ model originally proposed by
\citet{ogawa2024celestial} and developed in our work, the conformal
dimensions $\Delta$ associated with the celestial CFT and the scaling
dimensions $(h_{\mathrm{ws}},\overline{h}_{\mathrm{ws}})$ belonging
to the worldsheet-matter CFT of the $\mathrm{H}^{+}_{3}\text{-WZNW}$
model are distinct physical quantities: for isospin $j=-\frac{1}{2}+\mathtt{i}s$,
the celestial conformal dimension assigned by our dictionary is $\Delta_{j}=-2j=1-2\mathtt{i}s$,
whereas the conformal weights of the worldsheet-matter CFT are $h_{\mathrm{ws}}=\overline{h}_{\mathrm{ws}}=(\frac{1}{4}+s^{2})/(k-2)$.
Therefore, whenever $s\neq0$ and $k>2$, we have $h_{\mathrm{ws}},\overline{h}_{\mathrm{ws}}>0$
but $\Delta_{j}$ \emph{complex}, so that these celestial conformal
primaries cannot create non-zero radial physical states of finite,
positive norm in ordinary radial quantisation with a self-adjoint
dilatation operator.

Over the past decades, string theory has provided profound insights
into quantum gravity, inspiring us to undertake a broader research
programme, of which this paper constitutes the inaugural contribution.
Our objective is to investigate certain aspects of celestial holography
through the framework of string theory. The holographic dictionary
proposed herein opens multiple avenues for further research. Notably,
\citet{de2023celestial,de2023eikonal} have explored the derivation
of various aspects of celestial holography from the $AdS$/CFT correspondence
in a series of remarkable papers. We believe that our proposed holographic
dictionary between $AdS_{3}$ string theory and celestial CFT could
serve as a valuable toy model to support these efforts.

We further recall the well-known construction of black holes in $d=2+1$
gravity with a negative cosmological constant, introduced by \citet{banados1992black},
and subsequently shown by \citet{cangemi1993gauge} to arise as orbifold
solutions of string theory. \citet{strominger1998black} proposed
a unified framework for all ``black objects'' with near-horizon geometries
described by $AdS_{3}$, encompassing the BTZ solution. Building on
these works, we are optimistic that insights into black hole solutions
in string theory might be employed to study black holes in asymptotically
flat spaces, informed by the ideas presented in this paper. In particular,
these insights may intersect with the constructions recently discussed
by \citet{crawley2022black}. In our forthcoming note, we shall investigate
the connection between $\mathcal{N}=2$ string theory in Klein space
and celestial conformal blocks.

An important problem to be addressed in a forthcoming publication
concerns the description of quantum, namely loop-level, corrections
to celestial amplitudes in terms of the candidate models for perturbative
celestial CFTs constructed from $\mathrm{AdS}_{3}$ string theory
proposed in our work. Therefore, let us close our discussion by outlining
the strategy that we shall follow in order to incorporate quantum-theoretic
corrections to celestial amplitudes. In \citet{mol2024partial}, we
derived a set of partial differential equations (PDEs) governing gluonic
celestial amplitudes for a model of perturbative celestial CFT based
on Liouville theory. Following the proposal of \citet{stieberger2023yang},
this model may be referred to as celestial Liouville theory. These
PDEs are parameterised by the Liouville coupling constant $b$, and
we solved them perturbatively to obtain corrections to the gluonic
celestial leaf amplitudes in the MHV subsector of Yang-Mills theory
up to order $\mathcal{O}(b^{2})$. Then, following \citet{stieberger2023celestial},
these logarithmic corrections were identified with corrections expected
in one-loop gauge-theory amplitudes. We confirmed this identification
by showing that the celestial OPEs for gluons obtained from celestial
Liouville theory at order $\mathcal{O}(b^{2})$ match the expected
one-loop corrected celestial OPEs obtained by the method of collinear
singularities, employed by \citet{fan2019soft} for extracting celestial
OPEs.

Now, recall that, in Section III of the present manuscript, we derived
an analogous system of PDEs governing the celestial amplitudes for
the models of (perturbative) celestial CFTs proposed in our work based
on $\mathrm{AdS}_{3}$ string theory. These PDEs are parameterised
by a constant $b$ analogous to the Liouville coupling parameter.
Here, however, $b$ is determined by the $H^{+}_{3}$-WZNW level $k$
through $b^{2}=1/(k-2)$.

In our forthcoming publication, we shall demonstrate that the PDEs
obtained in Section III may also be solved perturbatively, and that
the corresponding corrections at order $\mathcal{O}(b^{2})$ may likewise
be identified with corrections associated with one-loop gauge-theory
and gravity amplitudes. In the case of $\mathrm{AdS}_{3}$ string
theory, however, these one-loop corrections admit a particularly compelling
physical interpretation. Remember that, in the present manuscript,
we showed that tree-level MHV celestial amplitudes for gluons and
gravitons can be expressed as the helicity projection of the leading-trace,
mini-superspace limit of correlators of appropriately dressed celestial
vertex operators. On the other hand, in the mini-superspace limit
$k\to\infty$, whose mathematical analysis was carried out in detail
by \citet{teschner1999mini}, the $H^{+}_{3}$-WZNW model reduces
to its zero-mode sector, which is identified with the quantum mechanics
of a point particle on $H^{+}_{3}$. Thus, if the finite-$k$ contributions
to the correlator can be interpreted, at least in the MHV sector,
as loop corrections to celestial amplitudes, these corrections may
be understood as arising from the stringy degrees of freedom of the
$H^{+}_{3}$-WZNW model. By contrast, the tree-level celestial amplitudes
arise when this model reduces, in the mini-superspace limit, to the
quantum mechanics of a particle on $H^{+}_{3}$.

To conclude, we would like to draw attention to the works of \citet{StiebergerTaylor2018Strings},
\citet{Jiang2022CelestialOPEs}, \citet{Donnay2023CelestialOpenStrings}
and \citet{KervynStieberger2025HighEnergy}, which, although differing
in the particular way in which string theory is employed to investigate
celestial CFT, belong, in our view, to the same broader and also ambitious
research programme of using string theory as an intellectual laboratory
for the study of celestial holography.

\section{Acknowledgement}

The author thanks Giorgio Torrieri for his important encouragement.
I also thank the referee for the careful consideration of my manuscript.

I gratefully acknowledge financial support from FAPEMIG (Minas Gerais)
and the Instituto de Ciências Exatas, Universidade Federal de Juiz
de Fora (UFJF). Part of this work was completed while I was hosted
by the Departamento de Matemática, Instituto de Ciências Exatas (ICEx),
Universidade Federal de Minas Gerais (UFMG).

\appendix

\section{Mini-Introduction to $H^{+}_{3}$-WZNW Model}

In this section, we develop a holographic dictionary that maps the
worldsheet vertex operators and generators of worldsheet Kac-Moody
and Virasoro current algebras to celestial conformal primaries, allowing
for the construction of the celestial conformal field theory (CFT)
entirely from the data provided by string theory. To this end, we
begin by reviewing the $H^{+}_{3}$-WZNW model, which we identify
as the analytic continuation of string theory on Lorentzian $AdS_{3}$
to its Euclidean counterpart, $H^{+}_{3}$.

\subsection{$H^{+}_{3}$-WZNW Model Review}

\citet{brown1986central} demonstrated that any theory of  $3d$ gravity
with a negative cosmological constant is endowed with an infinite-dimensional
symmetry group that includes two commuting copies of Virasoro algebras,
thus describing a $2d$ CFT. This observation was given a concrete
form in the context of $AdS_{3}$ string theory by \citet{giveon1998comments,de1999string,maldacena2001strings}.
In this work, for simplicity, we focus on the analytically continued
Euclidean $AdS_{3}$ string theory, though analogous considerations
apply to the $SL\left(2,\mathbf{R}\right)$ model studied by \citet{maldacena2001strings}.

\subsubsection{Classical Theory}

Let $\left(u^{0},\vec{u}\right)$ be Cartesian coordinates on $4d$
Minkowski space $\mathbf{R}^{1}_{3}$. Recall that Euclidean $AdS_{3}$
can be formulated as the hypersurface $H_{3}\subset\mathbf{R}^{1}_{3}$
defined by $-(u^{0})^{2}+\sum^{3}_{i=1}(u^{i})^{2}=-\ell^{2}$. Thus,
$H_{3}$ is a submanifold of $\mathbf{R}^{1}_{3}$ with constant negative
scalar curvature $R=-1/\ell^{2}$ and an isometry group given by $SL\left(2,\mathbf{C}\right)$.
This space can be parametrised by the coordinates $\left(r,\tau,\theta\right)$,
given by:
\begin{equation}
u^{0}=\sqrt{\ell^{2}+r^{2}}\cosh\tau,\,\,\,u^{1}=r\sin\theta,\,\,\,u^{2}=r\cos\theta,\,\,\,u^{3}=\sqrt{\ell^{2}+r^{2}}\sinh\tau,
\end{equation}
where $\theta\in[0,2\pi)$ and $r\geq0$. The first fundamental form
of $H_{3}$, inherited from the ambient space $\mathbf{R}^{1}_{3}$,
is given by the line element:
\begin{equation}
ds^{2}=\frac{1}{1+r^{2}/\ell^{2}}dr^{2}+\ell^{2}\left(1+\frac{r^{2}}{\ell^{2}}\right)d\tau^{2}+r^{2}d\theta^{2}.\label{eq:Line-Element}
\end{equation}
There are two disconnected components characterising the topology
of $H_{3}$, one corresponding to $u^{0}>0$ and the other to $u^{0}<0$.
We restrict our attention to the former, and the resulting hypersurface,
$H^{+}_{3}$, satisfies $\left|u^{0}\right|>u^{3}$.

To connect this with the $H^{+}_{3}$-WZNW model, it is more convenient
to introduce the coordinates $\left(\phi,\gamma,\bar{\gamma}\right)$,
defined as:
\begin{equation}
\phi=\log\ell^{-1}\left(u^{0}+u^{3}\right),\,\,\,\gamma=\frac{u^{2}+iu^{1}}{u^{0}+u^{3}},\,\,\,\bar{\gamma}=\frac{u^{2}-iu^{1}}{u^{0}+u^{3}}.
\end{equation}
In these coordinates, the line element becomes:
\begin{equation}
ds^{2}=\ell^{2}\left(d\phi^{2}+e^{2\phi}d\gamma d\bar{\gamma}\right).\label{eq:Line-Element-1}
\end{equation}
Notice that the boundary of Euclidean $AdS_{3}$ corresponds to $\phi\rightarrow\infty$,
which is homeomorphic to $\mathbf{CP}^{1}$, parametrised by $\left(\gamma,\bar{\gamma}\right)$.

As explained by \citet{maldacena1999ads3}, the equations of motion
for strings propagating in a background with the line element given
by Eq. (\ref{eq:Line-Element-1}) can only be satisfied by turning
on a Neveu-Schwarz two-form field $\mathbf{B}\coloneqq\left(1/2\right)B_{\mu\nu}dx^{\mu}\wedge dx^{\nu}$.
One way to understand this requirement is from the worldsheet perspective,
where a non-vanishing $\mathbf{B}$ is necessary to maintain conformal
invariance. It turns out, as shown by \citet{giveon1998comments,maldacena1999ads3,de1999string},
that the required background is characterised by $\mathbf{B}=\ell^{2}e^{2\phi}d\gamma\wedge d\bar{\gamma}$.
Consequently, the action integral for the worldsheet becomes\footnote{In these notes, we follow the conventions of \citet{francesco1997conformal}
regarding the integral measure on the Riemann sphere, such that:
\begin{equation}
d^{2}z\coloneqq\frac{1}{2i}d\bar{z}\wedge dz.
\end{equation}
As a result, the complex delta-function is represented as:
\begin{equation}
\partial_{\bar{z}}\frac{1}{z-w}=\pi\delta^{\left(2\right)}\left(z-w\right),\label{eq:Delta-Function}
\end{equation}
and the divergence theorem takes the form:
\begin{equation}
\int_{\Sigma}d^{2}z\left(\partial_{z}v^{z}+\partial_{\bar{z}}v^{\bar{z}}\right)=\frac{1}{2i}\oint_{\partial\Sigma}\left(dzv^{\bar{z}}-d\bar{z}v^{z}\right),\label{eq:Divergence-Theorem}
\end{equation}
where $\Sigma\subset\mathbf{CP}^{1}$ is a simply-connected region.}:
\begin{equation}
I\left[\phi,\gamma,\bar{\gamma}\right]=2\left(\frac{\ell}{\ell_{0}}\right)^{2}\int_{\Sigma}d^{2}z\left(\partial_{z}\phi\partial_{\bar{z}}\phi+e^{2\phi}\partial_{\bar{z}}\gamma\partial_{z}\bar{\gamma}\right).\label{eq:Worldsheet-Action}
\end{equation}
Here, $\ell_{0}$ denotes the fundamental string length, with $\ell^{2}_{0}=\alpha'$.
In this manuscript, we follow closely the notational conventions and
normalisations introduced by \citet{giveon1998comments}, according
to which the level $k$ of the $H^{+}_{3}$-WZNW model is related
to the $\mathrm{AdS}_{3}$ radius by
\begin{equation}
\ell^{2}=k\,\ell^{2}_{0}\,.
\end{equation}
Consequently, the large-level limit $k\to\infty$, also known as the
mini-superspace limit (see \citet{teschner1999mini}), may be identified
with the regime in which the $\mathrm{AdS}_{3}$ radius is much larger
than the fundamental string length, so that $\ell^{2}/\ell^{2}_{0}\to\infty$.
Let us further observe that the overall numerical factors in the sigma-model
action in Eq. (\ref{eq:Worldsheet-Action}) depend on the conventions
used for the worldsheet measure and for $\alpha'$.

The action in Eq. (\ref{eq:Worldsheet-Action}) can be identified
with the $H^{+}_{3}$-WZNW model, an example of a non-compact Wess-Zumino-Novikov-Witten
model. The Lagrangian formulation of this model was developed by \citet{gawedzki1989coset}
and later expanded by \citet{gawedzki1992non} with a focus on path-integral
methods. The pure WZNW action integral at level-$k$ is given by:
\begin{equation}
S\left[H\right]=-\frac{1}{2\pi}\int_{\Sigma}d^{2}z\text{tr}\left(H^{-1}\partial_{z}H\right)\left(H^{-1}\partial_{\bar{z}}H\right)+k\Gamma,\label{eq:WZNW-Action}
\end{equation}
where $\Gamma$ is the Wess-Zumino topological term. Parametrising
the coset space $SL\left(2,\mathbf{C}\right)/SU\left(2\right)$, which
consists of complex two-by-two Hermitian matrices with unit determinant,
by:
\begin{equation}
H\left(\phi,\gamma,\bar{\gamma}\right)=\begin{pmatrix}\left|\gamma\right|^{2}e^{\phi}+e^{-\phi} & \gamma e^{\phi}\\
e^{\phi}\bar{\gamma} & e^{\phi}
\end{pmatrix},
\end{equation}
one can see that the worldsheet action in Eq. (\ref{eq:WZNW-Action})
is \emph{classically} equivalent to the action in Eq. (\ref{eq:Worldsheet-Action})
when $H\left(\phi,\gamma,\bar{\gamma}\right)$ is restricted to this
form. As \citet{gawedzki1989quadrature} demonstrated using path-integral
techniques, these theories remain equivalent at the quantum level.
This identification justifies the equivalence between (bosonic) string
theory on Euclidean $AdS_{3}$ and the $H^{+}_{3}$-WZNW model.

\subsubsection{Quantum Theory}

Having reviewed the classical formulation of the $H^{+}_{3}$-WZNW
model, we now turn to the construction of the quantum state space,
following the approach of \citet{teschner1999mini}. In the Schrödinger
representation, the Hilbert space of the $H^{+}_{3}$-WZNW model is
given by $\mathcal{H}=L^{2}\left(H^{+}_{3},dh\right)$, the Lebesgue
space of square-integrable complex-valued functions on the hypersurface
$H^{+}_{3}$ with respect to the measure:
\begin{equation}
dh=d\phi d^{2}\gamma,\,\,\,h=\begin{pmatrix}e^{\phi}\sqrt{1+\gamma\bar{\gamma}} & u\\
\bar{u} & e^{-\phi}\sqrt{1+\gamma\bar{\gamma}}
\end{pmatrix}.\label{eq:Measure}
\end{equation}
We shall call the elements of $\mathcal{H}$ \emph{wavefunctions}
of the quantum theory.

We also define an action of the symmetry group $SL\left(2,\mathbf{C}\right)$
on wavefunctions $\Psi\in\mathcal{H}$, given by: 
\begin{equation}
g\in SL\left(2,\mathbf{C}\right)\mapsto T_{g}\Psi\left(h\right)=\Psi\left(g^{-1}h\left(g^{-1}\right)^{\dagger}\right),\label{eq:Action}
\end{equation}
for each point $h\in H^{+}_{3}$. 

The inner product on the quantum state space is defined by:
\begin{equation}
\left\langle \Psi_{1}\big|\Psi_{2}\right\rangle \coloneqq\int_{H^{+}_{3}}dh\,\Psi^{*}_{2}\left(h\right)\Psi_{1}\left(h\right).\label{eq:Inner-Product}
\end{equation}
Using the Iwasawa decomposition, it can be shown that for all pairs
of states $\Psi_{1},\Psi_{2}\in\mathcal{H}$, the inner product becomes:
\begin{equation}
\left\langle \Psi_{1}\big|\Psi_{2}\right\rangle =\frac{1}{\mathcal{V}_{SU\left(2\right)}}\int_{SL\left(2,\mathbf{C}\right)}dg\,\Psi^{*}_{2}(gg^{\dagger})\Psi_{1}(gg^{\dagger}).\label{eq:Iwasawa-Decomposed-Inner-Product}
\end{equation}
where $\mathcal{V}_{SU\left(2\right)}$ denotes the volume of $SU\left(2\right)$,
and $dg$ represents the $SL\left(2,\mathbf{C}\right)$-invariant
measure:
\begin{equation}
dg=\left(\frac{i}{2}\right)^{4}d^{2}\zeta_{1}d^{2}\zeta_{2}d^{2}\zeta_{3}d^{2}\zeta_{4}\,\delta^{\left(2\right)}\left(\zeta_{1}\zeta_{4}-\zeta_{2}\zeta_{3}\right),\,\,\,g=\begin{pmatrix}\zeta_{1} & \zeta_{2}\\
\zeta_{3} & \zeta_{4}
\end{pmatrix}.
\end{equation}

The rationale behind the choice of the inner product defined in Eq.
(\ref{eq:Inner-Product}) is that it ensures the unitary realisation
of the $SL\left(2,\mathbf{C}\right)$ action on the space $\mathcal{H}$
of wavefunctions, as introduced in Eq. (\ref{eq:Action}). This property
arises from Eq. (\ref{eq:Iwasawa-Decomposed-Inner-Product}) and the
invariance of the measure $dg$ on $SL\left(2,\mathbf{C}\right)$,
which satisfies $d\left(g_{0}g\right)=dg$.

We now introduce a key element in bridging the abstract mathematical
formalism of Hilbert spaces to physics through the specification of
an appropriate basis. In non-relativistic quantum mechanics, a convenient
choice of basis is provided by the plane waves of definite momentum,
which afford a clear physical interpretation. Analogously, in the
$H^{+}_{3}$-WZNW model, there exists a counterpart to the ``plane-wave''
basis, represented by the so-called conformal primaries with spin
$j$, denoted by $\Phi^{j}\left(x;z\right)$, and defined as:
\begin{equation}
\Phi^{j}\left(x;h\right)\coloneqq\frac{1}{\pi}\left(\begin{pmatrix}x & 1\end{pmatrix}\cdot h\cdot\begin{pmatrix}\bar{x}\\
1
\end{pmatrix}\right)^{2j},\,\,\,h\in SL\left(2;\mathbf{C}\right).
\end{equation}
Here, we adopt the normalisation introduced by \citet{giveon1998comments}.
Furthermore, by parametrising $h\in SL\left(2;\mathbf{C}\right)$
as in Eq. (\ref{eq:Measure}), and treating $\phi,\gamma,\bar{\gamma}$
as fields depending on the points $z,\bar{z}$ on the worldsheet,
we simplify the notation by writing $\Phi^{j}\left(x;z\right)\coloneqq\Phi^{j}\left(x;h\left(\phi,\gamma,\bar{\gamma}\right)\right)$.
Specifically, $x,\bar{x}$ will denote coordinates on the celestial
sphere, while $z,\bar{z}$ will serve as coordinates on the string
worldsheet. 

It was demonstrated by \citet[Appendix A]{teschner1999mini} through
spectral decomposition that the set of functions $\{\Phi^{j}\left(x;z\right):x\in\mathbf{C};i\in-1/2+i\mathbf{R}^{+}\}$
forms a complete and $\delta$-function normalisable basis, such that:
\begin{equation}
\left\langle \Phi^{j}\left(x;h\right),\Phi^{j'}\left(x';h\right)\right\rangle \propto\delta^{\left(2\right)}\left(x-x'\right)\delta\left(j-j'\right),
\end{equation}
for all $x,\bar{x}\in\mathbf{C}$ and $j,j'\in-1/2+i\mathbf{R}^{+}$.
In the following subsection, we will review how these wavefunctions
can be employed to construct a Fourier-type transform, further reinforcing
their role as analogues of ``plane-waves'' in the quantum theory of
the $H^{+}_{3}$-WZNW model.

\subsubsection{Symmetries\label{subsec:Symmetries}}

No discussion of quantum theory would be reasonable without emphasising
the role played by symmetries, and the case of the $H^{+}_{3}$-WZNW
model is no exception. Following \citet[Section 2]{giveon1998comments},
we observe that the classical theory described by the action integral
in Eq. (\ref{eq:Worldsheet-Action}) possesses an infinite-dimensional
affine $SL\left(2;\mathbf{C}\right)\times\overline{SL\left(2;\mathbf{C}\right)}$
symmetry, the global part of which admits the following generators:
\begin{equation}
J^{-}_{0}\coloneqq\frac{\partial}{\partial\gamma},\,\,\,J^{3}_{0}\coloneqq\gamma\frac{\partial}{\partial\gamma}-\frac{1}{2}\frac{\partial}{\partial\phi},\,\,\,J^{+}_{0}\coloneqq\gamma^{2}\frac{\partial}{\partial\gamma}-\gamma\frac{\partial}{\partial\phi}-e^{-2\phi}\frac{\partial}{\partial\bar{\gamma}},\label{eq:Currents}
\end{equation}
and similarly:
\begin{equation}
\bar{J}^{-}_{0}\coloneqq\frac{\partial}{\partial\bar{\gamma}},\,\,\,\bar{J}^{3}_{0}\coloneqq\bar{\gamma}\frac{\partial}{\partial\bar{\gamma}}-\frac{1}{2}\frac{\partial}{\partial\phi},\,\,\,\bar{J}^{+}_{0}\coloneqq\bar{\gamma}^{2}\frac{\partial}{\partial\bar{\gamma}}-\bar{\gamma}\frac{\partial}{\partial\phi}-e^{-2\phi}\frac{\partial}{\partial\gamma}.\label{eq:Currents-1}
\end{equation}
A significant class of observables in the $H^{+}_{3}$-WZNW model
consists of smooth functions on the hyperboloid $H^{+}_{3}$. A convenient
method to analyse the decomposition of these functions in terms of
representations of $SL\left(2;\mathbf{C}\right)\times\overline{SL\left(2;\mathbf{C}\right)}$
is to introduce a pair of complex variables $\left(x,\bar{x}\right)$
and realise the Lie algebra of $SL\left(2;\mathbf{C}\right)\times\overline{SL\left(2;\mathbf{C}\right)}$
by defining the following generators as linear differential operators
acting on the germ of smooth functions on $H^{+}_{3}$:
\begin{equation}
J^{-}_{0}=-\frac{\partial}{\partial x},\,\,\,J^{3}_{0}=-\left(x\frac{\partial}{\partial x}+p\right),\,\,\,J^{+}_{0}=-\left(x^{2}\frac{\partial}{\partial x}+2px\right),
\end{equation}
and:
\begin{equation}
\bar{J}^{-}_{0}=-\frac{\partial}{\partial\bar{x}},\,\,\,\bar{J}^{3}_{0}=-\left(\bar{x}\frac{\partial}{\partial\bar{x}}+2\bar{p}\right),\,\,\,\bar{J}^{+}_{0}=-\left(\bar{x}^{2}\frac{\partial}{\partial\bar{x}}+2\bar{p}\bar{x}\right),\label{eq:Currents-2}
\end{equation}
where $p$ is related to the ``spin'' of the $SL\left(2;\mathbf{C}\right)$
representation by $p=j+1$. Indeed, it should be recalled that the
value of the (quadratic) Casimir in this representation is $j\left(j+1\right)$.

We may now draw from the observation that, since $J^{-}_{0}=-\partial/\partial x$
and $\bar{J}^{-}_{0}=-\partial/\partial\bar{x}$, every observable
$\mathsf{T}\left(x,\bar{x}\right)$ is conjugate to $\mathsf{T}\left(0,0\right)$
by the relation:
\begin{equation}
\mathsf{T}\left(x,\bar{x}\right)=^{-xJ^{-}_{0}-\bar{x}\bar{J}^{-}_{0}}\mathsf{T}\left(0,0\right)e^{xJ^{-}_{0}+\bar{x}\bar{J}^{-}_{0}},
\end{equation}
leading us to define the currents:
\begin{equation}
J^{+}\left(x;z\right)\coloneqq e^{-xJ^{-}_{0}}J^{+}\left(z\right)e^{xJ^{-}_{0}},\,J\left(x;z\right)\coloneqq e^{-xJ^{-}_{0}}J^{3}\left(z\right)e^{xJ^{-}_{0}},\,J^{-}\left(x;z\right)=e^{-xJ^{-}_{0}}J^{-}\left(z\right)e^{xJ^{-}_{0}}.
\end{equation}
As observed by \citet{giveon1998comments}, since these currents are
related through differentiation, our attention may be focused on:
\begin{equation}
J\left(x;z\right)\coloneqq2xJ^{3}\left(z\right)-J^{+}\left(z\right)-x^{2}J^{-}\left(z\right),
\end{equation}
and similarly for:
\begin{equation}
\bar{J}\left(\bar{x};\bar{z}\right)\coloneqq2\bar{x}\bar{J}^{3}\left(\bar{z}\right)-\bar{J}^{+}\left(\bar{z}\right)-\bar{x}^{2}\bar{J}^{-}\left(\bar{z}\right).
\end{equation}

In the quantum theory, the currents in Eqs. (\ref{eq:Currents}, \ref{eq:Currents-1})
are generators of an $\widehat{SL}\left(2;\mathbf{C}\right)\times\widehat{\overline{SL}}\left(2;\mathbf{C}\right)$
algebra. Their operator product expansions may be succinctly expressed
as:
\begin{equation}
J\left(x;z\right)J\left(y;w\right)\sim k\frac{\left(y-x\right)^{2}}{\left(z-w\right)^{2}}+\frac{1}{z-w}\left[\left(y-x\right)^{2}\partial_{y}-2\left(y-x\right)\right]J\left(y;z\right),
\end{equation}
and:
\begin{equation}
J\left(x;z\right)\Phi^{j}\left(y;w\right)\sim\frac{1}{z-w}\left[\left(y-x\right)^{2}\partial_{y}+2j\left(y-x\right)\right]\Phi^{j}\left(y;w\right),
\end{equation}
with analogous expressions holding for the conjugates.

We conclude our discussion of the quantum theory of the $H^{+}_{3}$-WZNW
model by noting that $\bar{J}\left(\bar{x};\bar{z}\right)\Phi^{j=1}\left(x;z\right)$
is an observable that will appear frequently in the following subsections,
where the conformal primary $\Phi^{j=1}\left(x;z\right)$ will be
denoted simply as $\Phi\left(x;z\right)$. The observable $\bar{J}\left(\bar{x};\bar{z}\right)\Phi\left(x;z\right)$
satisfies the identity:
\begin{equation}
\bar{J}\left(\bar{x};\bar{z}\right)\Phi\left(x;z\right)=\frac{k}{\pi}\partial_{\bar{z}}\lambda\left(x,\bar{x};z,\bar{z}\right),
\end{equation}
where:
\begin{equation}
\lambda\coloneqq-\frac{1}{\gamma-x}\frac{\left(\gamma-x\right)\left(\bar{\gamma}-\bar{x}\right)e^{2\phi}}{1+\left(\gamma-x\right)\left(\bar{\gamma}-\bar{x}\right)e^{2\phi}}.
\end{equation}
Similarly, these objects satisfy the identity:
\begin{equation}
\Phi\left(x;z\right)=\frac{1}{\pi}\partial_{\bar{x}}\lambda\left(x,\bar{x};z,\bar{z}\right),
\end{equation}
with analogous expressions holding for the conjugates.

\subsubsection{The Mini-Superspace $\left(k\rightarrow\infty\right)$ Limit}

We are now in a position to discuss the mini-superspace limit of the
correlation functions within the framework of the $H^{+}_{3}$-WZNW
model. This limit holds particular significance for the developments
presented in these notes, as it is in this case that we can holographically
derive the tree-level MHV scattering amplitudes for gluons and gravitons
analytically continued to Klein space, within the framework of celestial
leaf amplitudes.

From a heuristic perspective, the results we are about to describe
can be motivated as follows. Recall that the conformal primaries $\Phi^{j}\left(x;z\right)$
in the $H^{+}_{3}$-WZNW model can be viewed as analogues of the ``plane-wave''
basis for the Hilbert space of the model. In this basis, we can define
the analogue of the Fourier transform as follows:
\begin{equation}
F\left(j\big|x,\bar{x}\right)\coloneqq\int_{H^{+}_{3}}\Phi^{j}\left(x,\bar{x};h\right)f\left(h\right),
\end{equation}
whose inverse is given by:
\begin{equation}
f\left(h\right)=\frac{i}{\left(4\pi\right)^{3}}\int_{-\frac{1}{2}+i\mathbf{R}^{+}}dj\left(2j+1\right)^{2}\int d^{2}x\left(\Phi^{j}\left(x,\bar{x};h\right)\right)^{*}F\left(j\big|x,\bar{x}\right).
\end{equation}
This transformation allows us, in particular, to decompose the correlation
function:
\begin{equation}
\mathcal{F}_{n}\coloneqq\left\langle \Phi^{j_{1}}\left(x_{1};z_{1}\right)...\Phi^{j_{n}}\left(x_{n},z_{n}\right)\right\rangle ,
\end{equation}
into the basis $\left\{ \Phi^{j}\left(x;z\right)\right\} $, which
now becomes the focus of our attention. However, as will be discussed
in more detail in Subsection \ref{subsec:Differential-Equations-for},
following an important observation by \citet{ribault2005h3+}, in
the so-called $\mu$-basis given by:
\begin{equation}
\Phi^{j}\left(\mu;z\right)\coloneqq\frac{\left|\mu\right|^{2j+2}}{\pi}\int d^{2}xe^{\mu x-\bar{\mu}\bar{x}}\Phi^{j}\left(x;z\right),
\end{equation}
the Knizhnik-Zamolodchikov (KZ) equations for the $H^{+}_{3}$-WZNW
model take the following form (\citet{ribault2005h3+}):
\begin{equation}
\frac{\partial\mathcal{F}_{n}}{\partial z_{k}}+b^{2}\sum_{k\neq\ell}\frac{\mu_{k}\mu_{\ell}}{z_{k}-z_{\ell}}\left(\frac{\partial}{\partial\mu_{k}}-\frac{\partial}{\partial\mu_{\ell}}\right)^{2}\mathcal{F}_{n}=b^{2}\sum_{k\neq\ell}\frac{\mu_{k}\mu_{\ell}}{z_{k}-z_{\ell}}\left[\frac{j_{k}\left(j_{k}+1\right)}{\mu^{2}_{k}}+\frac{j_{\ell}\left(j_{\ell}+1\right)}{\mu^{2}_{\ell}}\right]\mathcal{F}_{n},\label{eq:KZ-2}
\end{equation}
where $b^{2}\coloneqq1/\left(k-2\right)$. In the mini-superspace
($k\rightarrow\infty$) limit, Eq. (\ref{eq:KZ-2}) simplifies to:
\begin{equation}
\frac{\partial\mathcal{F}_{n}}{\partial z_{k}}=0.
\end{equation}
Thus, in the large-$k$ limit, $\mathcal{F}_{n}$ depends solely on
the ``boundary spacetime'' coordinates $x,\bar{x}$. It is therefore
natural to expect that, in this limit, the Fourier representation
of the correlation functions will be dominated by the convolution
of these ``plane waves,'' such that:
\begin{equation}
\lim_{k\rightarrow\infty}\mathcal{F}_{n}=\frac{\mathcal{N}}{2\pi^{2}}\int_{H^{+}_{3}}dh\prod^{n}_{i=1}\Phi^{j_{i}}\left(x_{i},\bar{x}_{i};h\right)=\mathcal{N}\int\frac{d\rho d\gamma d\bar{\gamma}}{\rho^{3}}\prod^{n}_{i=1}\left(\frac{\rho}{\rho^{2}+\left|x_{i}-\gamma\right|^{2}}\right)^{2j_{i}},\label{eq:Mini-Superspace-Limit}
\end{equation}
where we define the ``radial'' coordinate $\rho\coloneqq e^{-\phi}$.
Consequently, in the $k\rightarrow\infty$ limit, the correlation
functions of the primaries of the $H^{+}_{3}$-WZNW model can be interpreted
as a Feynman-Witten contact diagram for massless scalars propagating
on Euclidean $AdS_{3}$ (cf. \citet{penedones2017tasi}).

We shall now briefly review how one could rigorously establish these
arguments, drawing from the observation made by \citet{ribault2005h3+}
that the correlation functions of the $H^{+}_{3}$-WZNW model on the
sphere are related to the correlation functions of the Liouville vertex
operators $V_{\alpha}\left(z\right)=:e^{2\alpha\phi\left(z\right)}:$,
which include degenerate fields, through the following relation:
\begin{align}
 & \left\langle \Phi^{j_{1}}\left(\mu_{1};z_{1}\right)...\Phi^{j_{n}}\left(\mu_{n};z_{n}\right)\right\rangle \\
 & =\frac{\pi\left(-\pi\right)^{-n}}{2}b\delta^{\left(2\right)}\left(\sum^{n}_{i=1}\mu_{i}\right)\left|\Theta_{n}\right|^{2}\left\langle V_{\alpha_{1}}\left(z_{1}\right)...V_{\alpha_{n}}\left(z_{n}\right)V_{-1/2b}\left(y_{1}\right)...V_{-1/2b}\left(y_{n}\right)\right\rangle ,
\end{align}
where the function $\Theta$ is defined by:
\begin{equation}
\Theta_{n}\coloneqq u\prod_{r<s\leq n}z^{1/2b^{2}}_{rs}\prod_{k<\ell\leq n-2}y^{1/2b^{2}}_{k\ell}\prod^{n}_{r=1}\prod^{n-2}_{k=1}\frac{1}{\left(z_{r}-y_{k}\right)^{1/2b^{2}}},
\end{equation}
and the variables $y_{1},...,y_{n},u$ are related to $\mu_{1},...,\mu_{n}$
via the equation:
\begin{equation}
\frac{1}{u}\sum^{n}_{i=1}\frac{\mu_{i}}{t-z_{i}}=\frac{\prod^{n-2}_{j=1}\left(t-y_{j}\right)}{\prod^{n}_{i=1}\left(t-z_{i}\right)}.
\end{equation}
Using these results, it was demonstrated in Section 4 of \citet{ribault2005h3+}
that the large-$k$ asymptotics of the correlation functions in the
$H^{+}_{3}$-WZNW model are equivalent, up to a proportionality constant
that we shall denote by $\mathcal{N}$, to the semiclassical limit
$b\rightarrow0$ of the Liouville correlation functions. Furthermore,
recalling from Appendix A of \citet{melton2024celestial} that the
semiclassical limit of the Liouville vertex operators $V_{2\sigma_{1}}\left(x_{1}\right)$,
..., $V_{2\sigma_{n}}\left(x_{n}\right)$ can be represented as a
contact Feynman-Witten diagram for massless scalars on $AdS_{3}$,
\begin{equation}
\lim_{b\rightarrow0^{+}}\left\langle V_{2\sigma_{1}}\left(x_{1}\right)...V_{2\sigma_{n}}\left(x_{n}\right)\right\rangle \propto\int_{AdS_{3}/\mathbf{Z}}d^{3}\hat{y}\frac{\Gamma\left(2\sigma_{i}\right)}{\left(\varepsilon-iq\left(x_{i},\bar{x}_{i}\right)\cdot\hat{y}\right)},
\end{equation}
we see that Eq. (\ref{eq:Mini-Superspace-Limit}) follows from these
considerations.

\subsection{Kac-Moody Current Algebras\label{subsec:First-Entry:-Kac-Moody}}

We are now prepared to present the first entry of our holographic
dictionary. Our objective is to describe the construction of Kac-Moody
currents within the celestial CFT--henceforth referred to as \emph{celestial
Kac-Moody current algebras}--derived from worldsheet Kac-Moody currents
associated with gauge symmetries.

Let $\Sigma$ denote the worldsheet of a bosonic string embedded in
the hyperboloid $H^{+}_{3}$ (times an arbitrary compact manifold
$\mathcal{N}$), and let $G$ be a Lie group with structure constants
$if^{abc}$. Suppose we are given a set of generators $j^{a}\left(z\right)$
of a $G$ level-$\hat{k}_{G}$ Kac-Moody current algebra on the worldsheet
$\Sigma$, satisfying the operator product expansion:
\begin{equation}
j^{a}\left(z\right)j^{b}\left(w\right)\sim\frac{\hat{k}_{G}\delta^{ab}}{\left(z-w\right)^{2}}+\frac{if^{abc}j^{c}\left(w\right)}{z-w}.
\end{equation}

We recall from \citet{polchinski2001string} that, given such a set
of worldsheet currents $j^{a}\left(z\right)$ obeying the above OPE,
one obtains, in string theory, a gauge field $A^{a}_{\mu}\left(x\right)$
in the target space, whose associated vertex operator takes the form
$\int d^{2}zj^{a}\left(z\right)A^{a}_{\mu}\left(x\right)\partial_{\bar{z}}X^{\mu}\left(z,\bar{z}\right)$,
where $X^{\mu}\left(z,\bar{z}\right)$ represents the embedding of
the string in the target space. The gauge field transforms under infinitesimal
gauge transformations as $\delta A^{a}_{\mu}\left(x\right)=\partial_{\mu}\lambda^{a}\left(x\right)$.
Thus, a pure gauge field is described by a vertex operator of the
form $\int d^{2}zj^{a}\left(z\right)\partial_{\bar{z}}\lambda^{a}\left(x\right)$. 

However, as is well known from the study of asymptotic symmetries
and the infrared structure of gauge theories (cf. \citet{strominger2018lectures}),
an important class of vertex operators emerges from pure gauge configurations
whose gauge function $\lambda^{a}\left(x\right)$ does not have compact
support in the target space. In the present context, this implies
that $\lambda^{a}\left(x\right)$ does not vanish at the boundary
of $AdS_{3}$, thereby generating an algebra of large gauge transformations.
(Of course, in string theory, one must also impose worldsheet consistency
conditions, e.g., the integrands in the vertex operators must be primary
under both left- and right-moving Virasoro symmetries. Furthermore,
when the target space is flat, the gauge field must satisfy the gauge
condition $\partial_{\mu}A^{\mu a}=0$, in addition to the massless
Klein-Gordon equation $\partial_{\mu}\partial^{\mu}A^{a}_{\nu}=0$.)

Thus, we are led to define the generators of the celestial Kac-Moody
currents via the following vertex operators:
\begin{equation}
\mathcal{O}^{a}\left(x\right)\coloneqq-\frac{1}{\pi}\int_{\Sigma}d^{2}z\,j^{a}\left(z\right)\partial_{\bar{z}}\lambda\left(x,\bar{x};z,\bar{z}\right).\label{eq:Celestial-Kac-Moody-Current}
\end{equation}

Let $\mathsf{W}^{R,h}\left(x,\bar{x};z,\bar{z}\right)$ be a worldsheet
operator depending on the celestial coordinates $x,\bar{x}\in\mathbf{CP}^{1}$,
transforming in a representation $R$ of $G$ with weight $h$, such
that:
\begin{equation}
j^{a}\left(z\right)\mathsf{W}^{R,h}\left(x,\bar{x};w,\bar{w}\right)\sim\frac{1}{z-w}t^{a}\left(R\right)\mathsf{W}^{R,h}\left(x,\bar{x};w,\bar{w}\right),\label{eq:OPE}
\end{equation}
where $t^{a}\left(R\right)$ is the matrix associated with the representation
under which $\mathsf{W}^{R,h}$ transforms. To verify the consistency
of our definition of $\mathcal{O}^{a}\left(x\right)$, we must ensure
it satisfies the celestial current algebra Ward identities: 
\begin{equation}
\mathcal{O}^{a}\left(x\right)\mathsf{W}^{R,h}\left(y,\bar{y};z,\bar{z}\right)\sim\frac{1}{x-y}t^{a}\left(R\right)\mathsf{W}^{R,h}\left(x,\bar{x};z,\bar{z}\right),\label{eq:Ward}
\end{equation}
\begin{equation}
\mathcal{O}^{a}\left(x\right)\mathcal{O}^{b}\left(y\right)\sim\frac{\hat{k}_{G}\delta^{ab}}{\left(x-y\right)^{2}}+\frac{if^{abc}\mathcal{O}^{c}\left(y\right)}{x-y}.\label{eq:Ward-1}
\end{equation}

We begin our discussion of the Ward identities with Eq. (\ref{eq:Ward}),
which is useful for introducing our regularisation scheme, ensuring
that the celestial correlation functions are well-defined and finite.
Using Eqs. (\ref{eq:Celestial-Kac-Moody-Current}, \ref{eq:OPE}),
we arrive at the following expression for the correlation function
of the operators $\mathcal{O}^{a}\left(x\right)$ and $\mathsf{W}^{R,h}\left(y,\bar{y}\right)$:
\begin{align*}
 & \left\langle \mathcal{O}^{a}\left(x\right)\mathsf{W}^{R,h}\left(y,\bar{y}\right)...\right\rangle _{\text{CCFT}}\\
 & =-\frac{1}{\pi}\int_{\Sigma}d^{2}z_{1}\int_{\Sigma}d^{2}z\,\frac{t^{a}\left(R\right)}{z-z_{1}}\left\langle \partial_{\bar{z}}\lambda\left(x,\bar{x};z,\bar{z}\right)\mathsf{W}^{R,h}\left(y,\bar{y};z_{1},\bar{z_{1}}\right)...\right\rangle .
\end{align*}

Using integration by parts with respect to the anti-holomorphic variable
$\bar{z}$, and substituting the identity given by Eq. (\ref{eq:Delta-Function}),
we can rewrite the celestial correlation function as:
\begin{align}
 & \left\langle \mathcal{O}^{a}\left(x\right)\mathsf{W}^{R,h}\left(y,\bar{y}\right)...\right\rangle _{\text{CCFT}}\\
 & =-\frac{1}{\pi}\int_{\Sigma}d^{2}z_{1}\int_{\Sigma}d^{2}z\,\frac{\partial}{\partial\bar{z}}\left(\frac{t^{a}\left(R\right)}{z-w}\left\langle \lambda\left(x,\bar{x};z,\bar{z}\right)\mathsf{W}^{R,h}\left(y,\bar{y};z_{1},\bar{z}_{1}\right)...\right\rangle \right)\\
 & +\lim_{z\rightarrow z_{1}}\int_{\Sigma}d^{2}z_{1}\,\left\langle \lambda\left(x,\bar{x};z,\bar{z}\right)\mathsf{W}^{R,h}\left(y,\bar{y};z_{1},\bar{z}_{1}\right)\right\rangle .
\end{align}

As $z\rightarrow z_{1}$, the last term in this expression diverges,
as $\lambda$ and $\mathsf{W}^{R,h}$ are evaluated at the same insertion
on the worldsheet. To address this, we introduce the following regularisation
scheme:
\begin{align}
 & \left\langle \mathcal{O}^{a}\left(x\right)\mathsf{W}^{R,h}\left(y,\bar{y}\right)...\right\rangle _{\text{CCFT}}\\
 & \stackrel{\text{reg.}}{\longrightarrow}-\frac{1}{\pi}\int_{\Sigma}d^{2}z_{1}\int_{\Sigma}d^{2}z\,\frac{\partial}{\partial\bar{z}}\left(\frac{t^{a}\left(R\right)}{z-w}\left\langle \lambda\left(x,\bar{x};z,\bar{z}\right)\mathsf{W}^{R,h}\left(y,\bar{y};z_{1},\bar{z}_{1}\right)...\right\rangle \right).
\end{align}

Now, let $\varepsilon>0$ be a small real number, and split the integration
over $\Sigma$ as $\int_{\Sigma}d^{2}z=\int_{\Sigma\backslash D_{\varepsilon}\left(z_{1}\right)}d^{2}z+\int_{D_{\varepsilon}\left(z_{1}\right)}d^{2}z$.
As $\varepsilon\rightarrow0^{+}$, applying the complex divergence
theorem (see Eq. (\ref{eq:Divergence-Theorem})) yields:
\begin{align}
 & \left\langle \mathcal{O}^{a}\left(x\right)\mathsf{W}^{R,h}\left(y,\bar{y}\right)...\right\rangle _{\text{CCFT}}\\
 & =\frac{1}{\pi}\int_{\Sigma}d^{2}z_{1}\oint_{C_{\varepsilon}\left(z_{1}\right)}\frac{dz}{2\pi i}\left\langle j^{a}\left(z\right)\lambda\left(x,\bar{x};z,\bar{z}\right)\mathsf{W}^{R,h}\left(y,\bar{y};z_{1},\bar{z}_{1}\right)...\right\rangle .
\end{align}
Here, $C_{\varepsilon}\left(z_{1}\right)$ is a contour of radius
$\varepsilon$ centred at $z_{1}\in\Sigma$. The sign change arises
because, upon removing $D_{\varepsilon}\left(z_{1}\right)$ from $\Sigma$,
the contour integral traverses $C_{\varepsilon}\left(z_{1}\right)$
in the opposite orientation.

By repeating this procedure iteratively for all worldsheet operator
insertions at $z_{1},...,z_{n}$, we obtain the following Cauchy integral
representation of the celestial current $\mathcal{O}^{a}\left(x\right)$
around all vertices attached to the worldsheet:
\begin{equation}
\mathcal{O}^{a}\left(x\right)=\sum^{n}_{i=1}\oint_{C_{\varepsilon}\left(z_{i}\right)}\frac{dz}{2\pi i}j^{a}\left(z\right)\lambda\left(x,\bar{x};z,\bar{z}\right)
\end{equation}

\begin{rem*}
The celestial correlation function $\langle\mathcal{O}^{a}(x)\mathsf{W}^{R,h}(y,\bar{y})...\rangle_{\text{CCFT}}$
also admits another integral representation that offers insight into
the geometric nature of the regularisation scheme. Let $\hat{\Sigma}$
be the Riemann surface obtained from $\Sigma$ by removing small disks
around all vertex operator insertions. Since $j^{a}\left(z\right)$
is a holomorphic current on $\hat{\Sigma}$, we have $\partial_{\bar{z}}j^{a}\left(z\right)\big|_{\hat{\Sigma}}\equiv0$.
Therefore:
\begin{align*}
 & \left\langle \mathcal{O}^{a}\left(x\right)\mathsf{W}^{R,h}\left(y,\bar{y}\right)...\right\rangle _{\text{CCFT}}\\
 & =-\frac{1}{\pi}\int_{\Sigma}d^{2}z_{1}\int_{\hat{\Sigma}}d^{2}z\,\left\langle j^{a}\left(z\right)\partial_{\bar{z}}\lambda\left(x,\bar{x};z,\bar{z}\right)\mathsf{W}^{R,h}\left(y,\bar{y};z_{1},\bar{z}_{1}\right)...\right\rangle .
\end{align*}
Thus, the regularisation scheme can be understood as the removal of
small discs centred around each operator insertion point on the worldsheet,
which allow us to express the celestial current $\mathcal{O}^{a}\left(x\right)$
as:
\begin{equation}
\mathcal{O}^{a}\left(x\right)=-\frac{1}{\pi}\int_{\hat{\Sigma}}d^{2}z\,\frac{\partial}{\partial\bar{z}}\left[j^{a}\left(z\right)\lambda\left(z,\bar{z};x,\bar{x}\right)\right].
\end{equation}
\end{rem*}
We are now prepared to demonstrate that the celestial Kac-Moody current
$\mathcal{O}^{a}\left(x\right)$ satisfies the Ward identity, Eq.
(\ref{eq:Ward-1}). This can be established by analysing the behaviour
of $\mathcal{O}^{a}\left(x\right)$ within a correlation function
involving another operator $\mathcal{O}^{b}\left(y\right)$. A more
straightforward approach involves taking the partial derivative of
$\mathcal{O}^{a}\left(x\right)$ with respect to the anti-holomorphic
variable $\bar{x}$, yielding:
\begin{equation}
\partial_{\bar{x}}\mathcal{O}^{a}\left(x\right)=\oint\frac{dz}{2i}j^{a}\left(z\right)\Phi\left(x,\bar{x};z,\bar{z}\right),\label{eq:Partial-Derivative-Kac-Moody-Current}
\end{equation}
which follows from the identity $\partial_{\bar{x}}\lambda=\pi\Phi$.
Consequently, we find:
\begin{align}
 & \left\langle \partial_{\bar{x}}\mathcal{O}^{a}\left(x\right)\mathcal{O}^{b}\left(y\right)...\right\rangle _{\text{CCFT}}\\
 & =-\frac{1}{k}\int_{\Sigma}d^{2}w\oint_{C_{\varepsilon}\left(z_{1}\right)}\frac{dz}{2i}\left(\frac{\hat{k}_{G}\delta^{ab}}{\left(z-w\right)^{2}}+\frac{if^{abc}j^{c}\left(z\right)}{z-w}\right)\left\langle \Phi\left(x,\bar{x};z,\bar{z}\right)\bar{J}\left(\bar{y};\bar{w}\right)\Phi\left(y,\bar{y};w,\bar{w}\right)...\right\rangle .\label{eq:OO-Correlation-Function}
\end{align}

Two distinct contributions arise from the single pole in the above
correlation function. The first originates from the OPE $\Phi\Phi\bar{J}$,
and the second from $\bar{J}\Phi\Phi$. To compute the first, we use
the asymptotic expansion:
\begin{equation}
\lim_{z\rightarrow w}\Phi\left(x,\bar{x};z,\bar{z}\right)\Phi\left(y,\bar{y};w,\bar{w}\right)\sim\delta^{\left(2\right)}\left(x-y\right)\Phi\left(y,\bar{y};w,\bar{w}\right),\label{eq:Asymptotics}
\end{equation}
from which we derive:
\begin{align}
 & \lim_{z\rightarrow w}\Phi\left(x,\bar{x};z,\bar{z}\right)\Phi\left(y,\bar{y};w,\bar{w}\right)\bar{J}\left(\bar{y};\bar{w}\right)\\
 & \sim\delta^{\left(2\right)}\left(x-y\right)\Phi\left(y,\bar{y};w,\bar{w}\right)\bar{J}\left(\bar{y};\bar{w}\right)+\left(z-w\right)\lim_{z'\rightarrow w}\partial_{z'}\Phi\left(x,\bar{x};z',\bar{z}'\right)\Phi\left(y,\bar{y};w,\bar{w}\right)\bar{J}\left(\bar{y};\bar{w}\right).
\end{align}
To compute the second contribution, recall (cf. \citet[Eq. (2.25)]{giveon1998comments}):
\begin{equation}
\bar{J}\left(\bar{y};\bar{w}\right)\Phi\left(x,\bar{x};z,\bar{z}\right)=\frac{1}{\bar{w}-\bar{z}}\left[\left(\bar{x}-\bar{y}\right)^{2}\partial_{\bar{x}}+2\left(\bar{x}-\bar{y}\right)\right]\Phi\left(x,\bar{x};z,\bar{z}\right),
\end{equation}
so that Eq. (\ref{eq:Asymptotics}) implies:
\begin{align}
 & \lim_{z\rightarrow w}\bar{J}\left(\bar{y};\bar{w}\right)\Phi\left(x,\bar{x};z,\bar{z}\right)\Phi\left(y,\bar{y};w,\bar{w}\right)\\
 & \sim\frac{1}{\bar{w}-\bar{z}}\left[\left(\bar{x}-\bar{y}\right)^{2}\partial_{\bar{x}}+2\left(\bar{x}-\bar{y}\right)\right]\delta^{\left(2\right)}\left(x-y\right)\Phi\left(y,\bar{y};w,\bar{w}\right)=0.
\end{align}
Thus, the contribution to $\left\langle \partial_{\bar{x}}\mathcal{O}^{a}\left(x\right)\mathcal{O}^{b}\left(y\right)...\right\rangle _{\text{CCFT}}$
from the single pole in Eq. (\ref{eq:OO-Correlation-Function}) is:
\begin{align}
 & -\frac{if^{abc}}{k}\int_{\Sigma}d^{2}w\oint_{C_{\varepsilon\left(w\right)}}\frac{dz}{2i}\left\langle \frac{j^{c}\left(w\right)}{z-w}\delta^{\left(2\right)}\left(x-y\right)\Phi\left(y,\bar{y};w,\bar{w}\right)\bar{J}\left(\bar{y};\bar{w}\right)...\right\rangle \\
 & =\frac{\partial}{\partial\bar{x}}\left(\frac{if^{abc}}{x-y}\left\langle \mathcal{O}^{c}\left(y\right)...\right\rangle \right)\label{eq:Step}
\end{align}

To compute the contribution from the double pole, we first integrate
by parts with respect to the holomorphic variable $z$:
\begin{align*}
 & -\frac{1}{k}\int_{\Sigma}d^{2}w\oint_{C_{\varepsilon}\left(w\right)}\frac{dz}{2i}\frac{\hat{k}_{G}\delta^{ab}}{\left(z-w\right)^{2}}\left\langle \bar{J}\left(\bar{y};\bar{w}\right)\Phi\left(x,\bar{x};z,\bar{z}\right)\Phi\left(y,\bar{y};w,\bar{w}\right)...\right\rangle \\
 & =-\frac{\hat{k}_{G}\delta^{ab}}{k}\int_{\Sigma}d^{2}w\oint_{C_{\varepsilon}\left(w\right)}\frac{dz}{2i}\frac{1}{z-w}\left\langle \bar{J}\left(\bar{y};\bar{w}\right)\partial_{z}\Phi\left(x,\bar{x};z,\bar{z}\right)\Phi\left(y,\bar{y};w,\bar{w}\right)...\right\rangle \\
 & =-\frac{\pi\hat{k}_{G}\delta^{ab}}{k}\int_{\Sigma}d^{2}w\lim_{z\rightarrow w}\left\langle \bar{J}\left(\bar{y};\bar{w}\right)\partial_{z}\Phi\left(x,\bar{x};z,\bar{z}\right)\Phi\left(y,\bar{y};w,\bar{w}\right)...\right\rangle .
\end{align*}
We recall that the limit $z\rightarrow w$ arises because $\varepsilon\rightarrow0$,
where $\varepsilon$ is the radius of the small circle $C_{\varepsilon}\left(w\right)$
enclosing the insertion point $w$ on the worldsheet.

Using the identities from \citet{giveon1998comments}, namely $\partial_{z}\bar{\lambda}=\left(\pi/k\right)J\Phi$
and $\partial_{x}\bar{\lambda}=\pi\Phi$, we find the following result:
\begin{align}
 & \lim_{z\rightarrow w}\left\langle \bar{J}\left(\bar{y};\bar{w}\right)\partial_{z}\Phi\left(x,\bar{x};z,\bar{z}\right)\Phi\left(y,\bar{y};w,\bar{w}\right)...\right\rangle \\
 & =\frac{1}{\pi}\lim_{z\rightarrow w}\left\langle \bar{J}\left(\bar{y};\bar{w}\right)\partial_{z}\partial_{x}\bar{\lambda}\left(x,\bar{x};z,\bar{z}\right)\Phi\left(y,\bar{y};w,\bar{w}\right)...\right\rangle \\
 & =\frac{1}{k}\lim_{z\rightarrow w}\left\langle \bar{J}\left(\bar{y};\bar{w}\right)\partial_{x}\left[J\left(x;z\right)\Phi\left(x,\bar{x};z,\bar{z}\right)\right]\Phi\left(y,\bar{y};w,\bar{w}\right)...\right\rangle \\
 & \sim\frac{1}{k}\partial_{x}\delta^{\left(2\right)}\left(x-y\right)\left\langle J\left(y;w\right)\bar{J}\left(\bar{y};\bar{w}\right)\Phi\left(y,\bar{y};w,\bar{w}\right)...\right\rangle .
\end{align}

Defining a new operator $\mathcal{I}$ as:
\begin{equation}
\mathcal{I}\coloneqq\frac{1}{k^{2}}\int d^{2}z\,J\left(x;z\right)\bar{J}\left(\bar{x};\bar{z}\right)\Phi\left(x,\bar{x};z,\bar{z}\right),\label{eq:Central-Extension}
\end{equation}
the contribution from the double pole to $\left\langle \partial_{\bar{x}}\mathcal{O}^{a}\left(x\right)\mathcal{O}^{b}\left(y\right)...\right\rangle _{\text{CCFT}}$
can be expressed as:
\begin{equation}
\frac{\partial}{\partial\bar{x}}\left(\frac{\hat{k}_{G}\delta^{ab}}{\left(x-y\right)^{2}}\left\langle \mathcal{I}...\right\rangle _{\text{CCFT}}\right).\label{eq:Step-1}
\end{equation}

Finally, by combining Eqs. (\ref{eq:Step}, \ref{eq:Step-1}), we
arrive at:
\begin{equation}
\left\langle \partial_{\bar{x}}\mathcal{O}^{a}\left(x\right)\mathcal{O}^{b}\left(y\right)...\right\rangle _{\text{CCFT}}=\frac{\partial}{\partial\bar{x}}\left\langle \left(\frac{if^{abc}}{x-y}\mathcal{O}^{c}\left(y\right)+\frac{\hat{k}_{G}\mathcal{I}\delta^{ab}}{\left(x-y\right)^{2}}\right)...\right\rangle _{\text{CCFT}},
\end{equation}
which, upon integration with respect to $\bar{x}$, gives the celestial
Ward identity:
\begin{equation}
\mathcal{O}^{a}\left(x\right)\mathcal{O}^{b}\left(y\right)\sim\frac{\hat{k}_{G}\mathcal{I}\delta^{ab}}{\left(x-y\right)^{2}}+\frac{if^{abc}\mathcal{O}^{c}\left(y\right)}{x-y}.\label{eq:Ward-5}
\end{equation}

We are led to the conclusion that the operator $\mathcal{I}$ represents
the central extension of the celestial Kac-Moody current algebra generated
by $\mathcal{O}^{a}\left(x\right)$. To establish that the definition
of $\mathcal{I}$, as given in Eq. (\ref{eq:Central-Extension}),
is consistent, it is necessary to verify that $\mathcal{I}$ is independent
of the holomorphic $x$ and anti-holomorphic $\bar{x}$ celestial
coordinates.

In fact, employing the identities:
\begin{equation}
\partial_{x}\bar{\lambda}=\pi\Phi,\,\,\,J\Phi=\frac{k}{\pi}\partial_{z}\bar{\lambda},\,\,\,\bar{J}\Phi=\frac{k}{\pi}\partial_{\bar{z}}\lambda,
\end{equation}
established in \citet[Sec. 2]{giveon1998comments}, we deduce the
following result:
\begin{align}
\partial_{x}\mathcal{I} & =\frac{1}{k^{2}}\int_{\Sigma}d^{2}z\,\bar{J}\left(\bar{x};\bar{z}\right)\partial_{x}\left[J\left(x;z\right)\Phi\left(x,\bar{x};z,\bar{z}\right)\right]=\frac{1}{\pi k}\int_{\Sigma}d^{2}z\,\bar{J}\left(\bar{x};\bar{z}\right)\partial_{x}\partial_{z}\bar{\lambda}\left(x,\bar{x};z,\bar{z}\right)\\
 & =\frac{1}{k}\int_{\Sigma}d^{2}z\,\partial_{z}\left[\bar{J}\left(\bar{x};\bar{z}\right)\Phi\left(x,\bar{x};z,\bar{z}\right)\right]=-\frac{1}{2ik}\oint d\bar{z}\,\bar{J}\left(\bar{x};\bar{z}\right)\Phi\left(x,\bar{x};z,\bar{z}\right)\\
 & =-\oint\frac{d\bar{z}}{2\pi i}\partial_{\bar{z}}\lambda\left(x,\bar{x};z,\bar{z}\right)=0,
\end{align}
as claimed. This verifies that $\mathcal{I}$ is indeed independent
of the coordinates $x,\bar{x}$, confirming its consistency as a central
extension.

\section{Worldsheet Integrals}

\subsection{Worldsheet Integrals Related to the $2$-Point Function\label{subsec:Worldsheet-Integrals-Related}}

In this Appendix, we undertake the computation of the integral over
the worldsheet in Eq. (\ref{eq:Integral-3}), which is important for
analysing the celestial $2$-point function discussed in Subsection
\ref{subsec:Two-point-function}. We shall proceed with detail, as
this calculation exemplifies the basic reasoning employed to derive
the worldsheet integral relations that emerge in the computations
of the $3$-point function in Subsection \ref{subsec:Three-Point-Function}
and the structure constants of the celestial OPE in Subsection \ref{subsec:Operator-Product-Expansion}.
Additionally, we shall derive the distributional limit of the resulting
integral, thereby establishing the known $2d$ CFT identity (cf, \citet{simmons2014projectors,ketov1995conformal,vladimirov1971equations}):
\begin{equation}
\int d^{2}y\frac{1}{\left|x_{1}-y\right|^{2\tau}\left|y-x_{2}\right|^{2\left(2-\tau\right)}}=\frac{4\pi^{2}}{\nu^{2}}\delta^{\left(2\right)}\left(x_{1}-x_{2}\right),
\end{equation}
where $\tau_{1}\eqqcolon1+i\nu$, which shall prove helpful in the
derivations within Subsection \ref{subsec:Two-point-function}, while
also serving as a consistency check of our results.

We define the principal integral of interest in this Appendix as follows:
\begin{equation}
\mathcal{I}_{0}\left(\tau_{1},\tau_{2}\big|x_{1},x_{2}\right)\coloneqq\int d^{2}y\frac{1}{\left|x_{1}-y\right|^{2\tau_{1}}\left|y-x_{2}\right|^{2\tau_{2}}}.
\end{equation}
We begin by simplifying $\mathcal{I}_{0}$ through the translation
$y\mapsto\xi\coloneqq y-x_{2}$, yielding:
\begin{equation}
\mathcal{I}_{0}=\int d^{2}\xi\frac{1}{\left|\left(x_{2}-x_{1}\right)-\xi\right|^{2\tau_{1}}\left|\xi\right|^{2\tau_{2}}}.\label{eq:Step-13}
\end{equation}
We proceed by employing the method of Feynman $\alpha$-parameters.
For a modern exposition, we refer the reader to \citet{smirnov2004evaluating}
and \citet{weinzierl2022feynman}. Parametrising the integrand in
Eq. (\ref{eq:Step-13}) gives:
\begin{equation}
\frac{1}{\left|\left(x_{2}-x_{1}\right)-\xi\right|^{2\tau_{1}}}=\frac{1}{\Gamma\left(\tau_{1}\right)}\int^{\infty}_{0}d\alpha_{1}\alpha^{\tau_{1}-1}_{1}e^{-\alpha_{1}\left|\left(x_{2}-x_{1}\right)-\xi\right|^{2}},
\end{equation}
and:
\begin{equation}
\frac{1}{\left|\xi\right|^{2\tau_{2}}}=\frac{1}{\Gamma\left(\tau_{2}\right)}\int^{\infty}_{0}d\alpha_{2}\alpha^{\tau_{2}-1}_{2}e^{-\alpha_{2}\left|\xi\right|^{2}}.
\end{equation}
Substituting these into Eq. (\ref{eq:Step-13}) and denoting $x_{2}-x_{1}\eqqcolon u_{0}+iv_{0}$
and $\xi\eqqcolon u+iv$, we obtain:
\begin{equation}
\mathcal{I}_{0}=\frac{1}{\Gamma\left(\tau_{1}\right)\Gamma\left(\tau_{2}\right)}\int^{\infty}_{0}d\alpha_{1}\alpha^{\tau_{1}-1}_{1}\int^{\infty}_{0}d\alpha_{2}\alpha^{\tau_{2}-1}_{2}e^{-\alpha_{1}\left(u^{2}_{0}+v^{2}_{0}\right)}\mathcal{G}\left(\alpha_{1},\alpha_{2}\right),\label{eq:Step-14}
\end{equation}
where the Gaussian integral $\mathcal{G}\left(\alpha_{1},\alpha_{2}\right)$
is defined as:
\begin{equation}
\mathcal{G}\left(\alpha_{1},\alpha_{2}\right)\coloneqq\int dudve^{-\left(\alpha_{1}+\alpha_{2}\right)u^{2}+2\alpha_{1}u_{0}u}e^{-\left(\alpha_{1}+\alpha_{2}\right)v^{2}+2\alpha_{1}v_{0}v}.
\end{equation}
Evaluating the Gaussian integral yields:
\begin{equation}
\mathcal{G}=\frac{\pi}{\alpha_{1}+\alpha_{2}}e^{\left(u^{2}_{0}+v^{2}_{0}\right)\frac{\alpha^{2}_{1}}{\alpha_{1}+\alpha_{2}}}.
\end{equation}
Thus, Eq. (\ref{eq:Step-14}) becomes:
\begin{equation}
\mathcal{I}_{0}=\frac{\pi^{2}}{\Gamma\left(\tau_{1}\right)\Gamma\left(\tau_{2}\right)}\int^{\infty}_{0}d\alpha_{1}\alpha^{\tau_{1}-1}_{1}\int^{\infty}_{0}d\alpha_{2}\alpha^{\tau_{2}-1}_{2}\frac{1}{\alpha_{1}+\alpha_{2}}e^{-\left(u^{2}_{0}+v^{2}_{0}\right)\alpha_{1}}e^{\left(u^{2}_{0}+v^{2}_{0}\right)\frac{\alpha^{2}_{1}}{\alpha_{1}+\alpha_{2}}}.\label{eq:Step-15}
\end{equation}
To evaluate the integrals over $\alpha_{1}$ and $\alpha_{2}$, we
introduce the variables $t\coloneqq\alpha_{1}/\left(\alpha_{1}+\alpha_{2}\right)$
and $s\coloneqq\left(u^{2}_{0}+v^{2}_{0}\right)\alpha_{1}$. Recalling
that $u^{2}_{0}+v^{2}_{0}=\left|x_{1}-x_{2}\right|^{2}$, we reorganise
Eq. (\ref{eq:Step-15}) as:
\begin{equation}
\mathcal{I}_{0}=\frac{\pi}{\Gamma\left(\tau_{1}\right)\Gamma\left(\tau_{2}\right)}\frac{1}{\left|x_{12}\right|^{2\left(\tau_{1}+\tau_{2}-1\right)}}\int^{\infty}_{0}dss^{\left(\tau_{1}+\tau_{2}-1\right)-1}\int^{1}_{0}dtt^{-\tau_{2}}\left(1-t\right)^{\tau_{2}-1}e^{-\left(1-t\right)s}.
\end{equation}
Finally, applying the rescaling $s\mapsto r\coloneqq\left(1-t\right)s$,
we obtain:
\begin{equation}
\mathcal{I}_{0}=\frac{\pi}{\left|x_{12}\right|^{2\left(\tau_{1}+\tau_{2}-1\right)}}\frac{1}{\Gamma\left(\tau_{1}\right)\Gamma\left(\tau_{2}\right)}\int^{\infty}_{0}drr^{\left(\tau_{1}+\tau_{2}-1\right)-1}e^{-r}\int^{1}_{0}dtt^{-\tau_{2}}\left(1-t\right)^{-\tau_{1}},
\end{equation}
leading to the final expression:
\begin{equation}
\mathcal{I}_{0}=\frac{\pi}{\left|x_{12}\right|^{2\left(\tau_{1}+\tau_{2}-1\right)}}\frac{\Gamma\left(1-\tau_{1}\right)}{\Gamma\left(\tau_{1}\right)}\frac{\Gamma\left(1-\tau_{2}\right)}{\Gamma\left(\tau_{2}\right)}\frac{\Gamma\left(\tau_{1}+\tau_{2}-1\right)}{\Gamma\left(2-\tau_{1}-\tau_{2}\right)}.\label{eq:Integral-2}
\end{equation}

This expression must be interpreted in the sense of distributions;
that is, it should be understood as making sense when applied to a
test function. To clarify this concept and derive a useful $2d$ CFT
identity, which will be employed in Section \ref{subsec:Two-point-function},
we now utilise Eq. (\ref{eq:Integral-2}) to compute the limit $\tau_{1}+\tau_{2}\rightarrow2$
in the distributional sense. We begin by observing the following relation:
\begin{equation}
\frac{1}{\Gamma\left(2-\tau_{1}-\tau_{2}\right)}=\frac{2-\tau_{1}-\tau_{2}}{\Gamma\left(3-\tau_{1}-\tau_{2}\right)},
\end{equation}
which allows Eq. (\ref{eq:Integral-2}) to be rearranged in the form:
\begin{align}
 & \frac{\Gamma\left(\tau_{1}\right)}{\Gamma\left(1-\tau_{1}\right)}\frac{\Gamma\left(\tau_{2}\right)}{\Gamma\left(1-\tau_{2}\right)}\int d^{2}y\frac{1}{\left|x_{1}-y\right|^{2\tau_{1}}\left|y-x_{2}\right|^{2\tau_{2}}}\\
 & =2\pi\left(\frac{1}{2}\left(2-\tau_{1}-\tau_{2}\right)\left|x_{12}\right|^{2\left(1-\tau_{1}-\tau_{2}\right)}\right)\frac{\Gamma\left(\tau_{1}+\tau_{2}-1\right)}{\Gamma\left(3-\tau_{1}-\tau_{2}\right)}.
\end{align}
Introducing $\varepsilon\coloneqq2-\tau_{1}-\tau_{2}$ and $\tau_{1}\eqqcolon1+i\nu$,
we observe that:
\begin{equation}
\frac{\Gamma\left(\tau_{2}\right)}{\Gamma\left(1-\tau_{2}\right)}=-\left(\tau_{1}-1\right)\left(\tau_{1}+\varepsilon-1\right)\frac{\Gamma\left(1-\tau_{1}\right)}{\Gamma\left(\tau_{1}+\varepsilon\right)},
\end{equation}
yielding:
\begin{equation}
\frac{\Gamma\left(\tau_{1}\right)}{\Gamma\left(\tau_{1}+\varepsilon\right)}\int d^{2}y\frac{1}{\left|x_{1}-y\right|^{2\tau_{1}}\left|y-x_{2}\right|^{2\tau_{2}}}=\frac{2\pi}{\nu\left(\nu-i\varepsilon\right)}\left(\frac{1}{2}\varepsilon\left|x_{12}\right|^{2\left(\varepsilon-1\right)}\right)\frac{\Gamma\left(1-\varepsilon\right)}{\Gamma\left(1+\varepsilon\right)}.\label{eq:Step-16}
\end{equation}
Recalling that (cf. \citet{vladimirov1971equations}):
\begin{equation}
\lim_{\varepsilon\rightarrow0^{+}}\frac{1}{2}\varepsilon\left|x_{12}\right|^{2\left(\varepsilon-1\right)}=\delta\left(\left|x_{12}\right|^{2}\right)=\delta\left(x_{12}\right)\delta\left(\bar{x}_{12}\right)=2\pi\delta^{\left(2\right)}\left(x_{1}-x_{2}\right),
\end{equation}
we find that taking the limit $\varepsilon\rightarrow0^{+}$ of Eq.
(\ref{eq:Step-16}) finally yields:
\begin{equation}
\int d^{2}y\frac{1}{\left|x_{1}-y\right|^{2\tau}\left|y-x_{2}\right|^{2\left(2-\tau\right)}}=\frac{4\pi^{2}}{\nu^{2}}\delta^{\left(2\right)}\left(x_{1}-x_{2}\right).\label{eq:Integral-5}
\end{equation}

\subsection{Generalised Dirac Delta Function on the Worldsheet\label{subsec:Generalised-Dirac-Delta}}

In this Appendix, we recall how the generalised Dirac delta ``function,''
analytically continued in the complex plane, naturally arises in the
context of worldsheet integration. This result proves useful in deriving
the celestial $2$-point function, as described by our holographic
dictionary in Subsection \ref{subsec:Two-point-function}, particularly
Eq. (\ref{eq:Identity-4}). Moreover, it serves to illustrate some
fundamental methods of worldsheet integration.

We begin by considering the integral defined as:
\begin{equation}
\mathcal{F}\left(\beta\big|x_{1},x_{2}\right)\coloneqq\frac{1}{N}\int d^{2}z_{1}d^{2}z_{2}\frac{1}{\left|z_{1}-z_{2}\right|^{2\beta}},\label{eq:Step-17}
\end{equation}
where $N$ denotes the worldsheet ``volume,'' a regulator necessary
for the integral measure. The method of Feynman $\alpha$-parameters
instructs us to introduce the following representation:
\begin{equation}
\frac{1}{\left|z_{1}-z_{2}\right|^{2\beta}}=\frac{1}{\Gamma\left(\beta\right)}\int^{\infty}_{0}d\alpha\alpha^{\beta-1}e^{-\alpha\left|z_{1}-z_{2}\right|^{2}},
\end{equation}
so that Eq. (\ref{eq:Step-17}) can be rewritten as:
\begin{equation}
\mathcal{F}=\frac{1}{\Gamma\left(\beta\right)}\int^{\infty}_{0}d\alpha\alpha^{\beta-1}\frac{1}{N}\int d^{2}z_{1}d^{2}z_{2}e^{-\alpha\left|z_{1}-z_{2}\right|^{2}}.
\end{equation}
To perform the complex integrals, we introduce the parameterisations
$z_{1}\coloneqq x_{1}+iy_{1}$ and $z_{2}\coloneqq x_{2}+iy_{2}$,
leading to:
\begin{equation}
\mathcal{F}=\frac{1}{\Gamma\left(\beta\right)}\int^{\infty}_{0}d\alpha\alpha^{\beta-1}\frac{1}{N}\int dx_{1}dx_{2}e^{-\alpha\left(x_{1}-x_{2}\right)^{2}}\int dy_{1}dy_{2}e^{-\alpha\left(y_{1}-y_{2}\right)^{2}},
\end{equation}
which can be reorganised as:
\begin{equation}
\mathcal{F}=\frac{1}{\Gamma\left(\beta\right)}\int^{\infty}_{0}d\alpha\alpha^{\beta-1}\frac{1}{N}\left(\mathcal{G}\left(\alpha\right)\right)^{2},\label{eq:Step-18}
\end{equation}
where the Gaussian integral is:
\begin{equation}
\mathcal{G}\left(\alpha\right)\coloneqq\int dx_{1}dx_{2}e^{-\alpha\left(x_{1}-x_{2}\right)^{2}}=\frac{\sqrt{\pi N}}{\alpha^{1/2}}.
\end{equation}
Thus, Eq. (\ref{eq:Step-18}) yields:
\begin{equation}
\mathcal{F}=\frac{\pi}{4\Gamma\left(\beta\right)}\int^{\infty}_{0}d\alpha\alpha^{\left(\beta-1\right)-1}.
\end{equation}
Following \citet{donnay2020asymptotic}, this result can be expressed
using the analytic continuation of the Dirac delta function (cf. Eq.
(\ref{eq:Dirac-1})) as:
\[
\boldsymbol{\delta}\left(i\left(\Delta-z\right)\right)\coloneqq\frac{1}{2\pi}\int^{\infty}_{0}d\tau\,\tau^{\Delta-z-1},
\]
so that our final result reads:
\begin{equation}
\frac{1}{N}\int d^{2}z_{1}d^{2}z_{2}\frac{1}{\left|z_{1}-z_{2}\right|^{2\beta}}=\frac{\pi^{2}}{2}\boldsymbol{\delta}\left(\beta-1\right).\label{eq:Integral-4}
\end{equation}

\subsection{Triple Worldsheet Integral\label{subsec:Triple-Worldsheet-Integral}}

In this Appendix, we shall compute the worldsheet integral that arises
in the calculation of the $3$-point function of Euclidean $AdS_{3}$
bosonic string theory, as required for the continuation of the computations
in Subsection \ref{subsec:Triple-Worldsheet-Integral}. Let $\sigma_{1},\sigma_{2},\sigma_{3}$
represent a sequence of exponents, and define the triple complex integral
$\mathcal{I}\left(\sigma_{1},\sigma_{2},\sigma_{2}\right)$ as follows:
\begin{equation}
\mathcal{I}\left(\sigma_{1},\sigma_{2},\sigma_{3}\right)\coloneqq\frac{1}{N}\int d^{2}z_{1}\int d^{2}z_{2}\int d^{2}z_{3}\frac{1}{\left|z_{12}\right|^{2\sigma_{3}}\left|z_{23}\right|^{2\sigma_{1}}\left|z_{31}\right|^{2\sigma_{2}}},\label{eq:Definition}
\end{equation}
where $N$ denotes the worldsheet area.

To evaluate this integral, we will employ the method of Feynman $\alpha$-parameters,
as reviewed by \citet{smirnov2004evaluating,weinzierl2022feynman}.
Beginning with the expressions for the distance between points on
the complex plane, let $z_{1}\eqqcolon x_{1}+iy_{1}$, $z_{2}\eqqcolon x_{2}+iy_{2}$,
and $z_{3}\eqqcolon x_{3}+iy_{3}$. Consequently, we express the terms
in the integrand as:
\begin{equation}
\frac{1}{\left|z_{12}\right|^{2\sigma_{3}}}=\frac{1}{\Gamma\left(\sigma_{3}\right)}\int^{\infty}_{0}d\alpha_{3}\alpha^{\sigma_{3}-1}_{3}e^{-\alpha_{3}\left(x_{1}-x_{2}\right)^{2}-\alpha_{3}\left(y_{1}-y_{2}\right)^{2}},
\end{equation}
\begin{equation}
\frac{1}{\left|z_{23}\right|^{2\sigma_{1}}}=\frac{1}{\Gamma\left(\sigma_{1}\right)}\int^{\infty}_{0}d\alpha_{1}\alpha^{\sigma_{1}-1}_{1}e^{-\alpha_{1}\left(x_{2}-x_{3}\right)^{2}-\alpha_{1}\left(y_{2}-y_{3}\right)^{2}},
\end{equation}
\begin{equation}
\frac{1}{\left|z_{31}\right|^{2\sigma_{2}}}=\frac{1}{\Gamma\left(\sigma_{2}\right)}\int^{\infty}_{0}d\alpha_{2}\alpha^{\sigma_{2}-1}_{2}e^{-\alpha_{2}\left(x_{3}-x_{1}\right)^{2}-\alpha_{2}\left(y_{3}-y_{1}\right)^{2}}.
\end{equation}

The integral in Eq. (\ref{eq:Definition}) can now be expanded as:
\begin{align}
 & \mathcal{I}=\frac{1}{\Gamma\left(\sigma_{1}\right)\Gamma\left(\sigma_{2}\right)\Gamma\left(\sigma_{3}\right)}\frac{1}{N}\int dx_{1}dx_{2}dx_{3}\int dy_{1}dy_{2}dy_{3}\prod^{3}_{i=1}\int^{\infty}_{0}d\alpha_{i}\alpha^{\sigma_{i}-1}_{i}\\
 & e^{-\alpha_{1}\left(x_{2}-x_{3}\right)^{2}-\alpha_{2}\left(x_{3}-x_{1}\right)^{2}-\alpha_{3}\left(x_{1}-x_{2}\right)^{2}}e^{-\alpha_{1}\left(y_{2}-y_{3}\right)^{2}-\alpha_{2}\left(y_{3}-y_{1}\right)^{2}-\alpha_{3}\left(y_{1}-y_{2}\right)^{2}}.
\end{align}

Our next goal is to reorganise this integral into the $\alpha$-parameters
and the worldsheet coordinates. We rewrite the integral as:
\begin{equation}
\mathcal{I}=\frac{1}{\Gamma\left(\sigma_{1}\right)\Gamma\left(\sigma_{2}\right)\Gamma\left(\sigma_{3}\right)}\frac{1}{N}\prod^{3}_{i=1}\int^{\infty}_{0}d\alpha_{i}\alpha^{\sigma_{i}-1}_{i}\left(F\left(\alpha_{1},\alpha_{2},\alpha_{3}\right)\right)^{2},
\end{equation}
where:
\begin{equation}
F\left(\alpha_{1},\alpha_{2},\alpha_{3}\right)\coloneqq\int dx_{1}dx_{2}dx_{3}e^{-\alpha_{1}\left(x_{2}-x_{3}\right)^{2}-\alpha_{2}\left(x_{3}-x_{1}\right)^{2}-\alpha_{3}\left(x_{1}-x_{2}\right)^{2}}.
\end{equation}

Performing the Gaussian integrals, we find that:
\begin{equation}
F=\frac{\pi\sqrt{V}}{\left(\alpha_{1}\alpha_{2}+\alpha_{2}\alpha_{3}+\alpha_{3}\alpha_{1}\right)^{1/2}}.
\end{equation}
Thus, the integral $\mathcal{I}$ becomes:
\begin{equation}
\mathcal{I}=\frac{\pi^{2}}{\Gamma\left(\sigma_{1}\right)\Gamma\left(\sigma_{2}\right)\Gamma\left(\sigma_{3}\right)}\int^{\infty}_{0}d\alpha_{3}\alpha^{\sigma_{3}-1}_{3}\int^{\infty}_{0}d\alpha_{2}\alpha^{\sigma_{2}-1}_{2}\int^{\infty}_{0}d\alpha_{1}\frac{\alpha^{\sigma_{1}-1}_{1}}{\alpha_{1}\left(\alpha_{2}+\alpha_{3}\right)+\alpha_{2}\alpha_{3}}.\label{eq:Step-6}
\end{equation}
We employ the following identity from \citet{gradshteyn2014table}:
\begin{equation}
\int^{\infty}_{0}d\alpha\,\frac{\alpha^{\mu-1}}{\left(1+\beta\alpha\right)^{\nu}}=\beta^{-\mu}B\left(\mu,\nu-\mu\right),\label{eq:Identity-1}
\end{equation}
which allows us to evaluate the integral over $\alpha_{1}$, yielding:
\begin{equation}
\int^{\infty}_{0}d\alpha_{1}\frac{\alpha^{\sigma_{1}-1}_{1}}{\alpha_{1}\left(\alpha_{2}+\alpha_{3}\right)+\alpha_{2}\alpha_{3}}=\frac{\alpha^{\sigma_{1}-1}_{2}\alpha^{\sigma_{1}-1}_{3}}{\left(\alpha_{2}+\alpha_{3}\right)^{\sigma_{1}}}B\left(\sigma_{1},1-\sigma_{1}\right).
\end{equation}
Substituting this result into Eq. (\ref{eq:Step-6}) gives:
\begin{equation}
\mathcal{I}=\pi^{2}\frac{B\left(\sigma_{1},1-\sigma_{1}\right)}{\Gamma\left(\sigma_{1}\right)\Gamma\left(\sigma_{2}\right)\Gamma\left(\sigma_{3}\right)}\int^{\infty}_{0}d\alpha_{3}\alpha^{\left(\sigma_{1}+\sigma_{3}-1\right)-1}_{3}\int^{\infty}_{0}d\alpha_{2}\frac{\alpha^{\left(\sigma_{1}+\sigma_{2}-1\right)-1}_{2}}{\left(\alpha_{2}+\alpha_{3}\right)^{\sigma_{1}}}.\label{eq:Step-7}
\end{equation}

Using the identity from Eq. (\ref{eq:Identity-1}), we compute the
integral over the Feynman parameter $\alpha_{2}$, yielding:
\begin{equation}
\int^{\infty}_{0}d\alpha_{2}\frac{\alpha^{\left(\sigma_{1}+\sigma_{2}-1\right)-1}_{2}}{\left(\alpha_{2}+\alpha_{3}\right)^{\sigma_{1}}}=\alpha^{\sigma_{2}-1}_{3}B\left(\sigma_{1}+\sigma_{2}-1,1-\sigma_{2}\right).
\end{equation}
Substituting this into Eq. (\ref{eq:Step-7}), we find:
\begin{equation}
\mathcal{I}=\pi^{2}\frac{B\left(\sigma_{1},1-\sigma_{1}\right)}{\Gamma\left(\sigma_{1}\right)\Gamma\left(\sigma_{2}\right)\Gamma\left(\sigma_{3}\right)}B\left(\sigma_{1}+\sigma_{2}-1,1-\sigma_{2}\right)\int^{\infty}_{0}d\alpha_{3}\alpha^{\left(\sigma_{1}+\sigma_{2}+\sigma_{3}-2\right)-1}_{3}.
\end{equation}

We now recall from \citet{donnay2020asymptotic} the definition of
the \emph{generalised }Dirac delta ``function,'' analytically continued
to the complex plane:
\begin{equation}
\boldsymbol{\delta}\left(i\left(\Delta-z\right)\right)\coloneqq\frac{1}{2\pi}\int^{\infty}_{0}d\tau\,\tau^{\Delta-z-1},\label{eq:Dirac}
\end{equation}
such that the following identity holds:
\begin{equation}
\varphi\left(\Delta\right)=-i\int_{\mathcal{C}}dz\boldsymbol{\delta}\left(i\left(\Delta-z\right)\right)\varphi\left(z\right),
\end{equation}
for the contour $\mathcal{C}\coloneqq c+i\mathbf{R}$. Consequently,
the integral $\mathcal{I}$ can be written in the form:
\begin{equation}
\mathcal{I}=2\pi^{3}\frac{\Gamma\left(1-\sigma_{1}\right)\Gamma\left(1-\sigma_{2}\right)\Gamma\left(\sigma_{1}+\sigma_{2}-1\right)}{\Gamma\left(\sigma_{1}\right)\Gamma\left(\sigma_{2}\right)\Gamma\left(\sigma_{3}\right)}\boldsymbol{\delta}\left(2-\sigma_{1}-\sigma_{2}-\sigma_{3}\right).
\end{equation}

Finally, by choosing the contour $\mathcal{C}=c+i\mathbf{R}$, where
$c=\text{Re}\left(2-\sum_{i}\sigma_{i}\right)$, we can write the
following identity:
\begin{equation}
\Gamma\left(\sigma_{1}+\sigma_{2}-1\right)\boldsymbol{\delta}\left(2-\sigma_{1}-\sigma_{2}-\sigma_{3}\right)=\Gamma\left(1-\sigma_{3}\right)\boldsymbol{\delta}\left(2-\sigma_{1}-\sigma_{2}-\sigma_{3}\right).
\end{equation}
Thus, we arrive at the final form of the integral:
\begin{equation}
\mathcal{I}=2\pi^{3}\frac{\Gamma\left(1-\sigma_{1}\right)\Gamma\left(1-\sigma_{2}\right)\Gamma\left(1-\sigma_{3}\right)}{\Gamma\left(\sigma_{1}\right)\Gamma\left(\sigma_{2}\right)\Gamma\left(\sigma_{3}\right)}\boldsymbol{\delta}\left(2-\sigma_{1}-\sigma_{2}-\sigma_{3}\right).
\end{equation}
As we see, the triple integral evaluates to a product of gamma functions,
a common structure in worldsheet computations that exhibit conformal
symmetry and factorisation properties of vertex operator correlation
functions.

\subsection{Shadow Transform of Worldsheet Primaries\label{subsec:Shadow-Transform-of}}

In this Appendix, we compute an integral involving the conformal primary
$\Phi^{-1-j_{3}}\left(x_{3};z_{2}\right)$ of the $H^{+}_{3}$-model.
This calculation will prove helpful in deriving the operator product
expansion in Subsection \ref{subsec:Operator-Product-Expansion}.
We define the integral of interest as:
\begin{equation}
\mathcal{I}_{1}\left(j_{1},j_{2},j_{3}\big|x_{3}\right)\coloneqq\int d^{2}z_{2}\widetilde{\Phi}^{-1-j_{3}}\left(x_{3};z_{2}\right)=\int d^{2}z_{1}\int d^{2}z_{2}\frac{1}{\left|z_{12}\right|^{\sigma_{3}}}\Phi^{-1-j_{3}}\left(x_{3};z_{2}\right).\label{eq:Definition-1}
\end{equation}

Before delving into the technical details of the computation, it\textquoteright s
interesting to first observe a geometric interpretation of this integral.
The expression for $\mathcal{I}_{1}$ can be understood as representing
the worldsheet average of the shadow transform of the conformal primary
$\Phi^{-1-j_{3}}\left(x_{3};z_{2}\right)$. Recall that the shadow
transform is defined as:
\begin{equation}
\widetilde{\Phi}^{-1-j_{3}}\left(x_{3};z_{2}\right)\coloneqq\frac{1}{N}\int d^{2}z_{1}\frac{1}{\left|z_{12}\right|^{\sigma_{3}}}\Phi^{-1-j_{3}}\left(x_{3};z_{2}\right).
\end{equation}
Therefore, the integral $\mathcal{I}_{1}$ gives the averaged contribution
of the shadow-transformed conformal primary across the worldsheet.

To proceed with the calculation, we employ the method of Feynman $\alpha$-parameters.
First, we parametrise the complex coordinates as $z_{1}\eqqcolon\xi_{1}+i\xi_{2}$
and $z_{2}\eqqcolon\eta_{1}+i\eta_{2}$. Reorganising Eq. (\ref{eq:Definition-1}),
we obtain:
\begin{equation}
\mathcal{I}_{1}=\frac{1}{N}\int d^{2}z_{1}d^{2}z_{2}\Phi^{-1-j_{3}}\left(x_{3};z_{2}\right)\frac{1}{\Gamma\left(\sigma_{3}/2\right)}\int^{\infty}_{0}d\alpha\alpha^{\sigma_{3}/2-1}e^{-\alpha\left|z_{12}\right|^{2}},
\end{equation}
which simplifies to:
\begin{align}
\mathcal{I}_{1}= & \frac{1}{\Gamma\left(\sigma_{3}/2\right)}\int d\xi_{1}d\xi_{2}\int d\eta_{1}d\eta_{2}\Phi^{-1-j_{3}}\left(x_{3};\eta_{1}+i\eta_{2}\right)\int^{\infty}_{0}d\alpha\alpha^{\sigma_{3}/2-1}e^{-t\left(\xi_{1}-\eta_{1}\right)^{2}-t\left(\xi_{2}-\eta_{2}\right)^{2}}\\
 & =\frac{1}{\Gamma\left(\sigma_{3}/2\right)}\int d\eta_{1}d\eta_{2}\Phi^{-1-j_{3}}\left(x_{3};\eta_{1}+i\eta_{2}\right)\int^{\infty}_{0}d\alpha\alpha^{\sigma_{3}/2-1}\left(\int d\xi_{1}e^{-t\left(\xi_{1}-\eta_{1}\right)^{2}}\right)^{2}.
\end{align}
Performing the Gaussian integration yields:
\begin{equation}
\mathcal{I}_{1}=\frac{\pi}{\Gamma\left(\sigma_{3}/2\right)}\int d\eta_{1}d\eta_{2}\Phi^{-1-j_{3}}\left(x_{3};\eta_{1}+i\eta_{2}\right)\int^{\infty}_{0}d\alpha\alpha^{\left(\sigma_{3}/2-1\right)-1}.
\end{equation}
Using the definition of the generalised Dirac delta function, analytically
continued in the complex plane (as given in Eq. (\ref{eq:Dirac})),
we finally obtain:
\begin{equation}
\mathcal{I}_{1}=\pi^{2}\boldsymbol{\delta}\left(\sigma_{3}-2\right)\int d^{2}z_{2}\Phi^{-1-j_{3}}\left(x_{3};z_{2}\right).
\end{equation}
Substituting $\sigma_{3}=h_{1}+h_{2}-h_{3}$ gives the final expression
for the integral:
\begin{equation}
\mathcal{I}_{1}=\pi^{2}\boldsymbol{\delta}\left(h_{1}+h_{2}+h_{3}-2\right)\int d^{2}z_{2}\Phi^{-1-j_{3}}\left(x_{3};z_{2}\right).\label{eq:Integral-6}
\end{equation}

\subsection{Worldsheet Integral with Two Fixed Points\label{subsec:Worldsheet-Integral-with}}

In this Appendix, we shall undertake the computation of a worldsheet
integral which proves helpful for the evaluation of the operator product
expansion as discussed in Subsection \ref{subsec:Operator-Product-Expansion}:
\begin{equation}
\mathcal{I}_{2}\left(\lambda_{1},\lambda_{2},\lambda_{3}\big|x_{1},x_{2}\right)=\frac{1}{\left|x_{1}-x_{2}\right|^{2\lambda_{3}}}\int d^{2}x_{3}\frac{1}{\left|x_{2}-x_{3}\right|^{2\lambda_{1}}\left|x_{3}-x_{1}\right|^{2\lambda_{2}}}.
\end{equation}

To facilitate the computation, we introduce a translation $x_{3}\mapsto\xi\coloneqq x_{3}-x_{1}$,
so that $\mathcal{I}_{2}$ can be recast in the form:
\begin{equation}
\mathcal{I}_{2}=\frac{1}{\left|x_{1}-x_{2}\right|^{2\lambda_{3}}}\int d^{2}\xi\frac{1}{\left|\left(x_{2}-x_{1}\right)-\xi\right|^{2\lambda_{1}}\left|\xi\right|^{2\lambda_{2}}}.
\end{equation}
Moreover, we implement a rescaling $y\mapsto\xi=\left|x_{2}-x_{1}\right|y$,
which transforms our integral into:
\begin{equation}
\mathcal{I}_{2}=\frac{1}{\left|x_{1}-x_{2}\right|^{2\left(\lambda_{1}+\lambda_{2}+\lambda_{3}-1\right)}}\int d^{2}y\frac{1}{\left|y-\hat{x}\right|^{2\lambda_{1}}\left|y\right|^{2\lambda_{2}}},
\end{equation}
where $\hat{x}\coloneqq\left(x_{2}-x_{1}\right)/\left|x_{2}-x_{1}\right|$.
To proceed with the evaluation of this expression, we focus on the
integral:
\begin{equation}
\mathcal{F}\left(\lambda_{1},\lambda_{2}\right)\coloneqq\int d^{2}y\frac{1}{\left|y-\hat{x}\right|^{2\lambda_{1}}\left|y\right|^{2\lambda_{2}}},
\end{equation}
so that the final result for $\mathcal{I}_{2}$ will be expressed
as:
\begin{equation}
\mathcal{I}_{2}=\frac{\mathcal{F}\left(\lambda_{1},\lambda_{2}\right)}{\left|x_{1}-x_{2}\right|^{2\left(\lambda_{1}+\lambda_{2}+\lambda_{3}-1\right)}}.
\end{equation}

At this stage, we introduce the Feynman $\alpha$-parameters, leading
to the following representations:
\begin{equation}
\frac{1}{\left|y-\hat{x}\right|^{2\lambda_{1}}}=\frac{1}{\Gamma\left(\lambda_{1}\right)}\int^{\infty}_{0}d\alpha_{1}\alpha^{\lambda_{1}-1}_{1}e^{-\alpha_{1}\left|y-\hat{x}\right|^{2}},\,\,\,\frac{1}{\left|y\right|^{2\lambda_{2}}}=\frac{1}{\Gamma\left(\lambda_{2}\right)}\int^{\infty}_{0}d\alpha_{2}\alpha^{\lambda_{2}-1}_{2}e^{-\alpha_{2}\left|y\right|^{2}}.
\end{equation}
We then parametrise $y\eqqcolon y_{1}+iy_{2}$ over the complex plane,
with $\hat{x}\eqqcolon a_{1}+ia_{2}$ ($a^{2}_{1}+a^{2}_{2}=1$),
yielding:
\begin{equation}
\mathcal{F}=\frac{1}{\Gamma\left(\lambda_{1}\right)\Gamma\left(\lambda_{2}\right)}\int^{\infty}_{0}d\alpha_{1}\alpha^{\lambda_{1}-1}_{1}\int^{\infty}_{0}d\alpha_{2}\alpha^{\lambda_{2}-1}_{2}\int dy_{1}dy_{2}e^{-\alpha_{1}\left(y_{1}-a_{1}\right)^{2}-\alpha_{1}\left(y_{2}-a_{2}\right)^{2}-\alpha_{2}y^{2}_{1}-\alpha_{2}y^{2}_{2}}.
\end{equation}
Performing the Gaussian integrals gives:
\begin{equation}
\mathcal{F}=\frac{\pi}{\Gamma\left(\lambda_{1}\right)\Gamma\left(\lambda_{2}\right)}\int^{\infty}_{0}d\alpha_{1}\alpha^{\lambda_{1}-1}_{1}e^{-\alpha_{1}}\int^{\infty}_{0}d\alpha_{2}\alpha^{\lambda_{2}-1}_{2}\frac{1}{\alpha_{1}+\alpha_{2}}e^{\frac{\alpha^{2}_{1}}{\alpha_{1}+\alpha_{2}}}.
\end{equation}
By introducing the variable $\zeta\coloneqq\alpha_{1}/\left(\alpha_{1}+\alpha_{2}\right)$,
we obtain:
\begin{equation}
\mathcal{F}=\frac{\pi}{\Gamma\left(\lambda_{1}\right)\Gamma\left(\lambda_{2}\right)}\int^{1}_{0}d\zeta\zeta^{-\lambda_{2}}\left(1-\zeta\right)^{\lambda_{2}-1}\int^{\infty}_{0}d\alpha_{1}\alpha^{\left(\lambda_{1}+\lambda_{2}-1\right)-1}_{1}e^{-\alpha_{1}\left(1-\zeta\right)}.
\end{equation}
Finally, solving the integral over $\alpha_{1}$ by rescaling $\alpha'_{1}\coloneqq\left(1-\zeta\right)\alpha_{1}$,
we arrive at:
\begin{equation}
\mathcal{F}=\pi\frac{\Gamma\left(\lambda_{1}+\lambda_{2}-1\right)}{\Gamma\left(\lambda_{1}\right)\Gamma\left(\lambda_{2}\right)}\int^{1}_{0}d\zeta\zeta^{-\lambda_{2}}\left(1-\zeta\right)^{-\lambda_{1}}=\pi\frac{\Gamma\left(\lambda_{1}+\lambda_{2}-1\right)}{\Gamma\left(2-\lambda_{1}-\lambda_{2}\right)}B\left(1-\lambda_{1},1-\lambda_{2}\right).
\end{equation}
Thus, the final form of our integral is given by:
\begin{equation}
\mathcal{I}_{2}=\frac{\pi}{\left|x_{1}-x_{2}\right|^{2\left(\lambda_{1}+\lambda_{2}+\lambda_{3}-1\right)}}\frac{\Gamma\left(\lambda_{1}+\lambda_{2}-1\right)}{\Gamma\left(2-\lambda_{1}-\lambda_{2}\right)}\frac{\Gamma\left(1-\lambda_{1}\right)\Gamma\left(1-\lambda_{2}\right)}{\Gamma\left(\lambda_{1}\right)\Gamma\left(\lambda_{2}\right)}.\label{eq:Integral-7}
\end{equation}

\bibliographystyle{revtex-tds/bibtex/bst/revtex/aipnum4-2}
\bibliography{CCFT2}

\end{document}